\documentclass[12pt,preprint]{emulateapj}

\slugcomment{Astrophysical Journal}

\shorttitle{Buried AGNs as a function of galaxy IR luminosity}
\shortauthors{Imanishi}

\begin{document}


\title{Luminous buried AGNs as a function of galaxy infrared
luminosity revealed through Spitzer low-resolution infrared spectroscopy}   


\author{Masatoshi Imanishi\altaffilmark{1}}
\affil{National Astronomical Observatory, 2-21-1, Osawa, Mitaka, Tokyo
181-8588, Japan}
\email{masa.imanishi@nao.ac.jp}

\altaffiltext{1}{Department of Astronomy, School of Science, Graduate
University for Advanced Studies, Mitaka, Tokyo 181-8588}

\begin{abstract}
We present the results of {\it Spitzer} IRS infrared 5--35 $\mu$m 
low-resolution spectroscopic energy diagnostics of ultraluminous infrared
galaxies (ULIRGs) at $z >$ 0.15, classified optically as non-Seyferts.  
Based on the equivalent widths of polycyclic aromatic hydrocarbon
emission and the optical depths of silicate dust absorption features, 
we searched for 
signatures of intrinsically luminous, but optically elusive, buried 
AGNs in these optically non-Seyfert ULIRGs.  
We then combined the results with those of non-Seyfert ULIRGs at $z
<$ 0.15 and non-Seyfert galaxies with infrared luminosities 
L$_{\rm IR}$ $<$ 10$^{12}$L$_{\odot}$. 
We found that the energetic importance of buried AGNs clearly increases
with galaxy infrared luminosity, becoming suddenly discernible
in ULIRGs with L$_{\rm IR}$ $>$ 10$^{12}$L$_{\odot}$.
For ULIRGs with buried AGN signatures, a significant fraction of infrared 
luminosities can be accounted for by
detected buried AGN and modestly-obscured (A$_{\rm V}$ $<$ 20
mag) starburst activity. 
The implied masses of spheroidal stellar components in galaxies for
which buried AGNs become important roughly correspond to the value separating
red massive and blue, less-massive galaxies in the local universe.  
Our results may support the widely-proposed AGN-feedback scenario as the
origin of galaxy downsizing phenomena, where galaxies with currently 
larger stellar masses previously had higher 
AGN energetic contributions and star-formation-originating infrared
luminosities, and have finished their major star-formation more quickly,
due to stronger AGN feedback.
\end{abstract}


\section{Introduction}

Infrared sky surveys have discovered a large number of galaxies that
are bright in the infrared (L$_{\rm IR}$ $>$ 10$^{11}$L$_{\odot}$). They
are called luminous infrared galaxies (LIRGs), or ultraluminous infrared
galaxies (ULIRGs) when the infrared luminosity exceeds L$_{\rm IR}$
$>$ 10$^{12}$L$_{\odot}$ \citep{sam96}. 
The spectral energy distributions of (U)LIRGs are dominated by infrared
emission, which means that luminous energy sources are hidden
behind dust. The bulk of the energetic radiation from the energy sources
is absorbed by the surrounding dust. Then, the heated dust
grains emit this energy as infrared dust emission.  
The energy sources can be nuclear fusion inside rapidly formed stars
(starbursts), the release of gravitational energy produced by mass
accreting supermassive black holes (SMBHs) (i.e., AGN activity), or some
combination. 
The importance of (U)LIRGs to the cosmic energy output
increases rapidly with increasing redshift \citep{lef05,per05,cap07}. 
Thus, understanding the hidden energy sources of (U)LIRGs is closely
related to clarifying the history of star formation and SMBH growth 
in the dust-obscured portion of the universe. 
Since obtaining high quality data of distant ($z >$ 0.5) (U)LIRGs is
not simple with existing observational facilities, detailed studies
of nearby ($z <$ 0.3) (U)LIRGs continue to play an important role in
understanding the properties of the (U)LIRG population in the universe. 

Observations of nearby (U)LIRGs suggest that the properties of
dust-obscured energy sources differ between LIRGs with L$_{\rm IR}$ = 
10$^{11-12}$L$_{\odot}$ and ULIRGs with L$_{\rm IR}$ $>$ 
10$^{12}$L$_{\odot}$. 
First, the fraction of optical Seyferts 
\footnote{
We denote optical Seyferts as luminous AGNs whose accretion disk and
broad line regions are thought to be surrounded by torus-shaped dusty
obscuring medium, responsible for absorbing the nuclear radiation along
some lines of sight.} 
systematically increases with increasing infrared galaxy luminosity, and is
significantly higher in ULIRGs  than LIRGs \citep{vei99,got05}. 
Next, in LIRGs, a large fraction of infrared emission comes from spatially
extended regions, whereas the infrared dust emission from ULIRGs is
dominated by a spatially compact component \citep{soi00,soi01}.  
This suggests that much of the dust in LIRGs is heated by stars distributed
over spatially extended galactic regions, while in ULIRGs, a large
amount of dust is concentrated in the nuclear regions and heated by
spatially compact energy sources. 
Since a mass-accreting SMBH can produce high luminosity from a compact
area ($<$1 pc), an AGN is a plausible source for compact dust emission
in ULIRG nuclei, although very compact starbursts are still
a possibility.

The higher fraction of optical Seyferts and higher nuclear dust
concentration in ULIRGs than LIRGs are difficult to explain
using the scenario that the energetic role of starbursts and AGNs are
similar for LIRGs and ULIRGs.  
In optical Seyferts, lines of sight along the torus axis are relatively
transparent to AGN ionizing radiation, and narrow line regions,
photoionized by the central AGN radiation, should develop at the
10--1000 pc scale, above a torus scale height. 
Since narrow-line regions produce optical emission lines whose flux ratios
differ from those in clouds photoionized by stars, the presence of
such luminous AGNs surrounded by torus-shaped dust is easily
recognizable through optical spectroscopy \citep{vei87,kew06}. 
As a larger amount of dust concentrates in the nuclear regions of
ULIRGs, even the direction of the lowest dust column density can be
opaque to AGN ionizing radiation, blocking the radiation
at the inner part ($<$10 pc) in almost all directions.
Such {\it buried} AGNs lack well-developed narrow-line regions and 
so are classified optically as non-Seyferts \citep{ima07a,ima08}.
Thus, with increasing dust concentration around ULIRG nuclei, it is
expected that the fraction of buried AGNs (= optical non-Seyferts)
increases, while that of optical Seyferts decreases. This is contrary to
observations.  
The increasing fraction of detectable optical Seyferts in ULIRGs can
be explained if the energetic importance of AGNs
intrinsically increases in ULIRGs.
Specifically, even if the narrow-line regions under-develop,  
if the intrinsic AGN luminosities increase, then AGN-signature optical
detection relative to stellar emission becomes
easier, increasing the fraction of optical Seyferts.   
If this is the case, a large fraction of ULIRGs should contain 
luminous, but optically elusive, buried AGNs, and the buried AGN fraction
should increase substantially in ULIRGs, compared to LIRGs. 
Since such luminous buried AGNs can have strong feedback to the
surrounding dust and gas, it is very important to understand buried AGNs
in ULIRGs, not only to unveil the true nature of the ULIRG population, 
but also to observationally constrain the AGN-starburst connections in 
galaxies.

Low-resolution infrared spectroscopy is an effective tool for studying 
buried AGNs that lack well-developed narrow-line regions (or narrow-line
regions are obscured by foreground dust), for the following reasons.
First, emission from polycyclic aromatic hydrocarbons (PAHs), seen in
infrared spectra at $\lambda_{\rm rest}$ = 3--25 $\mu$m in the
rest-frame, can be used to distinguish between a buried AGN and a normal
starburst \citep{gen98,imd00}. In a normal starburst with moderate
metallicity ($>$0.3 solar), consisting of UV-emitting HII regions,
molecular gas and dust, and photo-dissociation regions (PDRs), PAHs are
excited by far-UV photons from stars, and strong PAH emission is
produced in PDRs \citep{sel81,wu06}. 
When stellar energy sources and dust are spatially well mixed, 
the flux of both PAH emission and the nearby continuum are similarly
attenuated, so that the equivalent width of the PAH emission is
insensitive to dust extinction. 
Thus, a normal starburst with PDRs should always show large equivalent
width PAH emission, regardless of the amount of dust extinction. 
In a pure buried AGN, PAHs are destroyed by strong X-ray radiation
from the AGN \citep{voi92,sie04}; thus, no PAH emission is seen.
Instead, a PAH-free continuum from hot, submicron-sized dust grains heated
by the AGN is observed. In a starburst/AGN composite galaxy, PAH
emission is seen if starbursts occur at locations that are sufficiently
shielded from the AGN X-ray radiation.
However, the PAH-equivalent width will be smaller than in a pure
starburst because of the dilution by PAH-free continuum produced by the
AGN. Thus, we can, in principle, disentangle a buried AGN from a normal
starburst based on infrared spectral shapes.

Second, a buried AGN and a normal starburst are distinguishable based
on the optical depths of dust absorption features at different
wavelengths. 
In a normal starburst, the energy sources (stars) and gas/dust are
spatially well mixed \citep{pux91,mcl93,for01}, whereas in a buried AGN
the energy source (= the central accreting SMBH) is very
compact and more centrally concentrated than the surrounding gas and
dust \citep{soi00,im03,sie04} (see also Figures 1a and 1b of Imanishi et
al. 2007). The difference in geometry is reflected in
two features in the observed low-resolution infrared spectra.
First, while the absolute optical depths of dust absorption
features in the 3--10 $\mu$m range cannot exceed a certain threshold 
in a normal starburst with mixed dust/source geometry, they can be
arbitrarily large in a buried AGN \citep{im03,idm06,ima07a,lev07}. 
Second, a buried AGN shows a strong dust temperature
gradient, in which inner dust, closer to the central energy source, has a
higher temperature than outer dust; a normal starburst does not.
The presence of this dust temperature gradient can be investigated 
observationally by comparing the
optical depths of dust absorption features at different infrared 
wavelengths \citep{dud97,ima00,idm06,ima07a}.

Since the PAH emission and dust absorption features are spectrally 
very broad, low-resolution (R = 50--100) infrared spectroscopy is adequate.  
We can thus examine fainter sources than high-resolution (R $>$ 500)
infrared spectroscopy \citep{far07},  in terms of sensitivity. 
\citet{ima07a} performed infrared 5--35 $\mu$m low-resolution 
spectroscopic investigations of ULIRGs at $z <$ 0.15, classified
optically as non-Seyferts, using {\it Spitzer} IRS, 
and found buried AGN signatures in a significant fraction of the
observed ULIRGs.  
However, the infrared luminosities of most of the observed ULIRGs are 
in a narrow range of L$_{\rm IR}$ = 10$^{12-12.3}$L$_{\odot}$,
hampering analysis of buried AGNs as a function of galaxy infrared
luminosity. 
It is known that the fraction of optical Seyferts increases with
ULIRGs with L$_{\rm IR}$ $\geq$ 10$^{12.3}$L$_{\odot}$, compared to those
with L$_{\rm IR}$ $=$ 10$^{12-12.3}$L$_{\odot}$ \citep{vei99}.
A similar analysis of buried AGNs in ULIRGs with L$_{\rm IR}$ $\geq$
10$^{12.3}$L$_{\odot}$ will help elucidate the nature of the ULIRG
population. 

In this paper, we present a systematic, uniform analysis of 
{\it Spitzer} IRS infrared 5--35 $\mu$m low-resolution spectra of ULIRGs
at $z >$ 0.15, optically classified as non-Seyferts. 
By extending our study to ULIRGs at $z >$ 0.15, many ULIRGs with 
L$_{\rm IR}$ $\geq$ 10$^{12.3}$L$_{\odot}$ will be included, enabling 
meaningful comparison of the buried AGN fraction between ULIRGs with 
L$_{\rm IR}$ $=$ 10$^{12-12.3}$L$_{\odot}$ and L$_{\rm IR}$ $\geq$
10$^{12.3}$L$_{\odot}$.  
Throughout this paper, H$_{0}$ $=$ 75 km s$^{-1}$ Mpc$^{-1}$,
$\Omega_{\rm M}$ = 0.3, and $\Omega_{\rm \Lambda}$ = 0.7 are adopted, to
be consistent with our previously published papers. 

\section{Targets}

We selected our targets from the {\it IRAS} 1 Jy sample \citep{kim98}. 
This 1 Jy sample lists 48 ULIRGs at $z >$ 0.15.  Based on the
optical spectral classifications by Veilleux et al. (1999 Table 2), 
33 ULIRGs are classified optically as non-Seyferts (i.e., LINERs, HII
regions, and unclassified), and the remaining 15 ULIRGs are classified
optically as Seyferts. 
Since our primary scientific goal is to study buried AGNs, these 33
optical non-Seyfert ULIRGs are our main targets.
Of the 33 ULIRGs, 15 and 12 ULIRGs are classified optically as
LINERs and HII-regions, respectively, and 6 ULIRGs are optically
unclassified.  
We analyzed the spectra of 10 LINER, 6 HII-region, and 4 unclassified
ULIRGs. 
These 20 (= 10 + 6 + 4) ULIRGs cover $>$60\% of 33 optically non-Seyfert 
ULIRGs at $z >$ 0.15, and should be unbiased in terms of their dominant
energy sources. 
Table 1 summarizes the basic information and {\it IRAS}-based infrared
properties of the observed optically non-Seyfert ULIRGs.  
Although only 8 out of 48 non-Seyfert ULIRGs at $z <$ 0.15 studied by
\citet{ima07a} have L$_{\rm IR}$ $\geq$ 10$^{12.3}$L$_{\odot}$, 17
out of 20 observed non-Seyfert ULIRGs at $z >$ 0.15 show L$_{\rm IR}$ 
$\geq$ 10$^{12.3}$L$_{\odot}$ (Table 1), substantially increasing the
number of ULIRGs with L$_{\rm IR}$ $\geq$ 10$^{12.3}$L$_{\odot}$.

\section{Observations and Data Analysis}

Observations of all 20 ULIRGs were performed using the Infrared
Spectrograph (IRS) \citep{hou04} onboard the Spitzer Space Telescope
\citep{wer04}.  All four modules, Short-Low 2 (SL2; 5.2--7.7 $\mu$m)
and 1 (SL1; 7.4--14.5 $\mu$m), and Long-Low 2 (LL2; 14.0--21.3 $\mu$m) and
1 (LL1; 19.5--38.0 $\mu$m) were used to obtain full 5--35 $\mu$m
low-resolution (R $\sim$ 100) spectra.  
Table 2 details the observing log.
The slit width was 3$\farcs$6 or 2 pixels for SL2 (1$\farcs$8
pixel$^{-1}$) and 3$\farcs$7 or $\sim$2 pixels for SL1 (1$\farcs$8
pixel$^{-1}$).  For LL2 and LL1, the slit widths were 10$\farcs$5 and
10$\farcs$7, respectively, corresponding to $\sim$2 pixels for both
LL2 (5$\farcs$1 pixel$^{-1}$) and LL1 (5$\farcs$1 pixel$^{-1}$).  

The latest pipeline-processed data products at the time of our data
analysis were used.  Frames taken at position A were subtracted from
those taken at position B to remove background emission, mostly the
zodiacal light. Spectra were then extracted in a standard manner. 
Apertures with 4--5 pixels were employed for SL and LL data, 
depending on the spatial extent of individual sources.  
Then, spectra extracted for the A and B positions were summed.
Wavelength calibration was made based on the files of 
the {\it Spitzer} pipeline processed data, named ``b0\_wavsamp.tbl''
and ``b2\_wavsamp.tbl'' for SL and LL, respectively.  These data are believed
to be accurate within 0.1 $\mu$m.  A small level of error in
wavelength calibration will not affect our main conclusions.  Since
emission from all ULIRGs is dominated by spatially compact sources at
the observed wavelength, flux calibration was performed using the 
{\it Spitzer} pipeline processed files ``b0\_fluxcon.tbl'' (SL) and
``b2\_fluxcon.tbl'' (LL).  
For SL1 spectra, data at $\lambda_{\rm obs}$ $>$ 14.5 $\mu$m in the
observed frame are invalid (Infrared Spectrograph Data Handbook Version
1.0) and were discarded.  For LL1 spectra, we used only data at
$\lambda_{\rm obs}$ $<$ 35 $\mu$m because the data point scatter was
large at $\lambda_{\rm obs}$ $>$ 35 $\mu$m and we did not need data at 
$\lambda_{\rm obs}$ $>$ 35 $\mu$m for our scientific discussions.
No defringing was attempted.

For flux calibration, we adopted the values of the pipeline processed
data.  We made no attempt to re-calibrate our spectra using {\it IRAS}
measurements at 12 $\mu$m and 25 $\mu$m, because only
upper limits are provided for {\it IRAS} 12 $\mu$m and/or 25 $\mu$m
photometry in many of the observed ULIRGs (Table 1).
Hence, the absolute flux calibration is dependent on the accuracy of the
pipeline processed data, which is taken to be $<$20\% 
for SL and LL (Infrared Spectrograph Data Handbook).  This level of
flux uncertainty will not significantly affect our main conclusions. 
In fact, in all cases, the {\it Spitzer} IRS 25 $\mu$m flux agrees
within 20\% to, or is smaller than, the {\it IRAS} 25 $\mu$m data.
For ULIRGs with {\it IRAS} non-detection at 25 $\mu$m, the measured
{\it Spizter} IRS 25 $\mu$m flux is always smaller than the 
{\it IRAS} upper limits.

For a fraction of the observed ULIRGs, slight flux discrepancies between
SL1 and LL2 were discernible, ranging from 30\% to 60\%.  When the
discrepancy was present, the SL1 flux (3$\farcs$7 wide slit) was always 
smaller than the LL2 flux (10$\farcs$5). 
In these cases, we adjusted the
smaller SL1 (and SL2) flux to match the larger LL2 flux.
Appropriate spectral binning with 2 or 4 pixels was applied to
reduce the scatter of data points at SL2 (5.2--7.7 $\mu$m) for some
faint ULIRGs, and at $\lambda_{\rm obs}$ $\sim$ 10 $\mu$m for ULIRGs
that display very strong 9.7 $\mu$m silicate dust absorption features.

\section{Results}

Figure 1 presents the infrared 5--35 $\mu$m low-resolution spectra of the
observed 20 ULIRGs. For most of the sources, full 5--35 $\mu$m spectra are
shown here for the first time.  

The spectra in Figure 1 are suitable for displaying the properties of
overall 5--35 $\mu$m spectral shapes and the 9.7 $\mu$m and 18 $\mu$m
silicate dust absorption features. However, 
they are not very useful for PAH emission features.  
Figure 2 presents enlarged spectra at $\lambda_{\rm obs}$ = 5.2--14.5
$\mu$m to better exhibit the properties of the PAH emission features.

\subsection{PAH emission}

The majority of ULIRGs in Figure 2 show clearly detectable PAH
emission features at $\lambda_{\rm rest}$ = 6.2 $\mu$m, 7.7 $\mu$m,
and 11.3 $\mu$m. To estimate the strengths of these PAH emission
features, we adopted a linear continuum, following \citet{ima07a} who made
systematic and detailed analysis of {\it Spitzer} IRS infrared 5--35
$\mu$m low-resolution spectra of optically non-Seyfert ULIRGs at 
$z <$ 0.15.  
For the 6.2 $\mu$m, 7.7 $\mu$m, and 11.3 $\mu$m PAH emission features,
data at $\lambda_{\rm rest}$ = 6.1 $\mu$m and 6.45 $\mu$m, 
7.3 and 8.1 $\mu$m, and 11.0 $\mu$m and 11.6 $\mu$m, were used,
respectively, to determine linear continuum levels, shown as
solid lines in Figure 2.
PAH emission features, above the adopted continuum levels, were fitted
with Gaussian profiles. 
The observed rest-frame equivalent widths (EW$_{\rm PAH}$) and
luminosities of the 6.2 $\mu$m, 7.7 $\mu$m, and 11.3 $\mu$m PAH emission
features, based on our adopted continuum levels, are summarized in Table 3. 
The uncertainties coming from the fittings are unlikely to exceed 30\%.

As noted by \citet{ima07a}, we estimate the strengths of the 7.7 $\mu$m
PAH emission feature in such a way that the uncertainties caused by 
the strong, broad 9.7 $\mu$m silicate dust absorption feature are
minimized. 
Since our continuum definition is significantly different from those
employed in previous papers (e.g., Genzel et al. 1998), readers must be
careful when our value is compared with other estimates in the
literature. For this reason, although the 7.7$\mu$m PAH emission
strength is shown for reference in this paper, it will not play an 
important part in our discussions, which will be based primarily 
on the 6.2 $\mu$m and 11.3 $\mu$m PAH emission strengths.  

\subsection{Silicate absorption}

To estimate the strengths of silicate dust absorption features, 
we used $\tau_{9.7}'$ and $\tau_{18}'$, defined by \citet{ima07a}.
The $\tau_{9.7}'$ value is the optical depth of the 9.7 $\mu$m silicate
absorption feature against a power-law continuum determined from data
points at $\lambda_{\rm rest}$ = 7.1 $\mu$m and 14.2 $\mu$m.  
The $\tau_{18}'$ value is the optical depth of the 18 $\mu$m silicate
absorption feature against a power-law continuum determined from data
points at $\lambda_{\rm rest}$ = 14.2 $\mu$m and 24 $\mu$m.  
These continua are shown as dotted lines in Figure 1. 
Since these continuum levels were determined using data points just outside 
the 9.7 $\mu$m and 18 $\mu$m features, close to the absorption peaks, 
the measured optical depths were well-defined as dips relative to the
nearby continuum emission. 
The $\tau_{9.7}'$ values for all ULIRGs are shown in Table 4 (column 2).
The $\tau_{18}'$ values are also shown in Table 4 (column 3) for ULIRGs
where the 18 $\mu$m silicate feature is clearly seen in absorption.

\subsection{Ice and CO absorption}

A significant fraction of ULIRGs display dips at the shorter
wavelength side of the 6.2 $\mu$m PAH emission feature. 
We ascribe the dips to the 6.0 $\mu$m H$_{2}$O ice absorption feature 
(bending mode). 
Figure 3 presents enlarged spectra at $\lambda_{\rm obs}$ = 5.2--9 $\mu$m
for ULIRGs that show this absorption feature clearly. 
The spectrum of IRAS 12018+1941 is also shown as an example of 
non-detection of this feature.
For ULIRGs with clearly detectable 6.0 $\mu$m H$_{2}$O ice
absorption features, observed optical depths ($\tau_{6.0}$) are
summarized in Table 5. 

The spectrum of IRAS 00397$-$1312 in Figure 3 displays a clear CO
absorption feature ($\lambda_{\rm rest}$ = 4.67 $\mu$m). 
Its optical depth is also shown in Table 5.

\section{Discussion}

To study buried AGNs in optically non-Seyfert ULIRGs at $z >$ 0.15 
and to investigate the buried AGN fraction as a function of ULIRG
infrared luminosity, we use the same criteria as applied to
non-Seyfert ULIRGs at $z <$ 0.15 \citep{ima07a}.   

\subsection{Magnitudes of detected starbursts}

The flux attenuation of continuum emission at $\lambda_{\rm rest}$ $>$ 5
$\mu$m, aside from the strong 9.7 $\mu$m silicate absorption peak, is small ($<$1 mag) for dust extinction with A$_{\rm V}$ $<$ 20 
mag \citep{rie85,lut96}.
Thus, observed PAH emission luminosities can roughly trace the intrinsic
luminosities of modestly-obscured (A$_{\rm V}$ $<$ 20 mag)
PAH-emitting normal starbursts (with PDRs).
Table 3 (columns 8 and 9) tabulates the 6.2 $\mu$m PAH to infrared
luminosity ratio, L$_{\rm 6.2PAH}$/L$_{\rm IR}$, and the 11.3 $\mu$m
PAH to infrared luminosity ratio, L$_{\rm 11.3PAH}$/L$_{\rm IR}$. The
ratios in normal starburst galaxies with modest dust obscuration
(A$_{\rm V}$ $<$ 20 mag) are estimated to be 
L$_{\rm 6.2PAH}$/L$_{\rm IR}$ $\sim$ 3.4 $\times$ 10$^{-3}$ \citep{pee04}
and L$_{\rm 11.3PAH}$/L$_{\rm IR}$ $\sim$ 1.4 $\times$ 10$^{-3}$
\citep{soi02}.  

The observed L$_{\rm 6.2PAH}$/L$_{\rm IR}$ ratios in the
observed ULIRGs (Table 3) range from $<$0.2 $\times$ 10$^{-3}$ to 
1.4 $\times$ 10$^{-3}$, 
or $<$6 \% to 41 \% of the value of 3.4 $\times$ 10$^{-3}$ for
modestly-obscured starburst galaxies. 
In the majority of the observed ULIRGs, the ratios are $<$1 $\times$
10$^{-3}$, or $<$30\% of 3.4 $\times$ 10$^{-3}$. 
Taken at face value, the detected modestly-obscured starbursts in these
ULIRGs can account for $<$6 \% to 41 \% (mostly $<$30\%) of their
infrared luminosities. 
The same argument can be applied to the observed 
L$_{\rm 11.3PAH}$/L$_{\rm IR}$ ratios in ULIRGs. 
The L$_{\rm 11.3PAH}$/L$_{\rm IR}$ ratios are 
(0.09--1.2) $\times$ 10$^{-3}$, or 6--86 \% of the 
value of 1.4 $\times$ 10$^{-3}$ for modestly-obscured starburst
galaxies.
In the majority of observed ULIRGs, the ratios are $<$0.7 $\times$ 
10$^{-3}$ ($<$50 \% of 1.4 $\times$ 10$^{-3}$). 
The observed L$_{\rm 11.3PAH}$/L$_{\rm IR}$ ratios suggest that
detected modestly-obscured starburst activity can account for 
5--86 \% ($<$50 \% in most cases) of the infrared luminosities of ULIRGs.
When we combine the L$_{\rm 6.2PAH}$/L$_{\rm IR}$ and 
L$_{\rm 11.3PAH}$/L$_{\rm IR}$ ratios, we can conclude that the detected
modestly-obscured starburst activity is energetically significant (say,
10--50 \%), but not dominant (say, $>$70--80 \%) in the observed ULIRGs.   
The remaining energy sources in these ULIRGs must therefore be 
(1) highly-obscured (A$_{\rm V}$ $>>$ 20 mag) starbursts, where
the fluxes of PAH emission are substantially attenuated by dust
extinction, and/or (2) buried AGNs that produce strong infrared
radiation, but virtually no PAH emission.

\subsection{ULIRG candidates with luminous buried AGNs}

\subsubsection{Low equivalent widths of PAH emission features}

Whether the dominant energy sources of ULIRG nuclei are 
highly-obscured normal starbursts or buried AGNs can be determined
using the equivalent width of the PAH emission.  Since the PAH
equivalent width (EW$_{\rm PAH}$) must always be large in a normal
starburst (with PDRs) regardless of the amount of dust extinction, a small
EW$_{\rm PAH}$ value suggests contribution from a PAH-free
continuum-emitting energy source, namely an AGN ($\S$1).

Following \citet{ima07a}, we classify ULIRGs with EW$_{\rm 6.2PAH}$  $<$
180 nm, EW$_{\rm 7.7PAH}$ $<$ 230 nm, and EW$_{\rm 11.3PAH}$ $<$ 200 nm
as sources displaying clear signatures of luminous AGNs.  
Since these equivalent width values are less than one-third of the typical
values for starburst galaxies \citep{bra06}, a substantial contribution
from AGN PAH-free continuum emission is indicated.
Table 6 (columns 2--4) presents detection or non-detection of buried
AGN signatures based on the PAH equivalent width threshold. 
The buried AGN fraction is much larger based on the small EW$_{\rm 6.2PAH}$
value (14/20; 70\%) than on the small EW$_{\rm 11.3PAH}$ value (3/20; 15\%). 
This is reasonable, because the 11.3 $\mu$m PAH emission feature is
inside the strong 9.7 $\mu$m silicate dust absorption feature; thus,  
buried AGN continuum emission at $\lambda_{\rm rest}$ $\sim$ 11.3 $\mu$m is  
severely attenuated, not strongly diluting the 11.3 $\mu$m PAH emission
from modestly-obscured starburst regions \citep{ima07a}. 

\subsubsection{Absolute optical depths of dust absorption features}

Based on the EW$_{\rm PAH}$ values, we can easily detect buried AGNs
with very weak starbursts. Even if strong starburst activity is present,
{\it weakly obscured} AGNs are detectable because weakly attenuated
PAH-free continua from the AGNs can dilute the PAH emission
considerably.  However, detecting {\it deeply buried} AGNs with
coexisting strong starbursts is not easy. Even if the
intrinsic luminosities of a buried AGN and surrounding 
less-obscured starbursts are similar, the AGN flux will be more highly
attenuated by dust extinction than the starburst emission, making 
the observed EW$_{\rm PAH}$ values apparently large. 

To determine whether a deeply buried AGN is present in addition to
strong starbursts, we use the optical depths of silicate dust absorption
features. As described in $\S$1 and in \citet{ima07a} in more detail,
these values can be used to distinguish whether the energy sources are
spatially well mixed with dust (a normal starburst), or are more
centrally concentrated than the dust (a buried AGN). 

In a normal starburst with mixed dust/source geometry, observed flux
is dominated by foreground, less-obscured, less-attenuated emission
(which shows only weak dust absorption features), with a small
contribution from highly-obscured, highly-attenuated emission 
at the background  side of the emitting regions (which shows strong dust
absorption features).
Thus, the observed optical depths of dust absorption features cannot
exceed a certain threshold, unless very unusual dust composition
patterns are assumed \citep{im03,idm06,ima07a}. In a buried AGN with 
centrally-concentrated energy source geometry, the 
foreground screen dust model is applicable, and the observed optical 
depths can be arbitrarily large. Hence, detection of strong dust absorption
features, whose optical depths substantially exceed the upper limit
achieved by the mixed dust/source geometry, argues for a foreground
screen dust geometry, as expected from a buried AGN
\citep{im03,idm06,ima07a}.   
\citet{ima07a} obtained a maximum value of $\tau_{9.7}'$ $<$ 1.7 for
a normal starburst with mixed dust/source geometry.
Considering possible uncertainties in the $\tau_{9.7}'$ estimate 
($\sim$10\%), we classify ULIRGs with
$\tau_{9.7}'$ $\geq$ 2 as candidates for harboring luminous
centrally-concentrated buried AGNs. 

In a buried AGN, silicate dust at the very inner part of the obscuring
material, close to the central energy source, can be heated to high
temperature, show silicate emission, and dilute the silicate
absorption feature. 
Although this dilution may have significant effects on the discussion of 
$\tau_{9.7}'$ and  $\tau_{18}'$ for weakly-obscured AGNs \citep{sir08},
it should not be significant in ULIRGs with large $\tau_{9.7}'$ (=
energy sources are highly dust obscured), as studied in this paper. 
Individual ULIRGs that show buried AGN signatures based
on large $\tau_{9.7}'$ values ($>$2; Table 4) are marked with open
circles in Table 6 (column 5).  

We have two notes on this method. 
First, this large $\tau_{9.7}$ method is sensitive to deeply buried AGNs
but obviously misses weakly obscured AGNs, which are more easily
detected with the above low EW$_{\rm PAH}$ method. Hence, this large
$\tau_{9.7}$ method plays a complementary role to the low EW$_{\rm PAH}$
method for the purpose of detecting buried AGN signatures.    
Second, a normal starburst nucleus with mixed dust/source geometry
can produce a large $\tau_{9.7}'$ value, {\it if} it is obscured by
a large amount of foreground screen dust in an edge-on host galaxy 
(Figure 1d of Imanishi et al. 2007). 
Although \citet{ima07a} argued that it is very unlikely that the
majority of ULIRGs with $\tau_{9.7}'$ $>$ 2 correspond to this non-AGN
case, some particular ULIRGs could so correspond.

\subsubsection{Strong dust temperature gradients}

A buried AGN with centrally-concentrated energy source geometry
should show a strong dust temperature gradient, in which inner dust,
close to the central energy source, has a higher temperature than outer
dust, whereas a normal starburst nucleus with mixed dust/source geometry
does not ($\S$1). As explained by \citet{ima07a} in detail, the presence
of a strong dust temperature gradient can be detected by comparing
the optical depths of 9.7 $\mu$m and 18 $\mu$m silicate dust absorption
features because the optical depths of dust absorption features 
at shorter wavelengths probe 
dust column density toward inner hotter dust than those at longer
wavelengths (Figure 2 of Imanishi et al. 2007). 
Following \citet{ima07a}, if an observed $\tau_{18}'$/$\tau_{9.7}'$ 
ratio is substantially smaller than $\tau_{18}'$/$\tau_{9.7}'$ = 0.3,
the most reasonable explanation is the presence of a strong dust
temperature gradient. 
This can provide additional evidence for centrally-concentrated buried
AGNs previously suggested by low EW$_{\rm PAH}$ or large
$\tau_{9.7}'$ values ($\geq$2).  

Table 4 (column 4) summarizes the $\tau_{18}'$/$\tau_{9.7}'$
ratios for ULIRGs showing clear 18 $\mu$m silicate absorption (mostly
$\tau_{9.7}'$ $>$ 2). 
Table 6 (column 6) displays the detection (or non-detection) of buried
AGN signatures, based on this small $\tau_{18}'$/$\tau_{9.7}'$ method. 

\subsubsection{Combination of energy diagnostic methods}

Table 6 (column 7) summarizes the strengths of the detected buried AGN
signatures in {\it Spitzer} IRS 5--35 $\mu$m spectra based on
three methods: (1) low PAH equivalent width; (2) large $\tau_{9.7}'$
value; and (3) small $\tau_{18}'$/$\tau_{9.7}'$ ratio.  
When buried AGN signatures in individual ULIRGs are consistently found
using all or most of these methods, the ULIRGs are classified as 
{\it very strong} buried AGN candidates, marked with open double
circles. When buried AGN signatures are seen only in the first method, or 
first and second methods,
then the ULIRGs are classified as {\it strong} AGN candidates (open
circles). When the signatures are detected only in the second method,
the ULIRGs are classified as {\it possible} buried AGN candidates (open
triangles), as a normal starburst nucleus obscured by foreground
dust in an edge-on host galaxy (Figure 1d of Imanishi et al. 2007) cannot be
ruled out completely in individual cases. 

\subsubsection{Absorption-corrected intrinsic luminosities of buried AGNs}

For ULIRGs that show buried AGN signatures and small PAH
equivalent widths, the observed fluxes are mostly ascribed to
AGN-heated PAH-free dust continuum emission.  
For these ULIRGs, we can estimate the absorption-corrected intrinsic
dust emission
luminosity at $\sim$10 $\mu$m ($\nu$F$_\nu$) heated by the AGN, based on
the observed fluxes at $\lambda_{\rm rest}$ $\sim$ 10 $\mu$m and the
dust extinction toward the 10 $\mu$m continuum emitting regions inferred
from $\tau_{9.7}'$ \citep{ima07a}.  
In a buried AGN with a strong dust temperature gradient, assuming a 
simple spherical dust distribution, dust emission luminosity is
conserved at each temperature from hot inside regions to cool outside
regions. 
Namely, the intrinsic luminosity of inner hot dust emission at 10 $\mu$m
($\nu$F$_\nu$) should be comparable to that of outer cool dust emission
at 60 $\mu$m, the wavelength which dominates the observed infrared
emission of ULIRGs \citep{san88a}.
Hence, from AGN-originating intrinsic $\nu$F$_\nu$(10 $\mu$m) values, we can
quantitatively estimate the energetic contribution from buried AGNs to
the infrared luminosities of ULIRGs.

Following \citet{ima07a}, we assume that $\tau_{9.7}'$ and the
extinction at $\lambda_{\rm rest}$ = 8 or 13 $\mu$m continuum just
outside the 9.7 $\mu$m silicate feature (A$_{\rm cont}$) are related
to $\tau_{9.7}'$/A$_{\rm cont}$ $\sim$ 2.3 \citep{rie85}. 
Based on a foreground screen dust absorption model applicable
to buried AGNs, we obtain, as seen in Table 7, absorption-corrected
intrinsic AGN luminosities for selected ULIRGs with low PAH equivalent
widths. 
The flux attenuation of the 8 or 13 $\mu$m continuum outside
the 9.7 $\mu$m silicate feature ranges from a factor of 1.7 (IRAS
12018+1941; $\tau_{9.7}'$ $\sim$ 1.3) to 3.1 (IRAS 04313$-$1649; 
$\tau_{9.7}'$ $\sim$ 2.8).  The absorption-corrected intrinsic AGN luminosities
could explain a significant fraction (15--60\%) of the luminosities of
these ULIRGs (Table 7). 

\subsection{Buried AGN fraction as a function of galaxy infrared
luminosity} 

Optically non-Seyfert ULIRGs at $z <$ 0.15 studied by
\citet{ima07a} mostly display L$_{\rm IR}$ $=$
10$^{12-12.3}$L$_{\odot}$.
The extension of our {\it Spitzer} low-resolution infrared
spectroscopic energy diagnostic to ULIRGs at $z >$ 0.15 gives a
large number ULIRGs with L$_{\rm IR}$ $\geq$ 10$^{12.3}$L$_{\odot}$.  
Specifically, only 8 of 48 observed non-Seyfert ULIRGs at $z <$ 0.15
have L$_{\rm IR}$ $\geq$ 10$^{12.3}$L$_{\odot}$ (Table 1 of Imanishi
et al. 2007), while 17 of 20 observed non-Seyfert ULIRGs at $z >$ 0.15
have L$_{\rm IR}$ $\geq$ 10$^{12.3}$L$_{\odot}$ (Table 1). We can
thus investigate the buried AGN fraction, separating ULIRGs into two 
categories: those with L$_{\rm IR}$ $=$ 10$^{12-12.3}$L$_{\odot}$ 
and those with L$_{\rm IR}$ $\geq$
10$^{12.3}$L$_{\odot}$.  

Figure 4 shows the distribution of EW$_{\rm 6.2PAH}$, EW$_{\rm 11.3PAH}$, 
and $\tau_{9.7}'$ as a function of galaxy infrared luminosity. 
In addition to ULIRGs, optically non-Seyfert galaxies with L$_{\rm IR}$ 
$<$ 10$^{12}$L$_{\odot}$ \citep{bra06} are plotted.
It is evident that in all plots, the fraction of galaxies which meet the
requirement of buried AGNs increases with increasing galaxy infrared
luminosity.   

To investigate the detectable buried AGN fraction as a function of
galaxy infrared luminosity in more detail, we use the combined method
($\S$5.2.4).  
Based on the classification of non-Seyfert ULIRGs at $z >$
0.15 in Table 6, 13 out of 17 ULIRGs with L$_{\rm IR}$ $\geq$
10$^{12.3}$L$_{\odot}$ show strong buried AGN signatures (open double
circles and open circles in column 7 of Table 6), in contrast to only 1 
out of 3 ULIRGs with L$_{\rm IR}$ $=$ 10$^{12-12.3}$L$_{\odot}$.
For non-Seyfert ULIRGs at $z <$ 0.15, 5 out of 8 ULIRGs with L$_{\rm
IR}$ $\geq$ 10$^{12.3}$L$_{\odot}$ show strong buried AGN signatures,
in contrast to only 11 out of 40 ULIRGs with L$_{\rm IR}$ $=$
10$^{12-12.3}$L$_{\odot}$ \citep{ima07a}.
When we combine these results, we obtain a fraction of strong
buried AGN signatures of 18/25 (72\%) for ULIRGs with 
L$_{\rm IR}$ $\geq$ 10$^{12.3}$L$_{\odot}$ and 12/43 (28\%) for
ULIRGs with L$_{\rm IR}$ $=$ 10$^{12-12.3}$L$_{\odot}$. 
When we include sources with {\it possible} buried AGN signatures 
(open triangles in Table 6 of this paper and Imanishi et al. 2007), 
the fraction of detectable buried AGNs signatures is 
22/25 (88\%) for ULIRGs with L$_{\rm IR}$ $\geq$
10$^{12.3}$L$_{\odot}$ and 19/43 (44\%) for ULIRGs with L$_{\rm IR}$ $=$
10$^{12-12.3}$L$_{\odot}$. 
For optically non-Seyfert galaxies with L$_{\rm IR}$ $<$ 
10$^{12}$L$_{\odot}$, although the sample size is limited, 
no sources are classified as buried AGNs in our criteria (Figure 4).  
Other various observations also suggest that the energetic 
importance of buried AGNs clearly decreases in galaxies with 
L$_{\rm IR}$ $<$ 10$^{12}$L$_{\odot}$, compared to ULIRGs 
\citep{soi00,soi01}.
Therefore, we clearly see the trend that {\it the detectable buried AGN
fraction increases with infrared luminosity of optically non-Seyfert
galaxies} (Figure 5), as has previously been suggested from 
{\it AKARI} infrared 2.5--5 $\mu$m low-resolution spectroscopy 
\citep{ima08}. 
Given that the fraction of ULIRGs with optical Seyfert signatures 
also increases with increasing galaxy infrared luminosity 
\citep{vei99,got05}, we can conclude that 
{\it AGN activity becomes more important with
increasing galaxy infrared luminosity}. 

The so-called galaxy downsizing phenomenon has recently been proposed;
it was found that galaxies with currently larger stellar 
masses finished their major star-formation in an earlier cosmic age
\citep{cow96,bun05}.   
AGN feedback is suggested to be responsible for the galaxy downsizing
phenomenon \citep{gra04,bow06,cro06}. 
Namely, in galaxies with currently large stellar masses, AGN feedback
was stronger in the past, heating or expelling gas in host galaxies and 
stopping star formation on a shorter time scale. 
Buried AGNs can have particularly strong feedback because the AGNs are
surrounded by a large amount of nuclear gas and dust.
In addition, galaxies with currently larger stellar masses should have
had higher star-formation-originating infrared luminosities in the past,
as more stars were formed.

We found that the buried AGN fraction increases with increasing galaxy
infrared luminosity. 
In ULIRGs with L$_{\rm IR}$ $>$ 10$^{12}$L$_{\odot}$, the importance of
buried AGNs suddenly becomes clear, compared to galaxies with 
L$_{\rm IR}$ $<$ 10$^{12}$L$_{\odot}$ (Imanishi et
al. 2007; this paper), suggesting that AGN feedback becomes significant
in ULIRGs.  
The absorption-corrected intrinsic luminosities of detected buried AGNs 
are $>$ a few $\times$ 10$^{45}$ ergs s$^{-1}$
($\gtrsim$10$^{12}$L$_{\odot}$), which could account for a significant
(15--100\%) fraction of ULIRG infrared luminosities (Table 7 of this paper 
and Imanishi et al. 2007).  
The infrared luminosities of detected modestly-obscured starbursts 
(Table 7, columns 3 and 4) are also a few $\times$ 10$^{45}$ ergs
s$^{-1}$, or $\sim$10$^{12}$L$_{\odot}$, which are still higher than
the total infrared luminosities of galaxies with 
L$_{\rm IR}$ $<$ 10$^{12}$L$_{\odot}$.
Thus, both {\it stronger AGN feedback} and {\it higher
star-formation-originating infrared luminosities} are suggested in
ULIRGs 
\footnote{The summed luminosities of detected buried AGNs and
modestly-obscured starbursts are generally smaller than the observed
infrared luminosities of ULIRGs. 
This discrepancy may be due to (1) possible underestimation of 
absorption-corrected intrinsic buried AGN luminosity, caused by a 
small starburst contribution to the observed infrared flux 
($\S$5.2.7 of Imanishi et al. 2007),
or by dust extinction curves in ULIRGs that differ from our assumption, 
or (2) the presence of 
highly dust-obscured ($A_{\rm V}$ $>>$ 20 mag) starbursts, or 
(3) intrinsically low PAH to infrared luminosity ratio in starburst
activity in ULIRGs, or some combination of factors.  
}
than in lower L$_{\rm IR}$ galaxies. 

\citet{kau03} found that galaxies above and below stellar masses of 
M$_{*}$ = 3 $\times$ 10$^{10}$M$_{\odot}$ show distinctly
different properties.
Galaxies with M$_{*}$ $>$ 3 $\times$ 10$^{10}$M$_{\odot}$ are dominated
by red galaxies with currently low star-formation rates, while those with
M$_{*}$ $<$ 3 $\times$ 10$^{10}$M$_{\odot}$ are mainly blue galaxies
with ongoing active star-formation.
If AGN feedback is responsible for the dichotomy, then
buried AGNs may become important in galaxies with  M$_{*}$ $>$ 3 $\times$
10$^{10}$M$_{\odot}$. 
Based on the velocity dispersion measurements of ULIRG host galaxies in
the infrared, \citet{das06} argued that ULIRGs will have spheroidal
stellar masses with several $\times$ 10$^{10}$M$_{\odot}$.  
Assuming the Eddington luminosity for detected buried AGNs 
($>$ a few $\times$ 10$^{45}$ ergs s$^{-1}$), SMBH masses are estimated
to be a few $\times$ 10$^{7}$M$_{\odot}$.  
If we adopt the correlation between the masses of SMBH and spheroidal
stars established in the local universe \citep{mag98}, similar
spheroidal stellar masses with $>$ a few
$\times$ 10$^{10}$M$_{\odot}$ are obtained.
These estimated spheroidal stellar masses of ULIRGs are similar to the
value separating red massive (= strong AGN feedbacks) and blue,
less-massive galaxies (= week AGN feedbacks). 
In summary, our discovery of increasing buried AGN fraction with 
galaxy infrared luminosity may observationally support the
widely-proposed AGN feedback scenario as the origin of the galaxy
downsizing phenomenon. 

\section{Summary}

We presented the results of {\it Spitzer} IRS infrared 5--35 $\mu$m
low-resolution spectroscopic energy diagnostic method of optically
non-Seyfert ULIRGs at $z >$ 0.15. 
The signatures of intrinsically luminous, but optically elusive, buried
AGNs were searched for, based on the  equivalent widths of PAH emission
and the optical depths of silicate dust absorption features in these
ULIRGs. Since most of the ULIRGs at $z >$ 0.15 have L$_{\rm IR} \geq$
10$^{12.3}$L$_{\odot}$, by combining results with our previous 
analysis of ULIRGs at $z <$ 0.15, we could investigate the detectable
buried AGN fraction by separating ULIRGs with L$_{\rm IR}$ = 
10$^{12-12.3}$L$_{\odot}$ and L$_{\rm IR}$ $\geq$ 10$^{12.3}$L$_{\odot}$.
We found that the detectable buried AGN fraction is clearly higher in
ULIRGs with L$_{\rm IR}$ $\geq$ 10$^{12.3}$L$_{\odot}$ than those with  
L$_{\rm IR}$ = 10$^{12-12.3}$L$_{\odot}$. 
Given that (1) the fraction of optical Seyferts is also higher in ULIRGs
with L$_{\rm IR}$ $\geq$ 10$^{12.3}$L$_{\odot}$ than in those with 
L$_{\rm IR}$ $=$ 10$^{12-12.3}$L$_{\odot}$, and (2) signatures of AGNs,
including both buried AGNs and optical Seyferts, are weaker in galaxies 
with L$_{\rm IR}$ $<$ 10$^{12}$L$_{\odot}$ than ULIRGs, we 
concluded that AGN importance increases as galaxy infrared luminosity 
increases.
Buried AGNs become clearly discernible in ULIRGs with implied stellar
masses of M$_{*}$ $>$ a few $\times$ 10$^{10}$M$_{\odot}$, the value
that separates red massive and blue, less-massive galaxies in the nearby 
universe. 
Our overall results support the AGN-feedback scenario as the origin
of the galaxy downsizing phenomenon.

\acknowledgments

We thank the anonymous referee for his/her very useful comments.
This work is based on observations made with the 
Spitzer Space Telescope, operated by the Jet Propulsion
Laboratory, California Institute of Technology under a contract with
NASA. Support for this work was provided by NASA and also by an award
issued by JPL/Caltech.  
M.I. is supported by Grants-in-Aid for Scientific Research
(19740109). 
This research made use of the SIMBAD database, operated at
CDS, Strasbourg, France, and the NASA/IPAC Extragalactic Database
(NED), which is operated by the Jet Propulsion Laboratory, California
Institute of Technology, under contract with NASA.

\begin{deluxetable}{lcrrrrcrc}
\tabletypesize{\scriptsize}
\tablecaption{Observed ULIRGs at $z >$ 0.15 and their {\it IRAS}-based
infrared emission properties
\label{tbl-1}}
\tablewidth{0pt}
\tablehead{
\colhead{Object} & \colhead{Redshift}   & 
\colhead{f$_{\rm 12}$}   & 
\colhead{f$_{\rm 25}$}   & 
\colhead{f$_{\rm 60}$}   & 
\colhead{f$_{\rm 100}$}  & 
\colhead{log L$_{\rm IR}$} & 
\colhead{f$_{25}$/f$_{60}$} & 
\colhead{Optical}   \\
\colhead{} & \colhead{}   & \colhead{(Jy)} & \colhead{(Jy)} 
& \colhead{(Jy)} & \colhead{(Jy)}  & \colhead{L$_{\odot}$} & \colhead{}
& \colhead{Class}   \\
\colhead{(1)} & \colhead{(2)} & \colhead{(3)} & \colhead{(4)} & 
\colhead{(5)} & \colhead{(6)} & \colhead{(7)} & \colhead{(8)} & 
\colhead{(9)}
}
\startdata
IRAS 03521+0028 & 0.152 & $<$0.11 & 0.20 & 2.52 & 3.62 & 12.5 & 0.08 (C) & LINER \\  
IRAS 09463+8141 & 0.156 & $<$0.07 & $<$0.07 & 1.43 & 2.29 & 12.3 & $<$0.05 (C)  & LINER  \\  
IRAS 10091+4704 & 0.246 & $<$0.06 & $<$0.08 & 1.18 & 1.55 & 12.6 & $<$0.07 (C) & LINER \\  
IRAS 11028+3130 & 0.199 & $<$0.09 & 0.09 & 1.02 & 1.44 & 12.4 & 0.09 (C) & LINER \\  
IRAS 11180+1623 & 0.166 & $<$0.08 & $<$0.19 & 1.19 & 1.60 & 12.2 & $<$0.16 (C) & LINER \\  
IRAS 11582+3020 & 0.223 & $<$0.10 & $<$0.15 & 1.13 & 1.49 & 12.5 & $<$0.14 (C) & LINER \\
IRAS 12032+1707 & 0.217 & $<$0.14 & 0.25 & 1.36 & 1.54 & 12.6 & 0.18 (C) & LINER \\
IRAS 16300+1558 & 0.242 & $<$0.07 & 0.07 & 1.48 & 1.99 & 12.7 & 0.05 (C) & LINER \\
IRAS 16333+4630 & 0.191 & $<$0.06 & 0.06 & 1.19 & 2.09 & 12.4 & 0.05 (C) & LINER \\  
IRAS 23129+2548 & 0.179 & $<$0.08 & 0.08 & 1.81 & 1.64 & 12.4 & 0.04 (C) & LINER \\  \hline
IRAS 00397$-$1312 & 0.261 & 0.14  & 0.33 & 1.83 & 1.90 & 12.9 & 0.18 (C) & HII-region \\   
IRAS 01199$-$2307 & 0.156 & $<$0.11 & $<$0.16 & 1.61 & 1.37 & 12.3 & $<$0.1 (C) & HII-region \\   
IRAS 01355$-$1814 & 0.192 & $<$0.06 & 0.12 & 1.40 & 1.74 & 12.4 & 0.09 (C) & HII-region  \\  
IRAS 08201+2801 & 0.168 & $<$0.09 & 0.15 & 1.17 & 1.43 & 12.3 & 0.13 (C) & HII-region  \\  
IRAS 13469+5833 & 0.158 & $<$0.05 & 0.04 & 1.27 & 1.73 & 12.2 & 0.03 (C) & HII-region \\  
IRAS 17068+4027 & 0.179 & $<$0.08 & 0.12 & 1.33 & 1.41 & 12.3 & 0.09 (C) & HII-region \\  \hline
IRAS 01494$-$1845 & 0.158 & $<$0.08 & $<$0.15 & 1.29 & 1.85 & 12.2 & $<$0.12 (C) & unclassified \\  
IRAS 04313$-$1649 & 0.268 & $<$0.07 & 0.07 & 1.01 & 1.10 & 12.6 & 0.07 (C) & unclassified \\  
IRAS 10035+2740 & 0.165 & $<$0.14 & $<$0.17 & 1.14 & 1.63 & 12.3 & $<$0.15 (C) & unclassified \\ 
IRAS 12018+1941 & 0.168 & $<$0.11 & 0.37 & 1.76 & 1.78 & 12.5 & 0.21 (W) & unclassified \\ \hline
\enddata

\tablecomments{
Col.(1): Object name.  
Col.(2): Redshift.
Col.(3)--(6): f$_{12}$, f$_{25}$, f$_{60}$, and f$_{100}$ are 
{\it IRAS} fluxes at 12 $\mu$m, 25 $\mu$m, 60 $\mu$m, and 100 $\mu$m,
respectively, taken from \citet{kim98}.
Col.(7): Decimal logarithm of infrared (8$-$1000 $\mu$m) luminosity
in units of solar luminosity (L$_{\odot}$), calculated with
$L_{\rm IR} = 2.1 \times 10^{39} \times$ D(Mpc)$^{2}$
$\times$ (13.48 $\times$ $f_{12}$ + 5.16 $\times$ $f_{25}$ +
$2.58 \times f_{60} + f_{100}$) ergs s$^{-1}$ \citep{sam96}.
Since the calculation is based on our adopted cosmology, the infrared
luminosities differ slightly ($<$10\%) from the values shown in Kim \&
Sanders (1998, Table 1, column 15).  
For sources with upper limits in some {\it IRAS} bands, 
we can derive upper and lower limits for infrared luminosity,
assuming that the actual flux is the {\it IRAS}-upper limit and zero
value, respectively.  
The difference in the upper and lower values is usually very small, less
than 0.2 dex.
We assume that the infrared luminosity is the average of these values. 
Col.(8): {\it IRAS} 25 $\mu$m to 60 $\mu$m flux ratio.
ULIRGs with f$_{25}$/f$_{60}$ $<$ 0.2 and $>$ 0.2 are
classified as cool and warm sources (denoted as ``C'' and ``W''),
respectively \citep{san88b}.
Col.(9): Optical spectral classification by \citet{vei99}.
}

\end{deluxetable}

\begin{deluxetable}{lclcccc}
\tabletypesize{\small}
\tablecaption{{\it Spitzer} IRS observing log
\label{tbl-2}}
\tablewidth{0pt}
\tablehead{
\colhead{Object} & \colhead{PID} &\colhead{Date} & \colhead{Integration
time [sec]} \\  
\colhead{} & \colhead{} & \colhead{[UT]} & \colhead{SL2} & \colhead{SL1} &
\colhead{LL2} & \colhead{LL1} \\ 
\colhead{(1)} & \colhead{(2)} & \colhead{(3)} & \colhead{(4)} & 
\colhead{(5)} & \colhead{(6)} & \colhead{(7)} 
}
\startdata 
IRAS 03521+0028 & 105 & 2004 Feb 27 & 240 & 240 & 180 & 180 \\
IRAS 09463+8141 & 105 & 2004 Mar 23 & 240 & 240 & 120 & 120 \\
IRAS 10091+4704 & 105 & 2004 Apr 19 & 240 & 240 & 120 & 120 \\
IRAS 11028+3130 & 30407 & 2007 Jun 9 & 240 & 240 & 180 & 180 \\
IRAS 11180+1623 & 30407 & 2007 Jun 8 & 196 & 196 & 180 & 180 \\
IRAS 11582+3020 & 105 & 2005 Dec 16 & 240 & 240 & 120 & 120 \\
IRAS 12032+1707 & 105 & 2004 Jan 4  & 240 & 240 & 120 & 120 \\
IRAS 16300+1558 & 105 & 2005 Aug 13 & 240 & 240 & 300 & 300 \\
IRAS 16333+4630 & 105 & 2004 Mar 4  & 240 & 240 & 120 & 120 \\
IRAS 23129+2548 & 105 & 2003 Dec 17 & 360 & 360 & 300 & 300 \\
IRAS 00397$-$1312 & 105 & 2004 Jan 4 & 240 & 240 & 180 & 180 \\
IRAS 01199$-$2307 & 105 & 2004 Jul 18 & 240 & 240 & 180 & 180 \\
IRAS 01355$-$1814 & 105 & 2005 Jul 10 & 240 & 240 & 120 & 120 \\
IRAS 08201+2801 & 30407 & 2007 May 3 & 196 & 196 & 180 & 180 \\  
IRAS 13469+5833 & 20589 & 2006 Apr 26 & 240 & 112 & 480 & 960 \\
IRAS 17068+4027 & 105 & 2004 Apr 16 & 240 & 240 & 180 & 180 \\
IRAS 01494$-$1845 & 105 & 2005 Jul 14 & 240 & 240 & 120 & 120 \\
IRAS 04313$-$1649 & 105 & 2004 Mar 1  & 240 & 240 & 300 & 300 \\
IRAS 10035+2740 & 30407 & 2007 Jun 8 & 196 & 196 & 180 & 180 \\
IRAS 12018+1941 & 105 & 2004 May 15 & 120 & 120 & 180 & 180 \\ \hline
\enddata

\tablecomments{
Col.(1): Object name.
Col.(2): PID number. 
Col.(3): Observing date in UT. 
Col.(4): Net on-source integration time for SL2 spectroscopy in sec.
Col.(5): Net on-source integration time for SL1 spectroscopy in sec.
Col.(6): Net on-source integration time for LL2 spectroscopy in sec
Col.(7): Net on-source integration time for LL1 spectroscopy in sec.
}

\end{deluxetable}

\begin{deluxetable}{lccccccccc}
\tabletypesize{\scriptsize}
\tablecaption{Observed properties of PAH emission features 
\label{tbl-3}}
\tablewidth{0pt}
\tablehead{
\colhead{Object} & \colhead{EW$_{\rm 6.2PAH}$} & 
\colhead{EW$_{\rm 7.7PAH}$ \tablenotemark{a}} & 
\colhead{EW$_{\rm 11.3PAH}$}  & \colhead{L$_{\rm 6.2PAH}$} & 
\colhead{L$_{\rm 7.7PAH}$ \tablenotemark{a}} &
\colhead{L$_{\rm 11.3PAH}$} & \colhead{L$_{\rm 6.2PAH}$/L$_{\rm IR}$} & 
\colhead{L$_{\rm 11.3PAH}$/L$_{\rm IR}$} \\
\colhead{} & \colhead{[nm]} & \colhead{[nm]} & \colhead{[nm]} &
\colhead{10$^{42}$ [ergs s$^{-1}$]} & \colhead{10$^{42}$ [ergs s$^{-1}$]} & 
\colhead{10$^{42}$ [ergs s$^{-1}$]} & \colhead{[$\times$ 10$^{-3}$]} & 
\colhead{[$\times$ 10$^{-3}$]} \\  
\colhead{(1)} & \colhead{(2)} & \colhead{(3)} & \colhead{(4)} &
\colhead{(5)} & \colhead{(6)} & \colhead{(7)} & \colhead{(8)} & 
\colhead{(9)} 
}
\startdata 
IRAS 03521+0028 & 245 & 855 & 450 & 6.1 & 24.3 & 5.2 & 0.5 & 0.4 \\
IRAS 09463+8141 & 225 & 765 & 530 & 2.9 & 11.3 & 1.7 & 0.4 & 0.3 \\
IRAS 10091+4704 & $<$50 & 480 & 560 & $<$1.7 & 29.1 & 3.5 & $<$0.2 & 0.2\\
IRAS 11028+3130 & 160 & 870 & 460 & 1.9 & 12.1 & 1.7 & 0.2 & 0.2 \\
IRAS 11180+1623 & 165 & 855 & 340 & 4.1 & 25.6 & 2.7 & 0.6 & 0.4 \\
IRAS 11582+3020 & 70 & 235 & 325  & 6.4 & 63.1 & 5.9 & 0.5 & 0.5 \\
IRAS 12032+1707 & 70 & 195 & 315  & 5.2 & 26.1 & 5.9 & 0.3 & 0.4 \\
IRAS 16300+1558 & 85 & 410 & 400  & 7.4 & 40.6 & 6.0 & 0.4 & 0.3 \\
IRAS 16333+4630 & 275 & 680 & 485 & 13.1 & 42.6 & 11.6 & 1.4 & 1.2 \\
IRAS 23129+2548 & 110 & 320 & 330 & 5.9 & 30.8 & 4.1 & 0.6 & 0.4 \\
IRAS 00397$-$1312 & 25 & 160 & 35 & 21.4 & 135.5 & 2.9 & 0.6 & 0.09 \\
IRAS 01199$-$2307 & 130 & 440 & 215 & 4.3 & 21.8 & 1.9 & 0.6 & 0.3 \\
IRAS 01355$-$1814 & 170 & 605 & 195 & 4.2 & 19.1 & 1.7 & 0.4 & 0.2 \\
IRAS 08201+2801 & 105 & 425 & 280 & 5.5 & 33.8 & 4.2 & 0.8 & 0.6 \\  
IRAS 13469+5833 & 235 & 675 & 460 & 4.0 & 15.7 & 3.4 & 0.7 & 0.6 \\
IRAS 17068+4027 & 110  & 405 & 185 & 8.5 & 40.1 & 5.0 & 1.0 & 0.6 \\
IRAS 01494$-$1845 & 365 & 810 & 485 & 8.0 & 25.0 & 5.0 & 1.3 & 0.8 \\
IRAS 04313$-$1649 & $<$120 & 305 & 240 & $<$3.2 & 15.3 & 2.1 & $<$0.3 & 0.1 \\
IRAS 10035+2740 & 660 & 900 & 275 & 5.0 & 9.8 & 1.1 & 0.8 & 0.2 \\
IRAS 12018+1941 & 120 & 220 & 35  & 9.3 & 24.5 & 2.3 & 0.9 & 0.2 \\ \hline
\enddata

\tablenotetext{a}{We regard flux excess at $\lambda_{\rm rest}$ =
7.3--8.1 $\mu$m above an adopted continuum level as 7.7 $\mu$m PAH 
emission, to reduce the effects of the strong 9.7 $\mu$m silicate dust
absorption feature. 
Our definition is different from those presented in previous papers.}

\tablecomments{
Col.(1): Object name.  
Col.(2): Rest-frame equivalent width of the 6.2 $\mu$m PAH emission.
Col.(3): Rest-frame equivalent width of the 7.7 $\mu$m PAH emission.
Col.(4): Rest-frame equivalent width of the 11.3 $\mu$m PAH emission.
Col.(5): Luminosity of the 6.2 $\mu$m PAH emission in units 
of 10$^{42}$ ergs s$^{-1}$.
Col.(6): Luminosity of the 7.7 $\mu$m PAH emission in units 
of 10$^{42}$ ergs s$^{-1}$.
Col.(7): Luminosity of the 11.3 $\mu$m PAH emission in units 
of 10$^{42}$ ergs s$^{-1}$.
Col.(8): The 6.2 $\mu$m PAH to infrared luminosity ratio in units of 
10$^{-3}$. 
The ratio for normal starbursts with modest dust obscuration 
(A$_{\rm V}$ $<$ 20 mag) is $\sim$3.4 $\times$ 10$^{-3}$ \citep{pee04}.
Col.(9): The 11.3 $\mu$m PAH to infrared luminosity ratio in units of 
10$^{-3}$.
The ratio for normal starbursts with modest dust obscuration 
(A$_{\rm V}$ $<$ 20 mag) is $\sim$1.4 $\times$ 10$^{-3}$ \citep{soi02}.
}

\end{deluxetable}

\begin{deluxetable}{lccc}
\tabletypesize{\scriptsize}
\tablecaption{9.7 $\mu$m and 18 $\mu$m silicate
dust absorption feature optical depths
\label{tbl-4}}
\tablewidth{0pt}
\tablehead{
\colhead{Object} & \colhead{$\tau_{9.7}'$} & \colhead{$\tau_{18}'$} & 
\colhead{$\tau_{18}'$/$\tau_{9.7}'$}  \\
\colhead{(1)} & \colhead{(2)} & \colhead{(3)} & \colhead{(4)} 
}
\startdata 
IRAS 03521+0028 & 1.3 & 0.4 & 0.31 \\
IRAS 09463+8141 & 2.0 & 0.6 & 0.30 \\
IRAS 10091+4704 & 2.5 & 1.0 & 0.40 \\
IRAS 11028+3130 & 2.5 & 1.1 & 0.44 \\
IRAS 11180+1623 & 2.0 & 0.6 & 0.30 \\
IRAS 11582+3020 & 2.7 & 0.8 & 0.30 \\
IRAS 12032+1707 & 2.6 & 0.6 & 0.23 \\
IRAS 16300+1558 & 2.6 & 0.9 & 0.35 \\
IRAS 16333+4630 & 1.3 & \nodata & \nodata \\
IRAS 23129+2548 & 2.6 & 0.9 & 0.35 \\
IRAS 00397$-$1312 & 2.7 & 0.6 & 0.22 \\
IRAS 01199$-$2307 & 2.4 & 0.7 & 0.29 \\
IRAS 01355$-$1814 & 2.4 & 0.7 & 0.29 \\
IRAS 08201+2801 & 2.2 & 0.5 & 0.23 \\  
IRAS 13469+5833 & 1.7 & 0.5 & 0.29 \\
IRAS 17068+4027 & 1.8 & 0.5 & 0.28 \\
IRAS 01494$-$1845 & 1.6 & \nodata & \nodata \\
IRAS 04313$-$1649 & 2.8 & 0.7 & 0.25 \\
IRAS 10035+2740 & 2.0 & 0.7 & 0.35 \\
IRAS 12018+1941 & 1.3 & 0.3 & 0.23 \\ \hline
\enddata

\tablecomments{
Col.(1): Object name.  
Col.(2): $\tau_{9.7}'$ is optical depth of the 9.7 $\mu$m silicate
dust absorption feature, against a power-law continuum shown
as dotted lines in Figure 1.  
Once the continuum levels are fixed, the uncertainty of $\tau_{9.7}'$ 
is $<$5\% for ULIRGs with large $\tau_{9.7}'$ values ($>$2) and 
can be $\sim$10\% for ULIRGs with small $\tau_{9.7}'$.
Col.(3): $\tau_{18}'$ is optical depth of the 18 $\mu$m silicate dust
absorption feature, against a power-law continuum shown as
dotted lines in Figure 1. 
Once the continuum is fixed, the uncertainty of $\tau_{18}'$
is $<$10\% because the value is estimated only for ULIRGs with clearly
detectable 18 $\mu$m silicate absorption features. 
Col.(4): $\tau_{18}'$/$\tau_{9.7}'$ ratio for ULIRGs with clearly
detectable 18 $\mu$m silicate absorption.
}

\end{deluxetable}

\begin{deluxetable}{lcc}
\tablecaption{Ice and CO absorption features
\label{tbl-5}}
\tablewidth{0pt}
\tablehead{
\colhead{Object} & \colhead{$\tau_{6.0}$} & \colhead{$\tau_{CO}$} \\
\colhead{(1)} & \colhead{(2)} & \colhead{(3)} 
}
\startdata 
IRAS 09463+8141 & 0.7 & \nodata \\
IRAS 11582+3020 & 0.3 & \nodata \\
IRAS 12032+1707 & 0.5 & \nodata \\
IRAS 16300+1558 & 0.3 & \nodata \\
IRAS 16333+4630 & 0.4 & \nodata \\
IRAS 23129+2548 & 0.4 & \nodata \\
IRAS 00397$-$1312 & 0.2 & 1.0 (P), 1.2 (R)\\
IRAS 01355$-$1814 & 0.3 & \nodata \\
IRAS 08201+2801 & 0.5 & \nodata \\  
IRAS 13469+5833 & 0.9 & \nodata \\
IRAS 17068+4027 & 0.3 & \nodata \\ \hline
\enddata

\tablecomments{
Col.(1): Object name.  
Col.(2): Optical depth of the 6.0 $\mu$m H$_{2}$O ice absorption feature for
clearly detected sources. 
Col.(3): Optical depth of the 4.67 $\mu$m CO absorption feature. 
``P'' means the P-branch of the CO absorption feature (the sub-peak at
longer wavelength).  
``R'' means the R-branch of the CO absorption feature (the sub-peak at
shorter wavelength).
}

\end{deluxetable}

\begin{deluxetable}{lcccccc}
\tabletypesize{\scriptsize}
\tablecaption{Buried AGN signatures \label{tbl-6}}
\tablewidth{0pt}
\tablehead{
\colhead{Object} & \colhead{EW$_{\rm 6.2PAH}$} & \colhead{EW$_{\rm 7.7PAH}$} & 
\colhead{EW$_{\rm 11.3PAH}$} & \colhead{$\tau_{9.7}'$} & 
\colhead{T-gradient} &\colhead{Total} \\
\colhead{(1)} & \colhead{(2)} & \colhead{(3)} &  
\colhead{(4)} & \colhead{(5)} & \colhead{(6)} & 
\colhead{(7)} 
}
\startdata 
IRAS 03521+0028   & X          & X          & X          & X          &
X & X \\
IRAS 09463+8141   & X          & X          & X          & $\bigcirc$ &
X & $\bigtriangleup$ \\
IRAS 10091+4704   & $\bigcirc$ & X          & X          & $\bigcirc$ &
X & $\bigcirc$ \\
IRAS 11028+3130   & $\bigcirc$ & X          & X          & $\bigcirc$ &
X & $\bigcirc$ \\
IRAS 11180+1623   & $\bigcirc$ & X          & X          & $\bigcirc$ &
X & $\bigcirc$ \\
IRAS 11582+3020   & $\bigcirc$ & X          & X          & $\bigcirc$ &
X & $\bigcirc$ \\
IRAS 12032+1707   & $\bigcirc$ & $\bigcirc$ & X          & $\bigcirc$ &
$\bigcirc$ & {\large $\circledcirc$} \\ 
IRAS 16300+1558   & $\bigcirc$ & X          & X          & $\bigcirc$ &
X & $\bigcirc$ \\ 
IRAS 16333+4630   & X          & X          & X          & X          &
X & X \\
IRAS 23129+2548   & $\bigcirc$ & X          & X          & $\bigcirc$ &
X & $\bigcirc$ \\
IRAS 00397$-$1312 & $\bigcirc$ & $\bigcirc$ & $\bigcirc$ & $\bigcirc$ &
$\bigcirc$ & {\large $\circledcirc$} \\
IRAS 01199$-$2307 & $\bigcirc$ & X          & X          & $\bigcirc$ &
X & $\bigcirc$ \\
IRAS 01355$-$1814 & $\bigcirc$ & X          & $\bigcirc$ & $\bigcirc$ &
X & $\bigcirc$ \\
IRAS 08201+2801   & $\bigcirc$ & X          & X          & $\bigcirc$ &
$\bigcirc$ & {\large $\circledcirc$} \\  
IRAS 13469+5833   & X          & X          & X          & X          & X & X \\
IRAS 17068+4027   & $\bigcirc$ & X          & $\bigcirc$ & X          &
X & $\bigcirc$ \\
IRAS 01494$-$1845 & X          & X          & X          & X          &
X & X \\
IRAS 04313$-$1649 & $\bigcirc$ & X          & X          & $\bigcirc$ &
$\bigcirc$ & {\large $\circledcirc$} \\
IRAS 10035+2740   & X          & X          & X          & $\bigcirc$ &
X & $\bigtriangleup$ \\
IRAS 12018+1941   & $\bigcirc$ & $\bigcirc$ & $\bigcirc$ & X          &
$\bigcirc$ & {\large $\circledcirc$} \\ \hline
\enddata

\tablecomments{
Col.(1): Object name.  
Col.(2): Buried AGN signatures based on the low equivalent width of the
         6.2 $\mu$m PAH emission (EW$_{\rm 6.2PAH}$ $<$ 180 nm) ($\S$5.2.1).
         $\bigcirc$: present.  X: none. 
Col.(3): Buried AGN signatures based on the low equivalent width of the
         7.7 $\mu$m PAH emission (EW$_{\rm 7.7PAH}$ $<$ 230 nm) ($\S$5.2.1).
         $\bigcirc$: present.  X: none. 
Col.(4): Buried AGN signatures based on the low equivalent width of the
         11.3 $\mu$m PAH emission (EW$_{\rm 11.3PAH}$ $<$ 200 nm) ($\S$5.2.1).
         $\bigcirc$: present.  X: none. 
Col.(5): Buried AGN signatures based on the large $\tau_{9.7}'$ value 
         ($>$2) ($\S$5.2.2).
         $\bigcirc$: present.  X: none. 
Col.(6): Buried AGN signatures based on the small
         $\tau_{18}'$/$\tau_{9.7}'$ ratio ($\S$5.2.3).   
         $\bigcirc$: present. X: none. 
Col.(7): Buried AGN signatures from combined methods in Col. (2)--(6). 
         {\large $\circledcirc$}: very strong.
         $\bigcirc$: strong.
         $\bigtriangleup$: possible. 
         X: none.
         Please see $\S$5.2.4 for more details. 
}

\end{deluxetable}

\begin{deluxetable}{lcccc}
\tablecaption{Absorption-corrected AGN luminosity \label{tbl-7}}
\tablewidth{0pt}
\tablehead{
\colhead{Object} & \colhead{L(AGN)} 
& \colhead{L(SB-6.2PAH)} & \colhead{L(SB-11.3PAH)} 
& \colhead{L$_{\rm IR}$} \\
\colhead{(1)} & \colhead{10$^{45}$ [ergs s$^{-1}$]} & 
\colhead{10$^{45}$ [ergs s$^{-1}$]} & 
\colhead{10$^{45}$ [ergs s$^{-1}$]} & 
\colhead{10$^{45}$ [ergs s$^{-1}$]} \\
\colhead{(1)} & \colhead{(2)} & \colhead{(3)} & \colhead{(4)} & \colhead{(5)} 
}
\startdata 
IRAS 10091+4704   & 2   & $<$0.5 & 2.5 & 15 \\
IRAS 11582+3020   & 4   & 1.9    & 4.2 & 12 \\
IRAS 12032+1707   & 3.5 & 1.5    & 4.2 & 15 \\
IRAS 16300+1558   & 3   & 2.2    & 4.3 & 18 \\
IRAS 00397$-$1312 & 20  & 6.3    & 2.0 & 34 \\
IRAS 17068+4027   & 2   & 2.5    & 3.6 & 8  \\
IRAS 04313$-$1649 & 2   & $<$1.0 & 1.5 & 15 \\
IRAS 12018+1941   & 2   & 2.7    & 1.7 & 11 \\ \hline
\enddata

\tablecomments{
Col.(1): Object name.  
Col.(2): Absorption-corrected intrinsic luminosity of buried AGN 
in units of 10$^{45}$ ergs s$^{-1}$.
Col.(3): Infrared (8--1000 $\mu$m) luminosity of detected
modestly-obscured (A$_{\rm V}$ $<$ 20 mag) starbursts, estimated from
the 6.2 $\mu$m PAH emission luminosity, in units of 10$^{45}$ ergs
s$^{-1}$.  
Col.(3): Infrared (8--1000 $\mu$m) luminosity of detected
modestly-obscured (A$_{\rm V}$ $<$ 20 mag) starbursts, estimated from
the 11.3 $\mu$m PAH emission luminosity, in units of 10$^{45}$ ergs
s$^{-1}$. 
Col.(3): Observed infrared luminosity in units of
10$^{45}$ ergs s$^{-1}$. 
}

\end{deluxetable}

\clearpage 

\begin{figure}
\includegraphics[angle=-90,scale=.35]{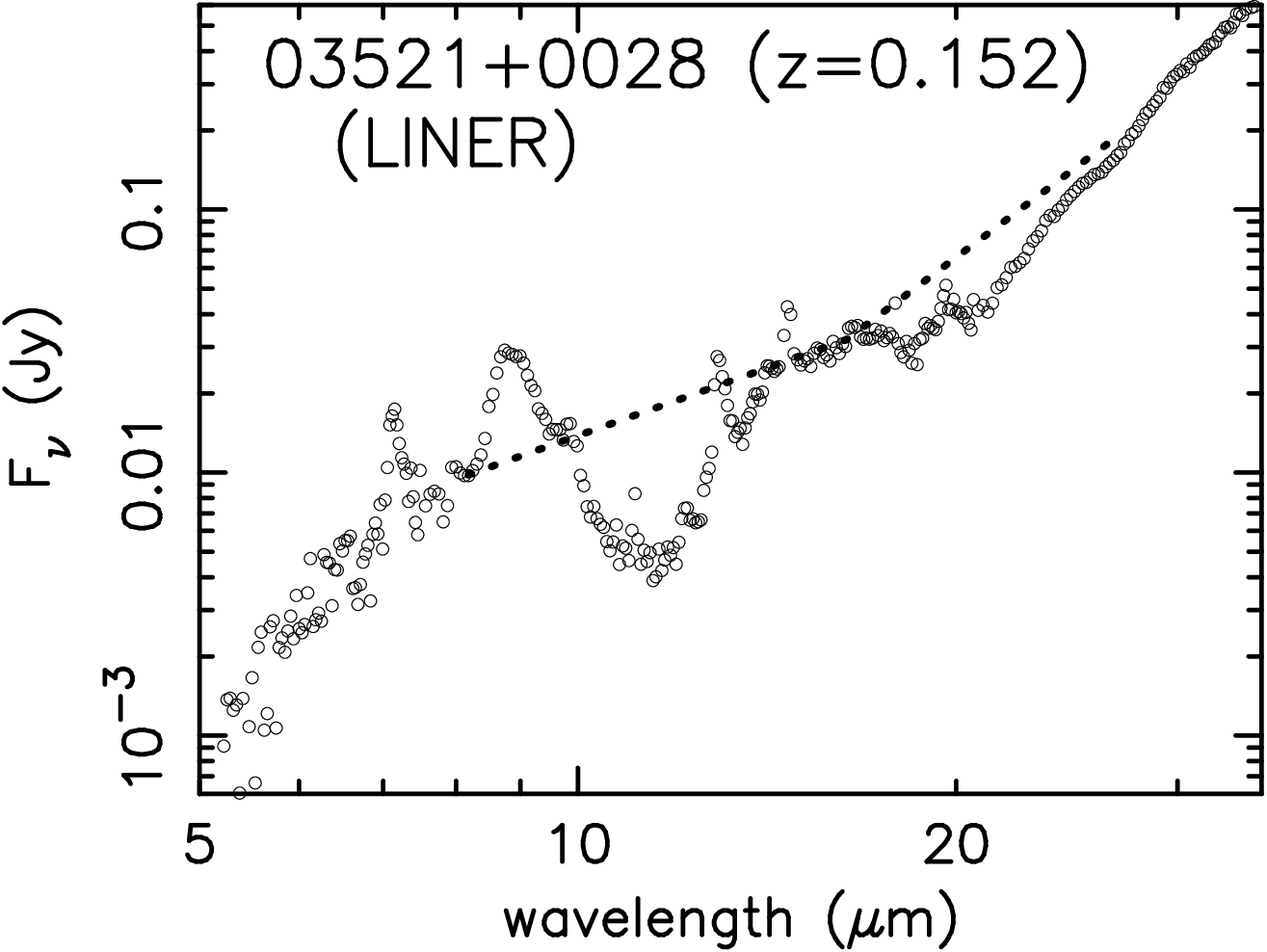} \hspace{0.3cm}
\includegraphics[angle=-90,scale=.35]{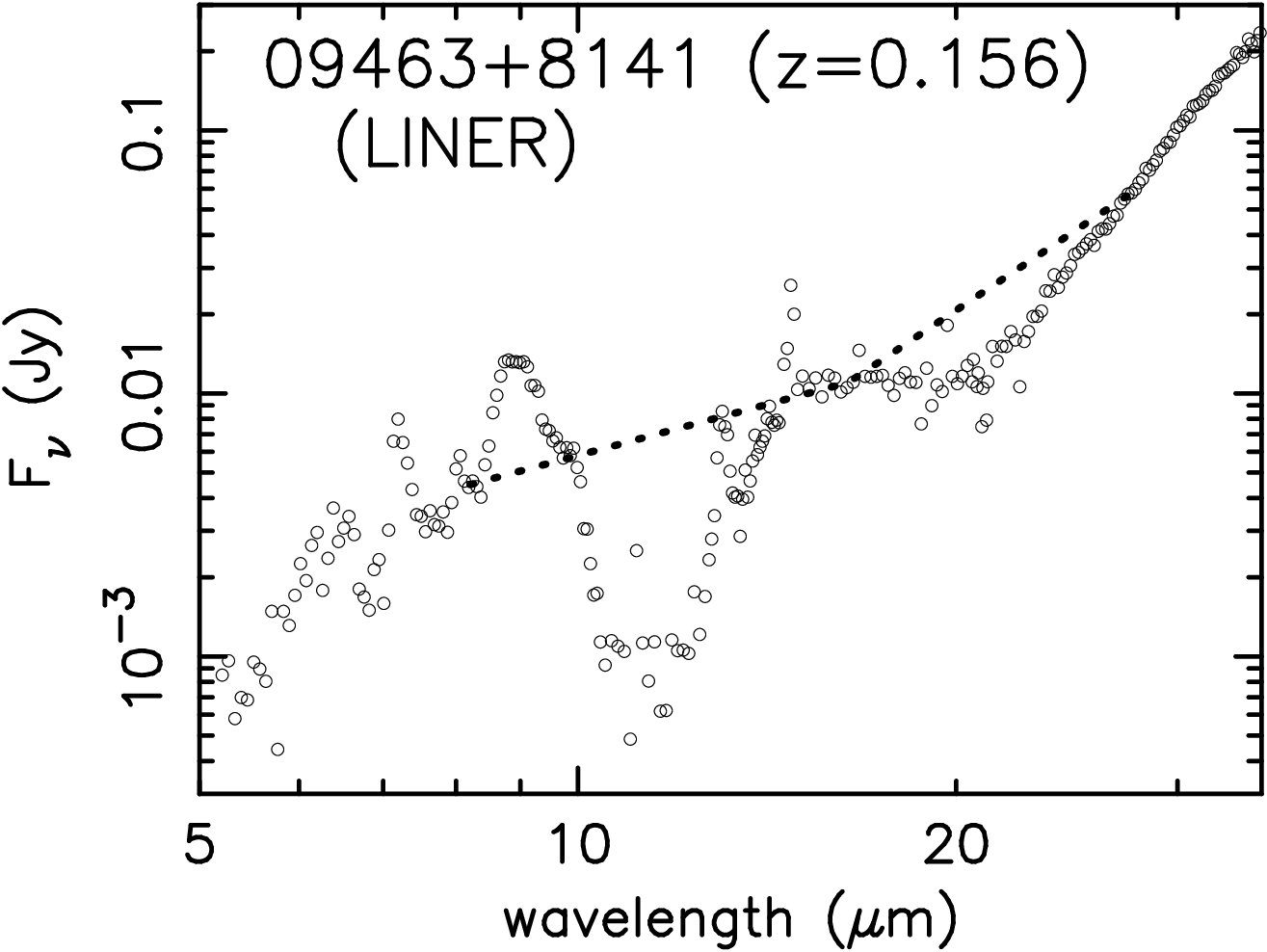} \\
\includegraphics[angle=-90,scale=.35]{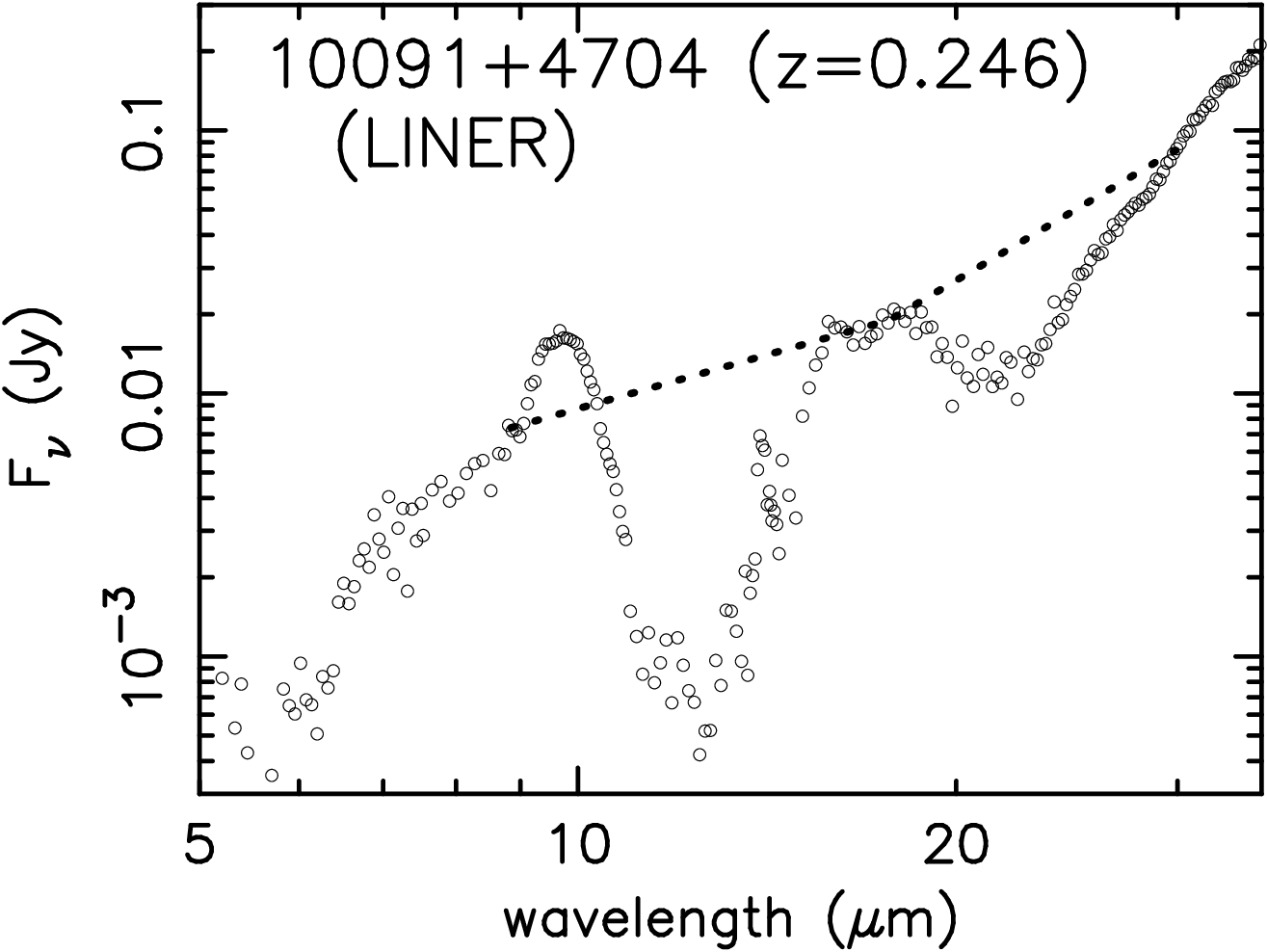} \hspace{0.3cm}
\includegraphics[angle=-90,scale=.35]{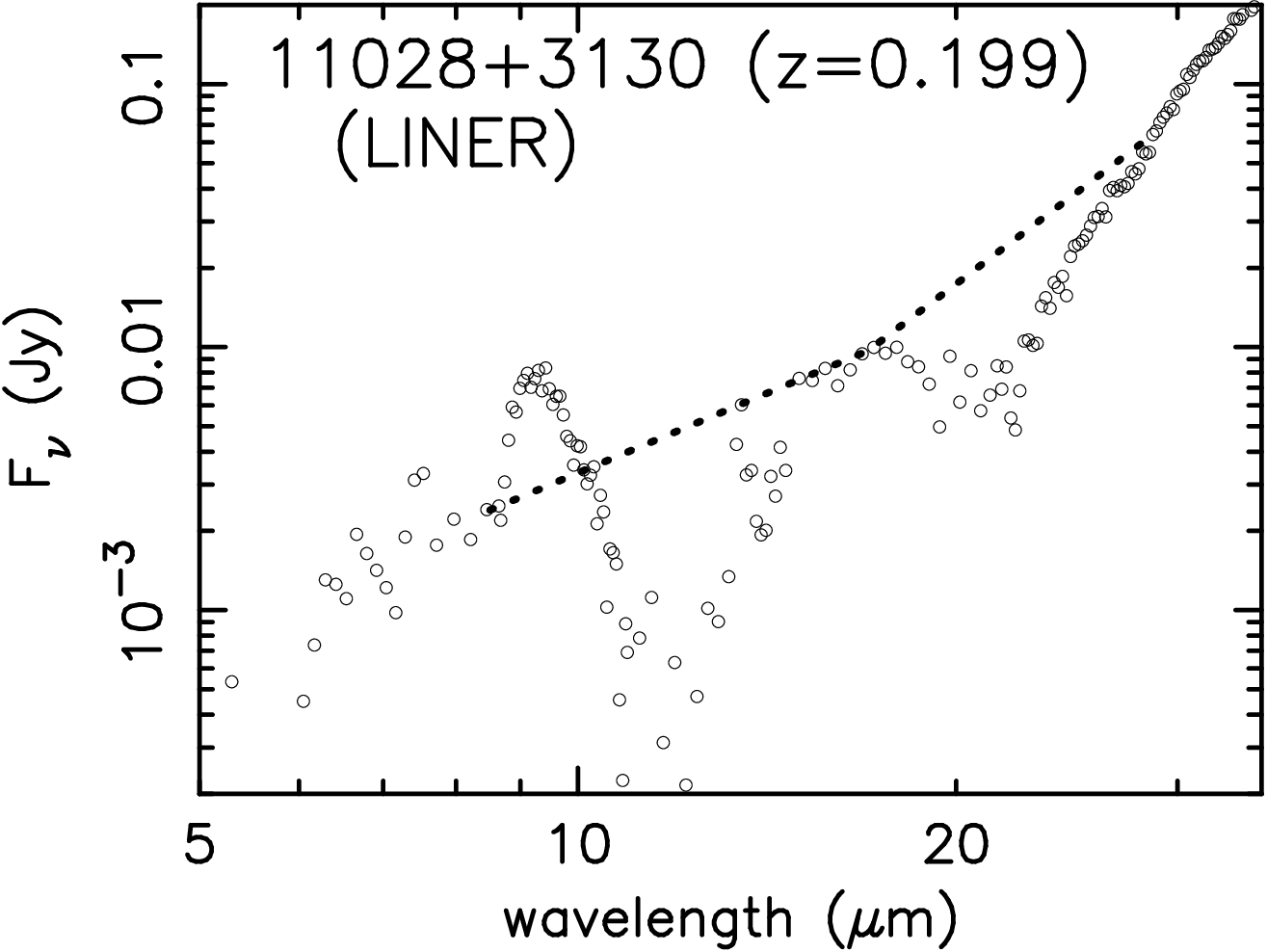} \\
\includegraphics[angle=-90,scale=.35]{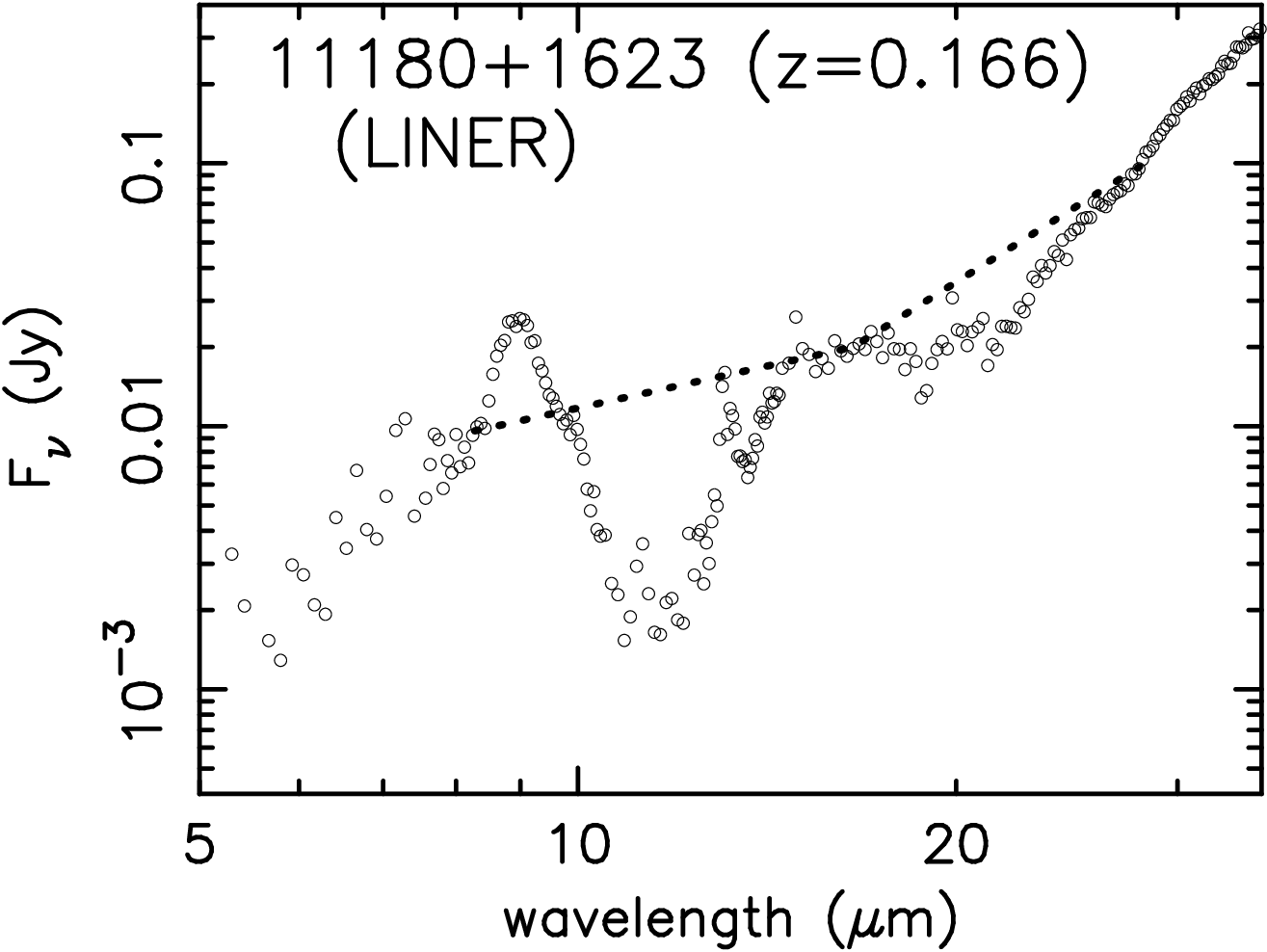} \hspace{0.3cm} 
\includegraphics[angle=-90,scale=.35]{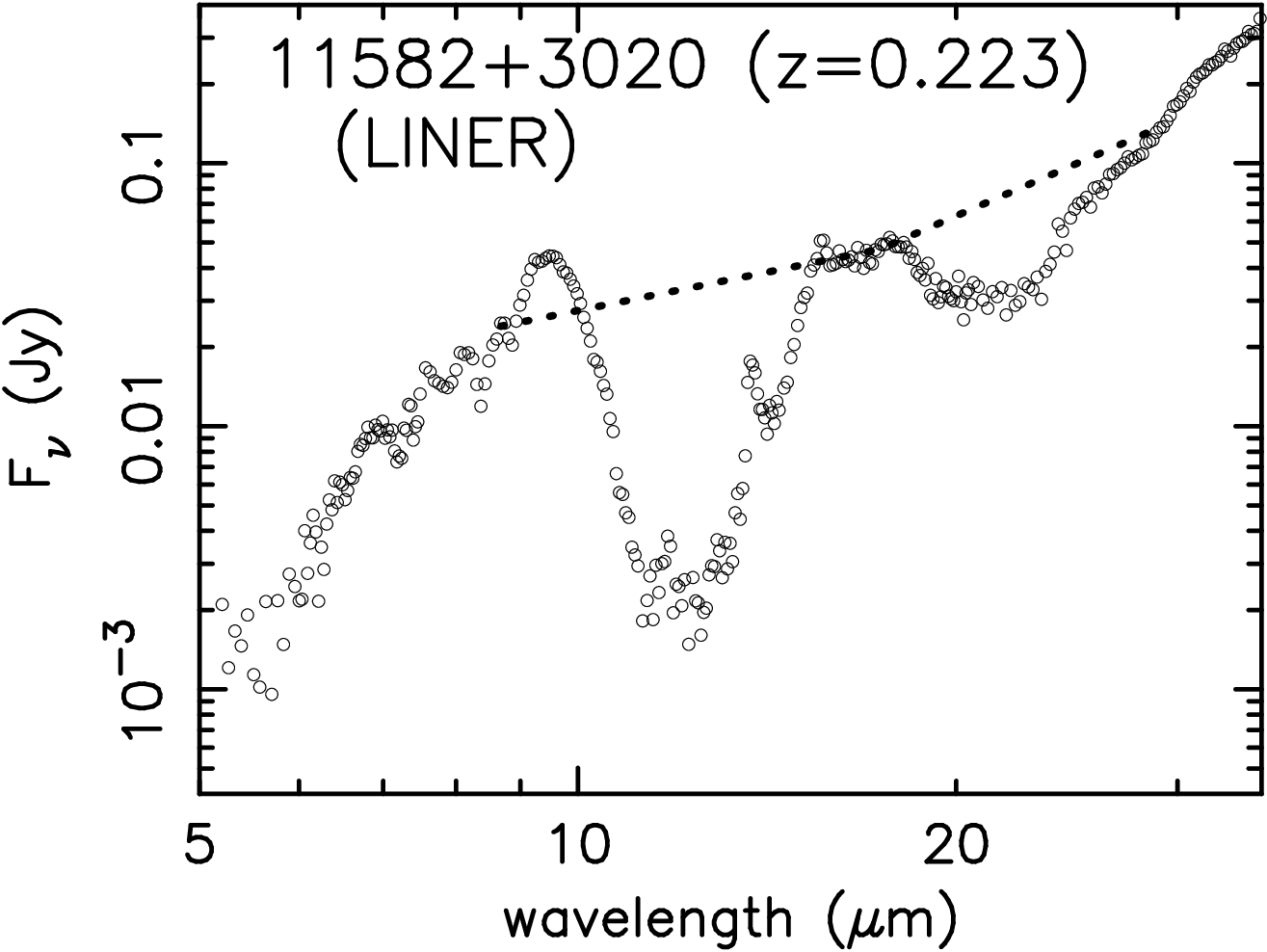} \\
\includegraphics[angle=-90,scale=.35]{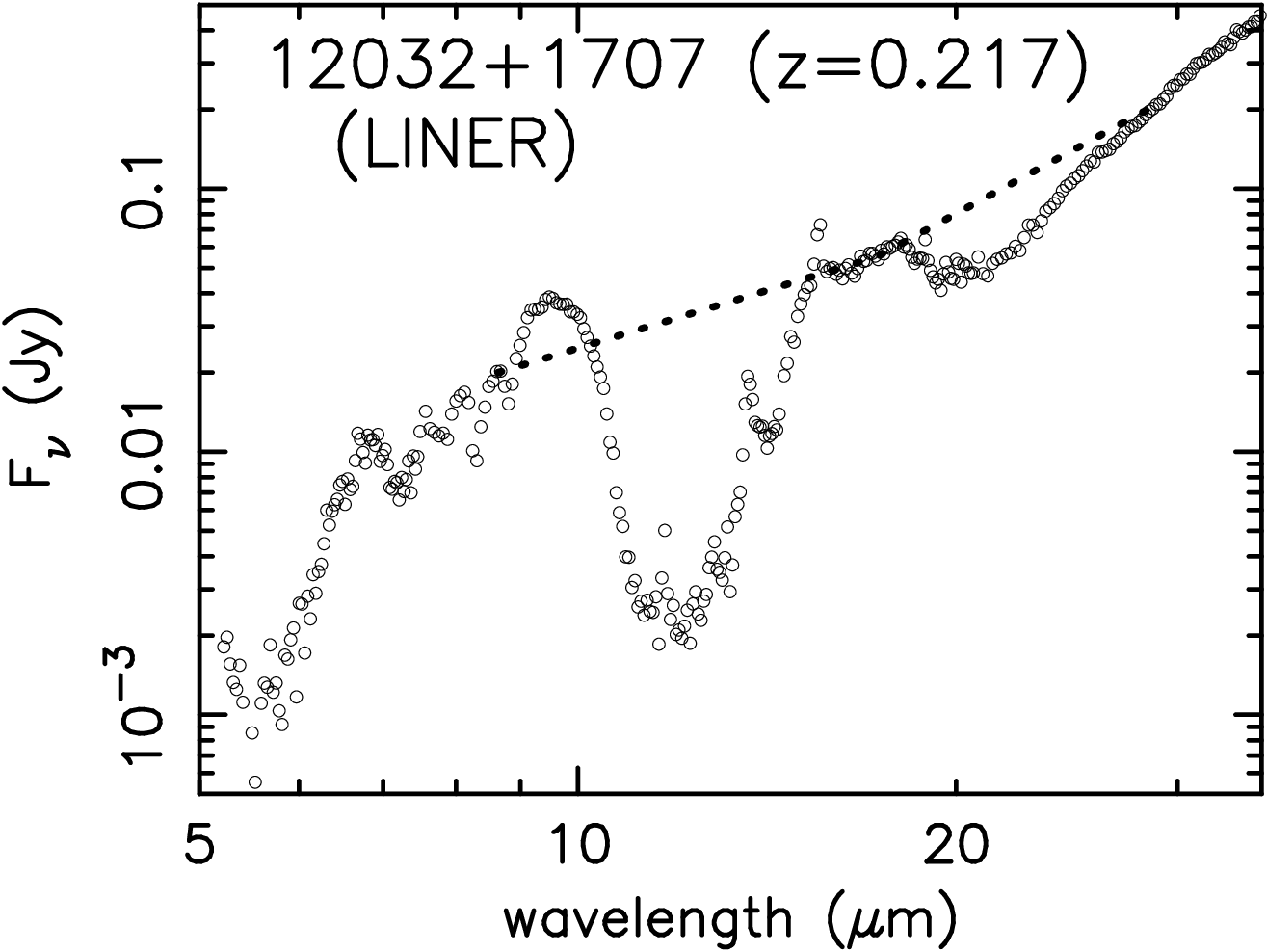} \hspace{0.3cm} 
\includegraphics[angle=-90,scale=.35]{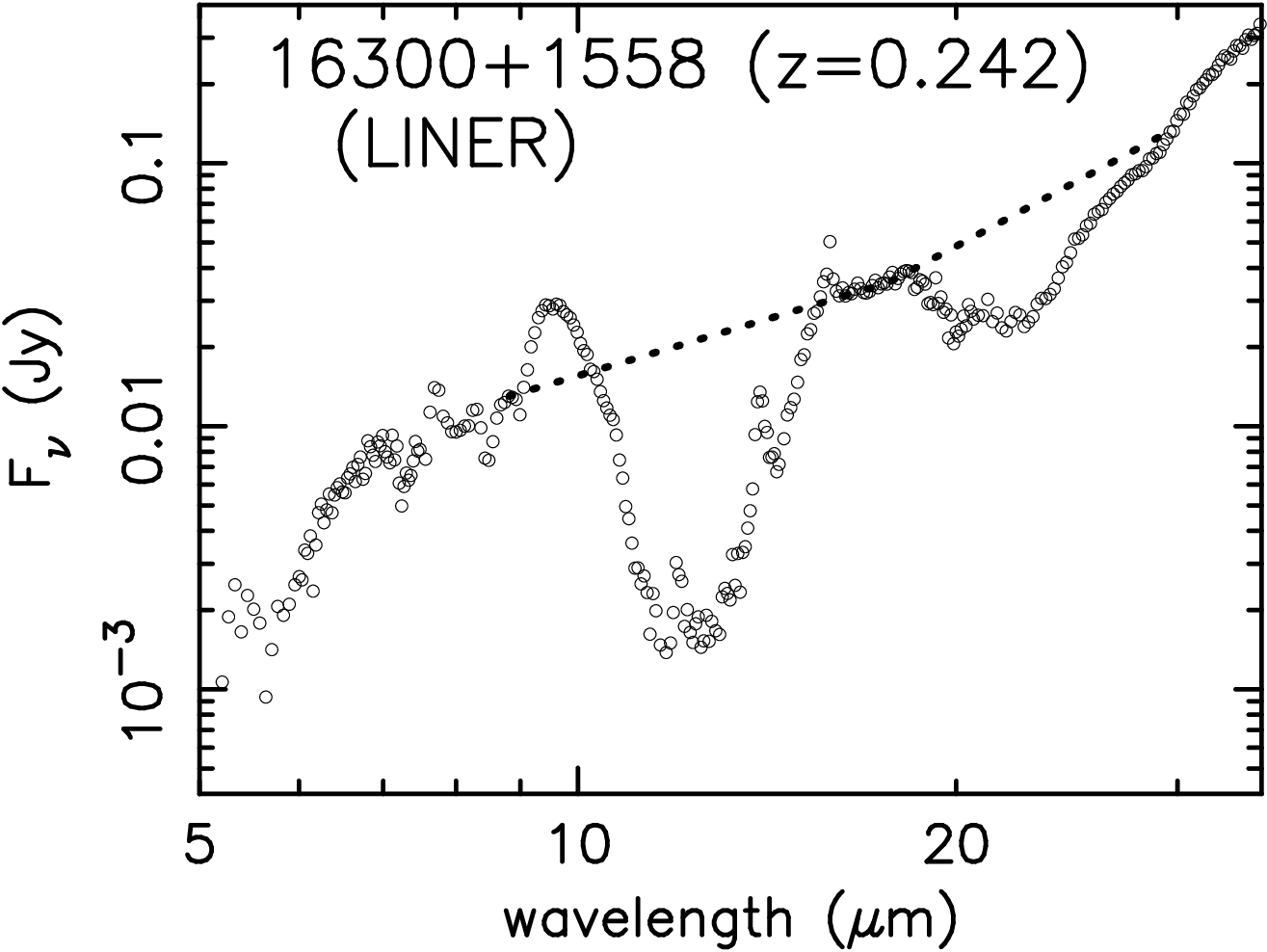} \\
\end{figure}

\begin{figure}
\includegraphics[angle=-90,scale=.35]{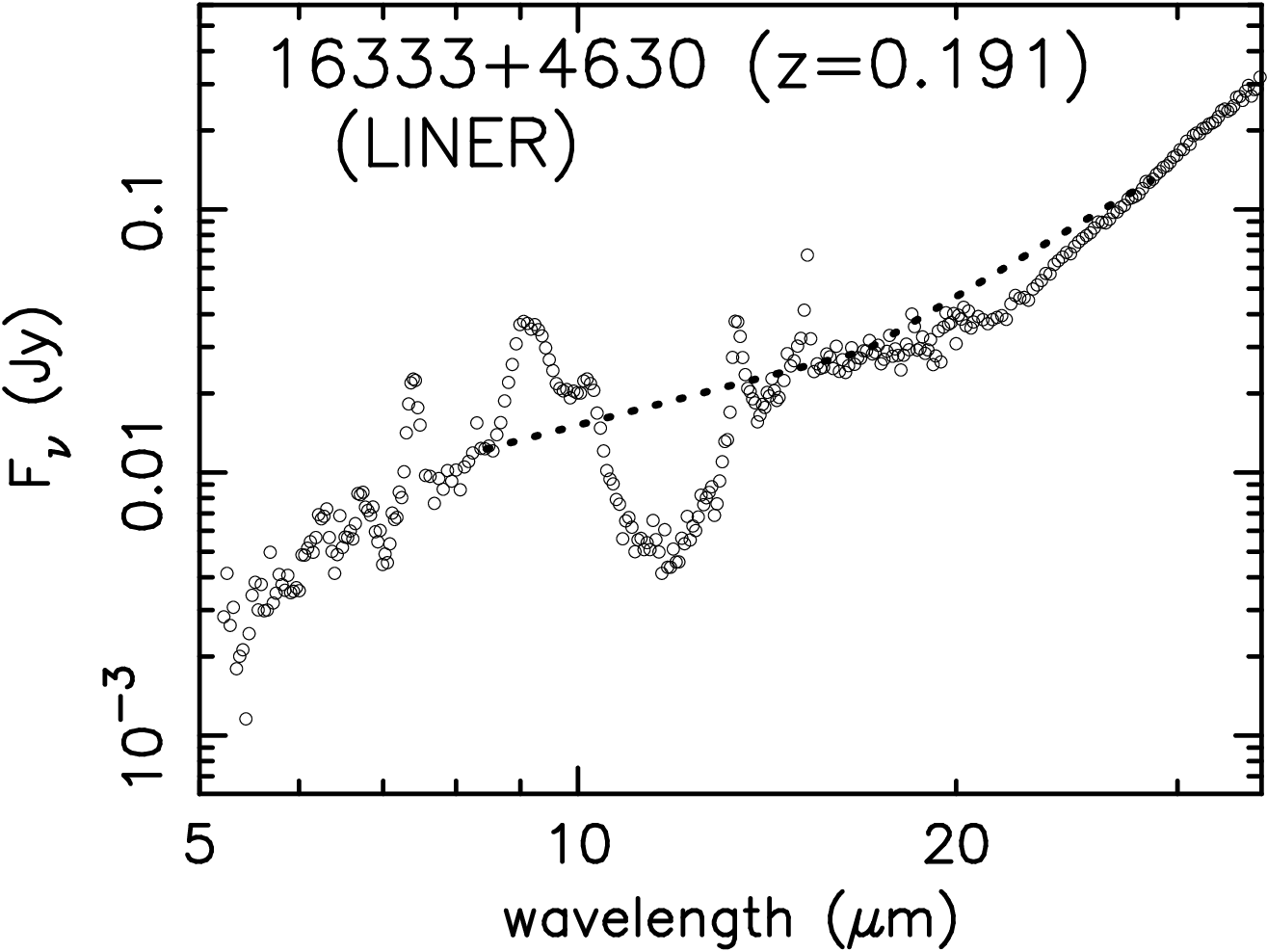} \hspace{0.3cm} 
\includegraphics[angle=-90,scale=.35]{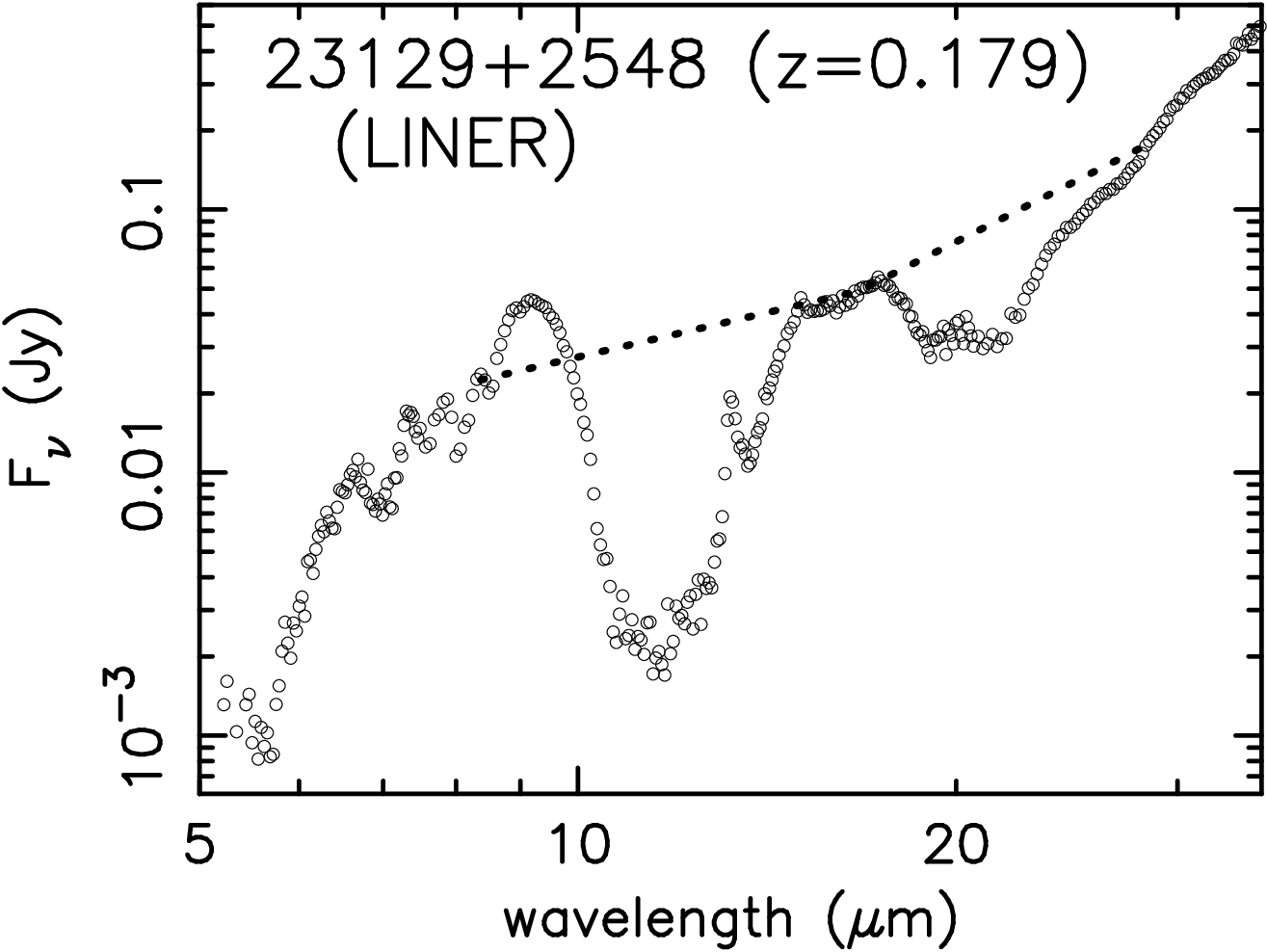} \\ 
\includegraphics[angle=-90,scale=.35]{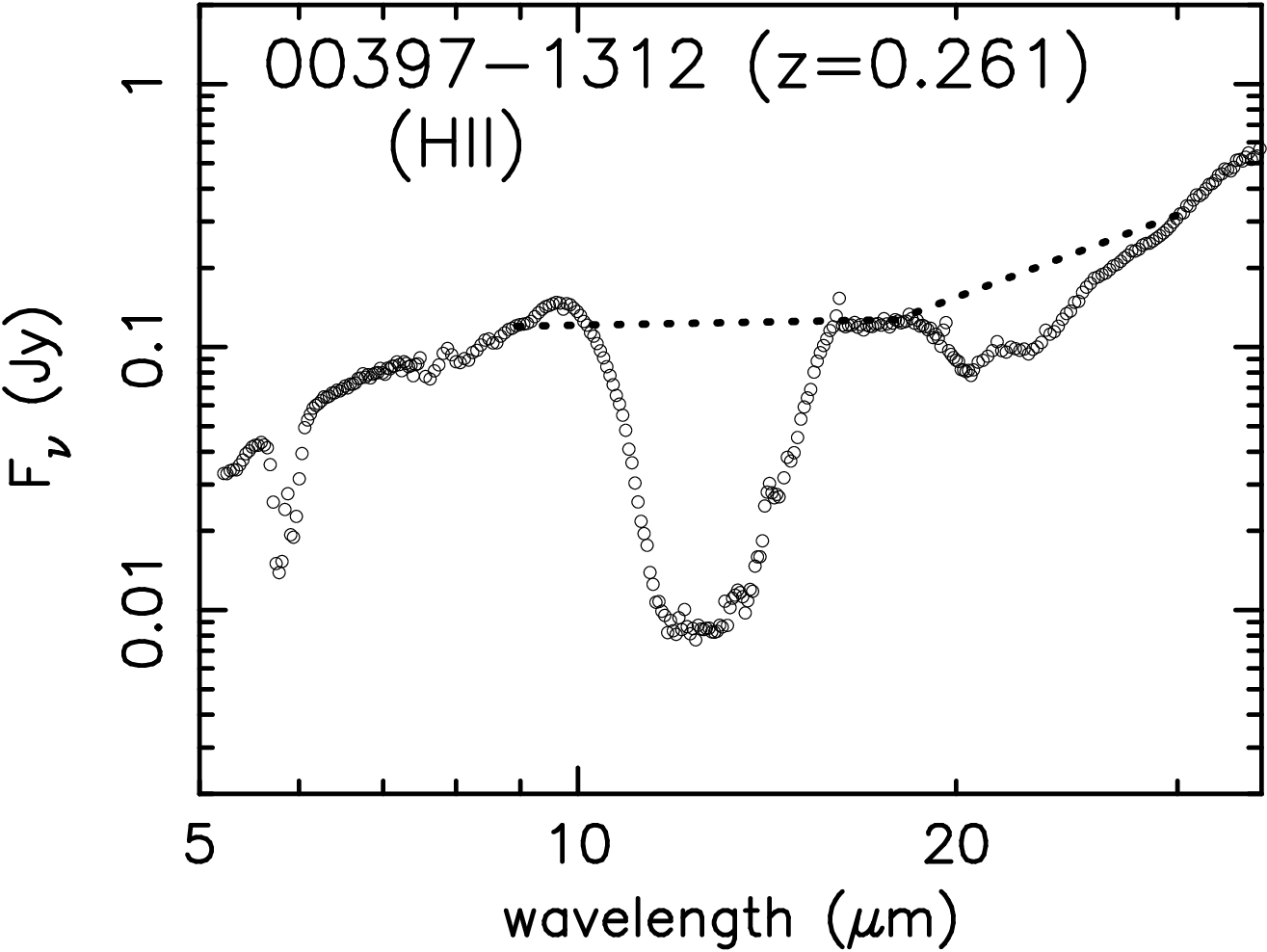} \hspace{0.3cm}
\includegraphics[angle=-90,scale=.35]{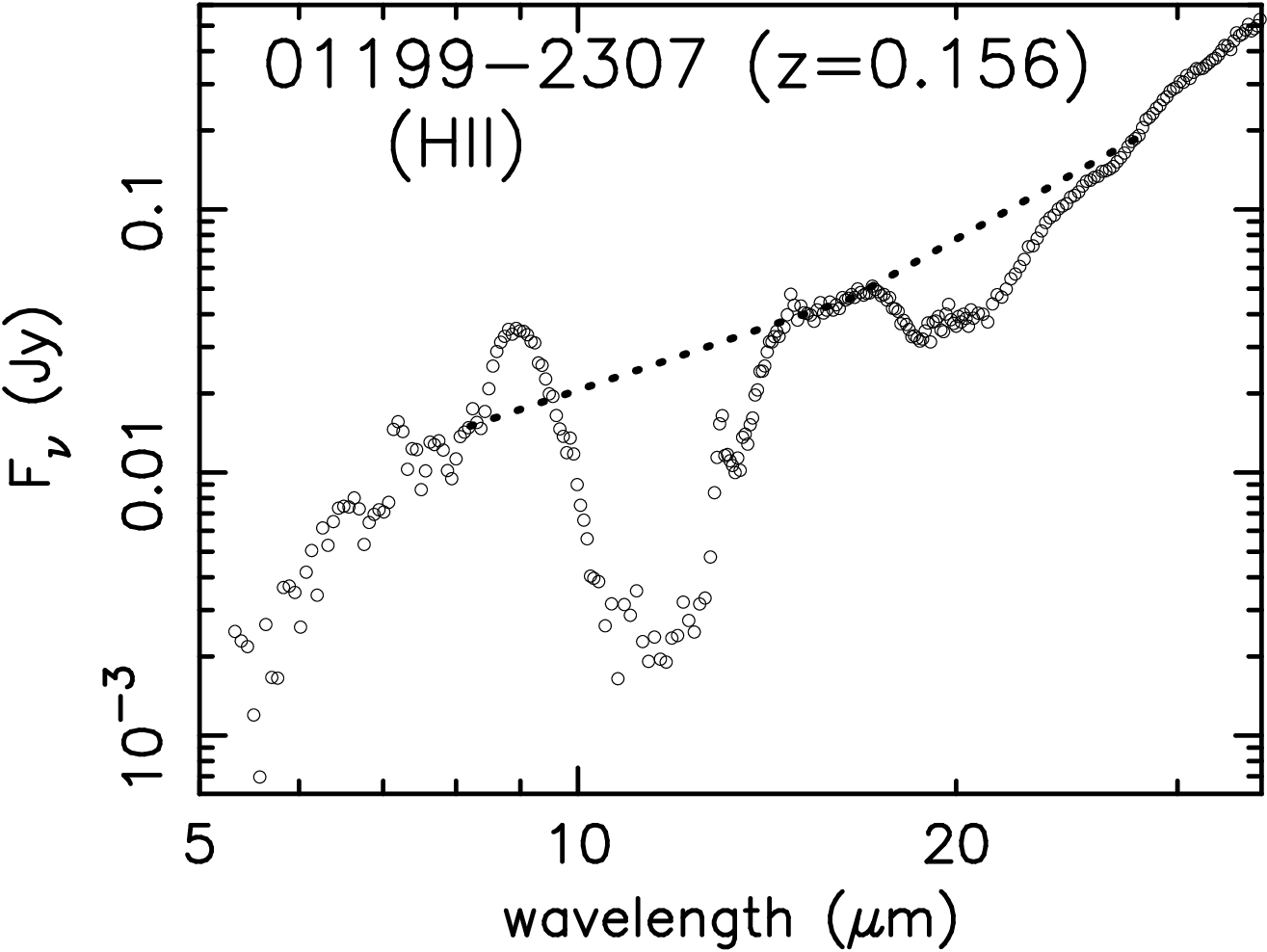} \\
\includegraphics[angle=-90,scale=.35]{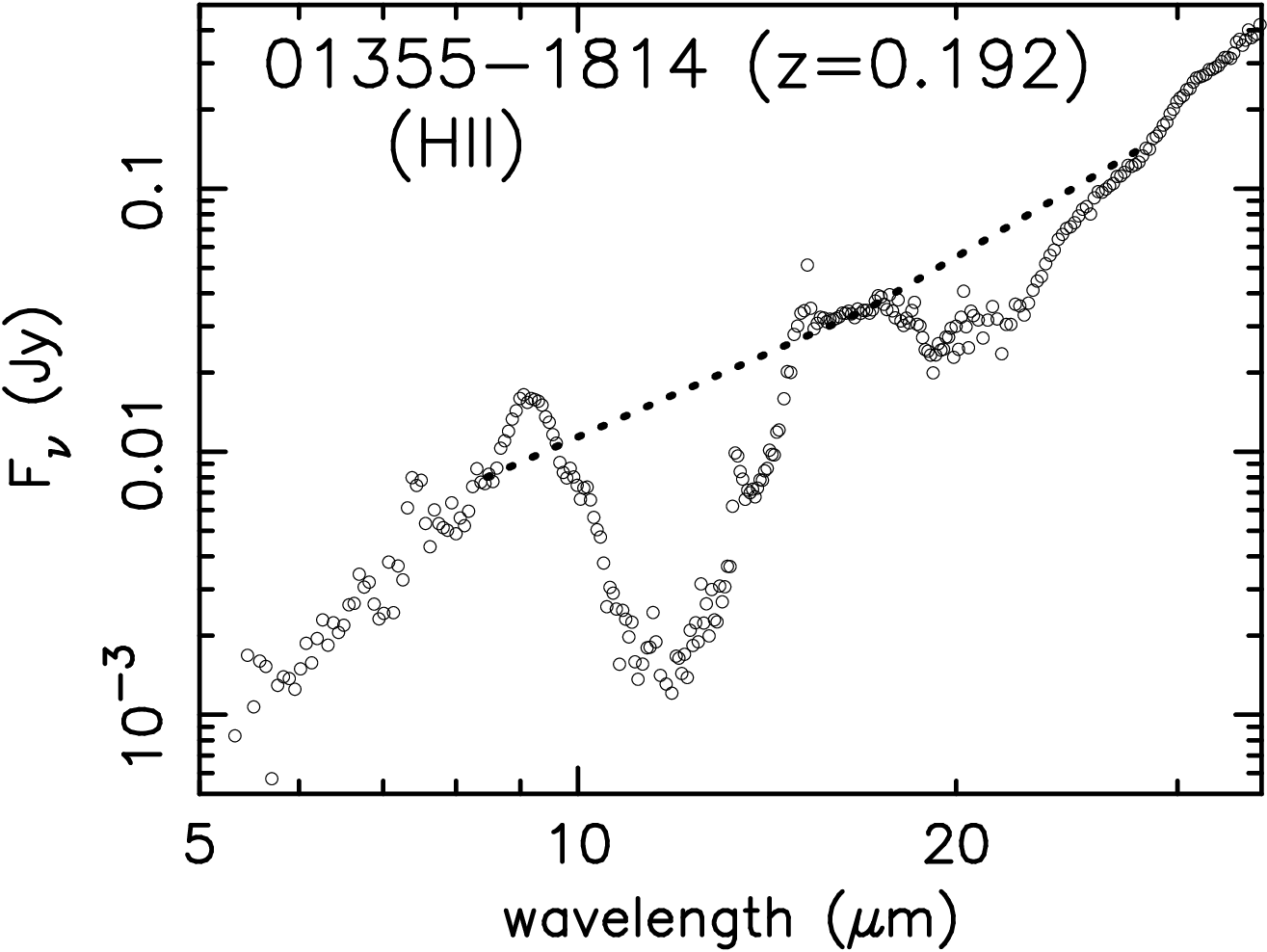} \hspace{0.3cm}
\includegraphics[angle=-90,scale=.35]{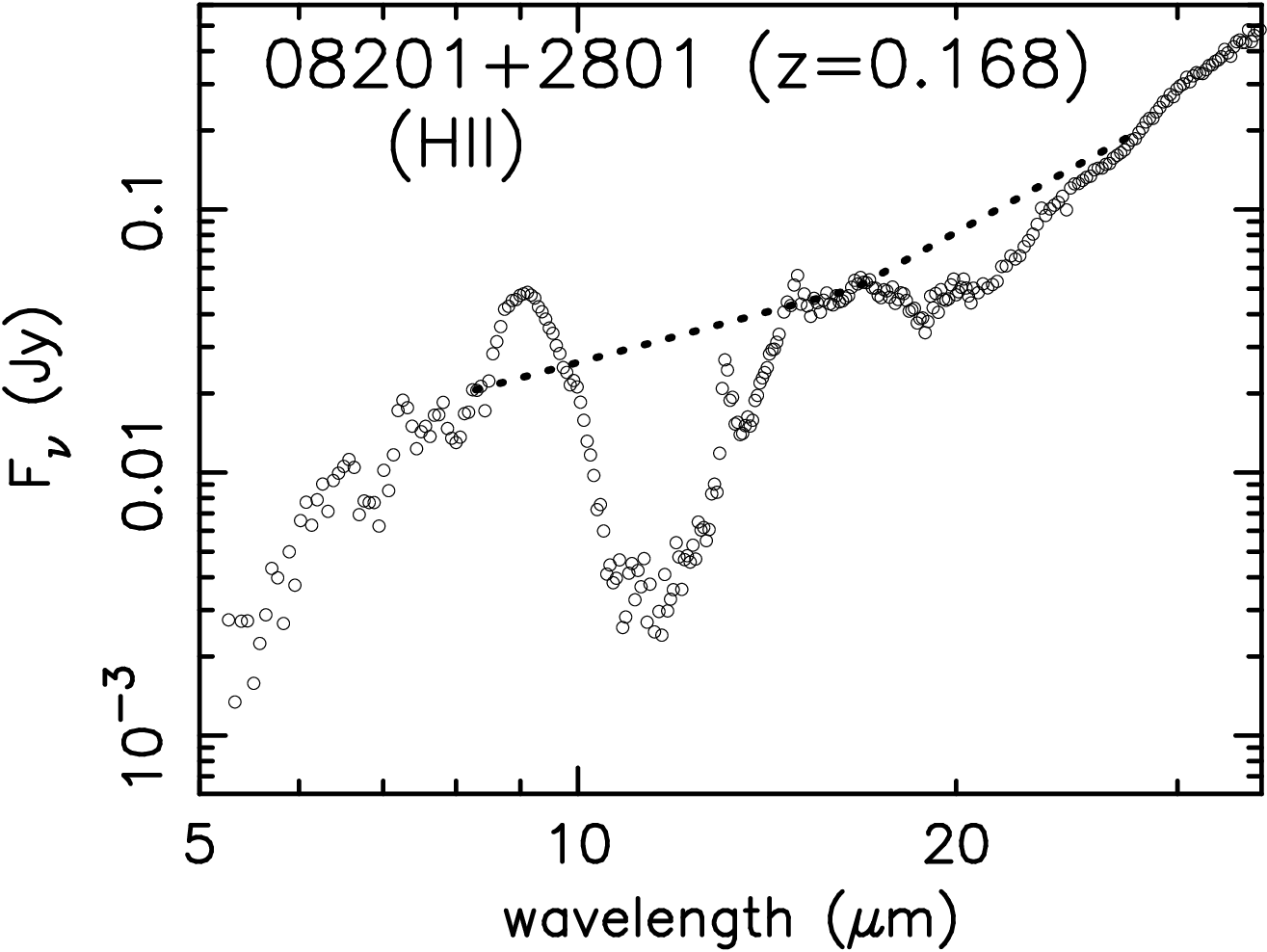} \\
\includegraphics[angle=-90,scale=.35]{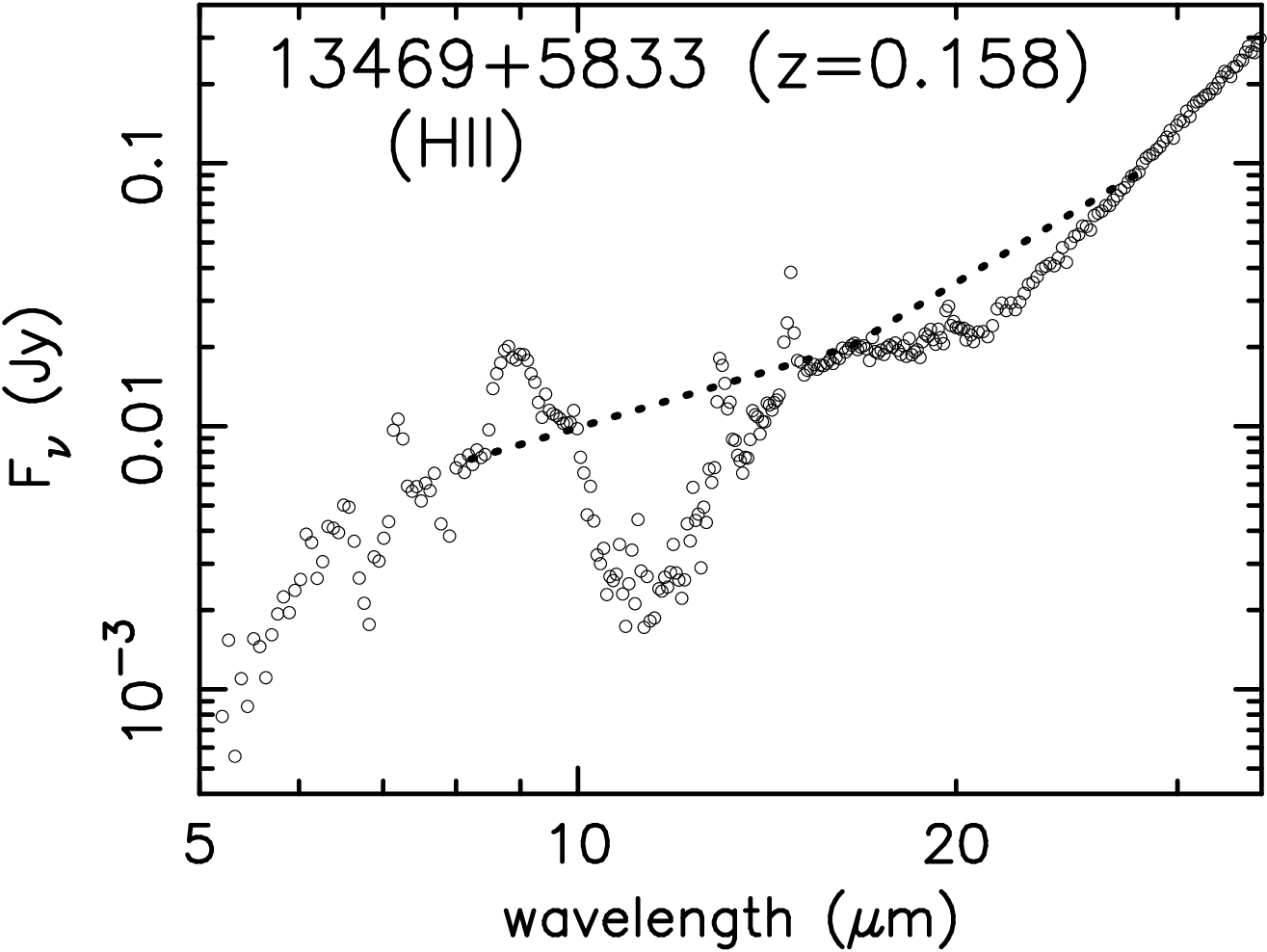} \hspace{0.3cm}
\includegraphics[angle=-90,scale=.35]{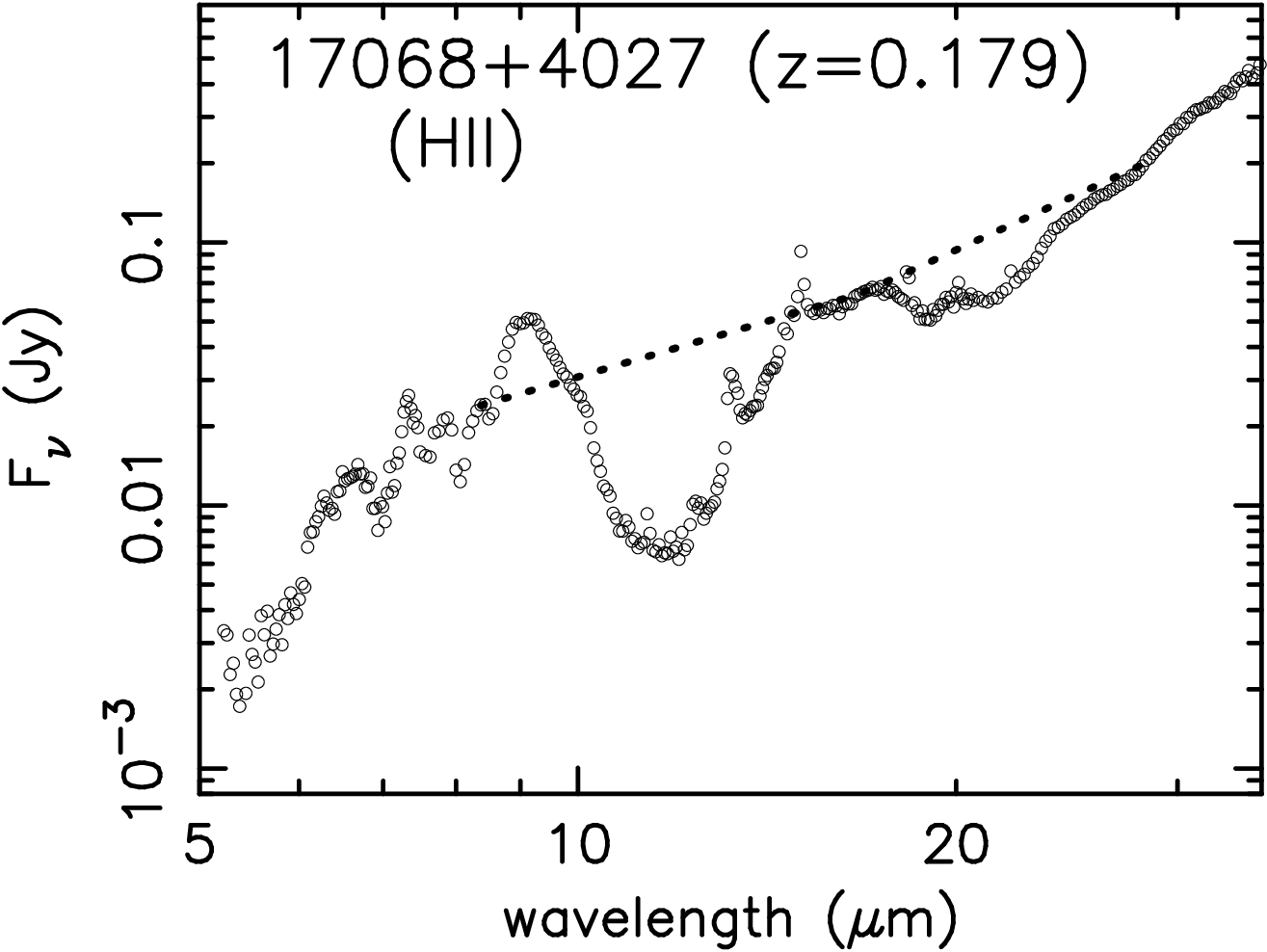} \\
\end{figure}

\clearpage

\begin{figure}
\includegraphics[angle=-90,scale=.35]{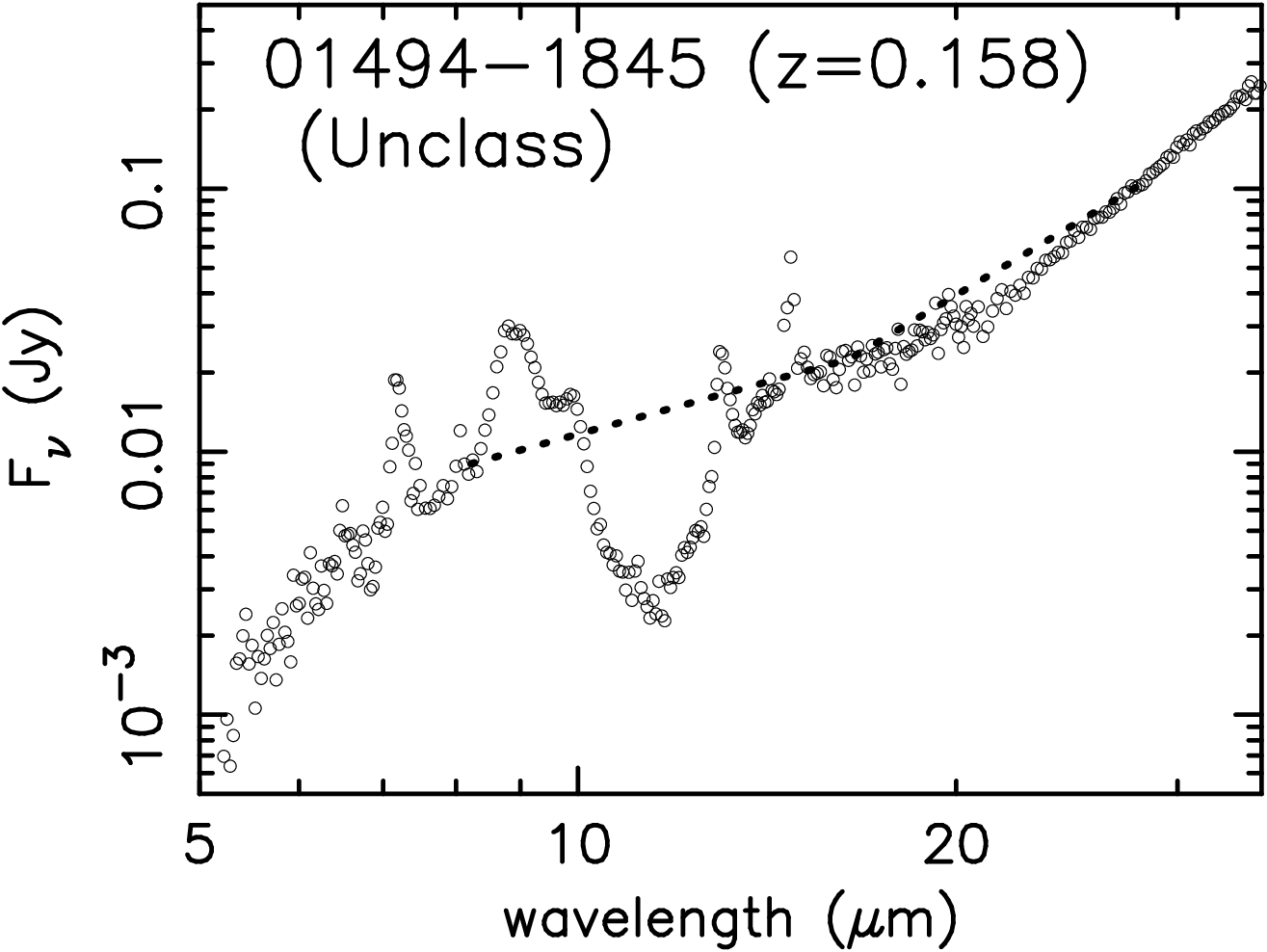} \hspace{0.3cm}
\includegraphics[angle=-90,scale=.35]{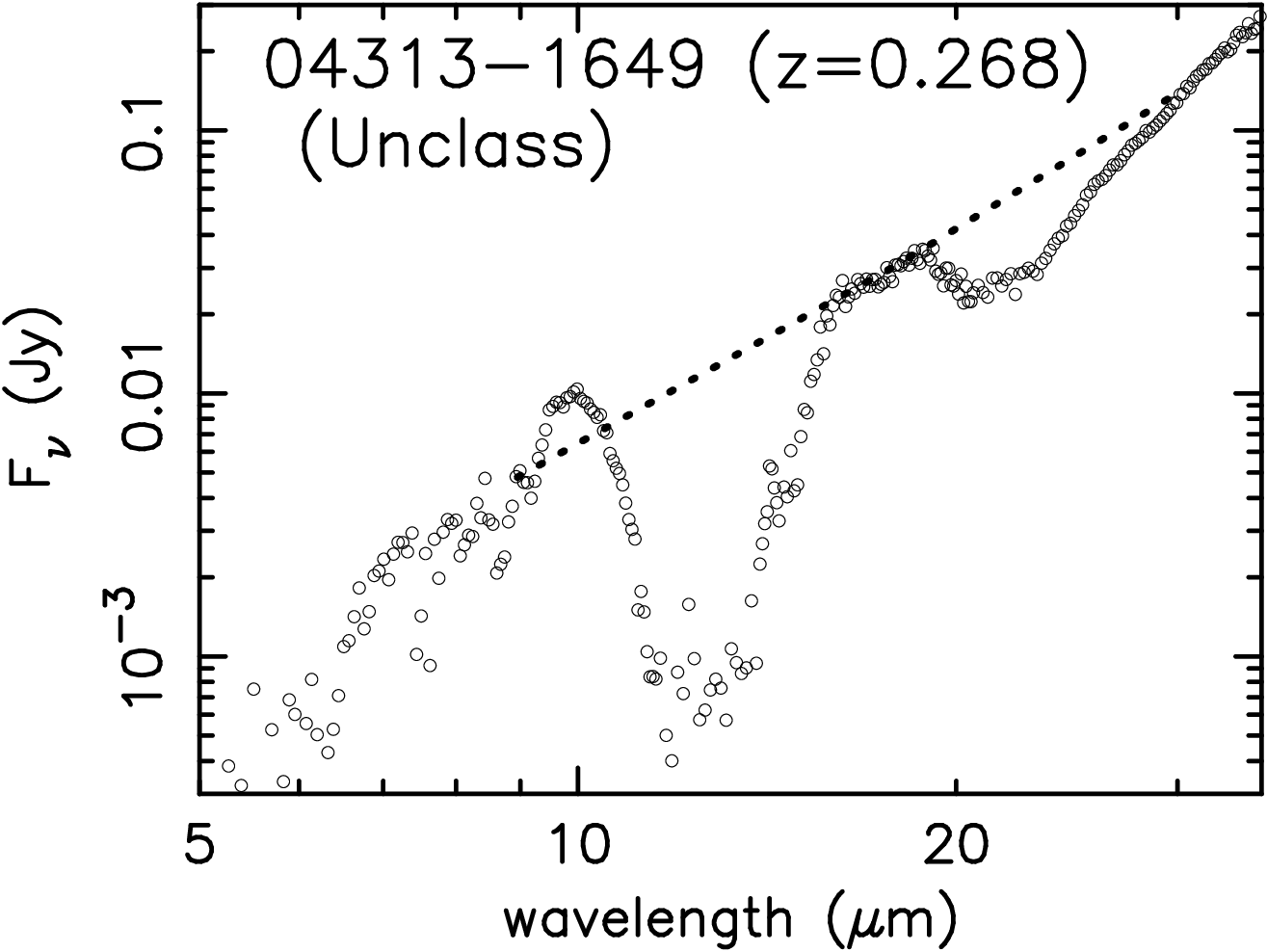} \\
\includegraphics[angle=-90,scale=.35]{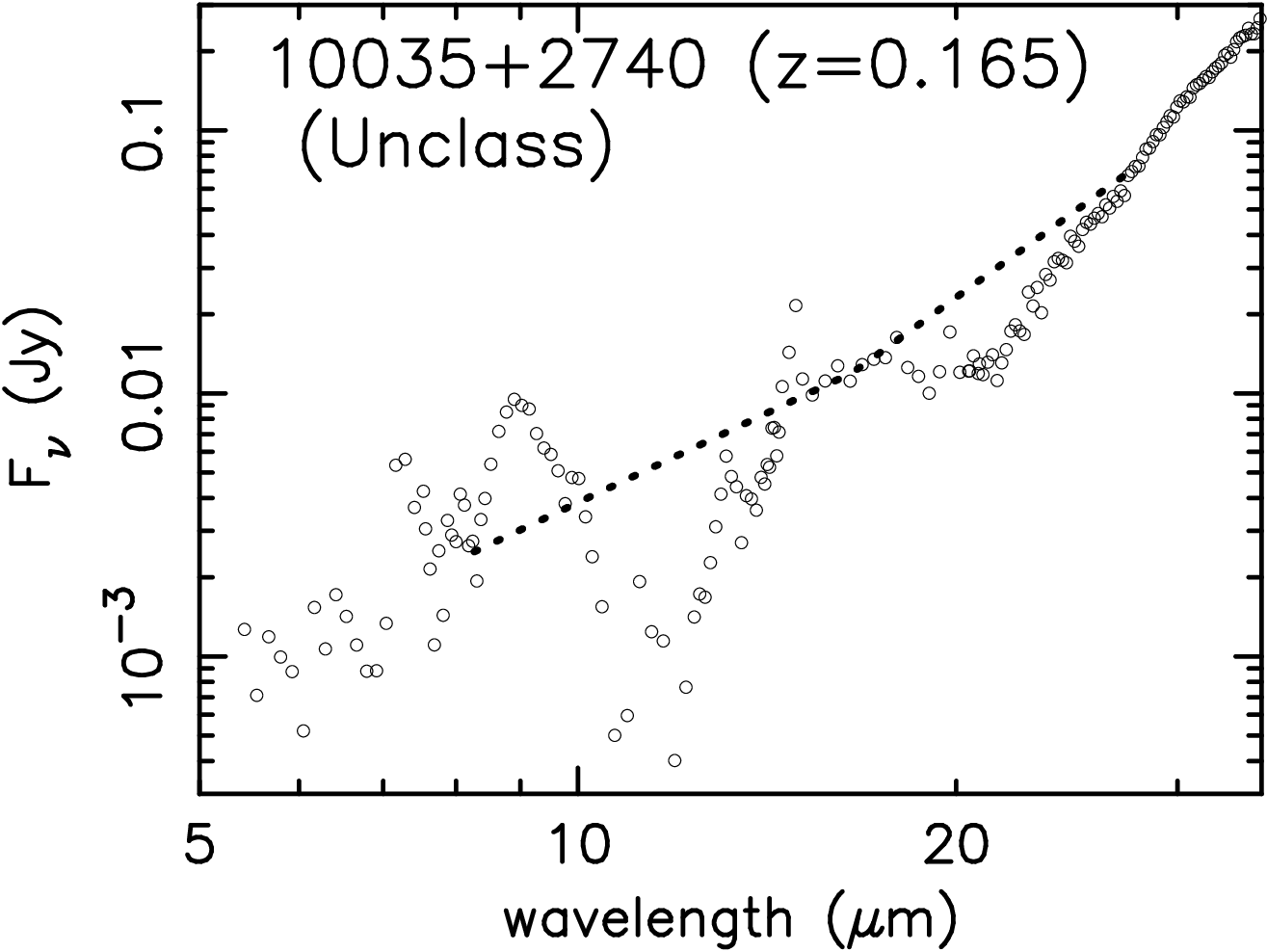} \hspace{0.3cm}
\includegraphics[angle=-90,scale=.35]{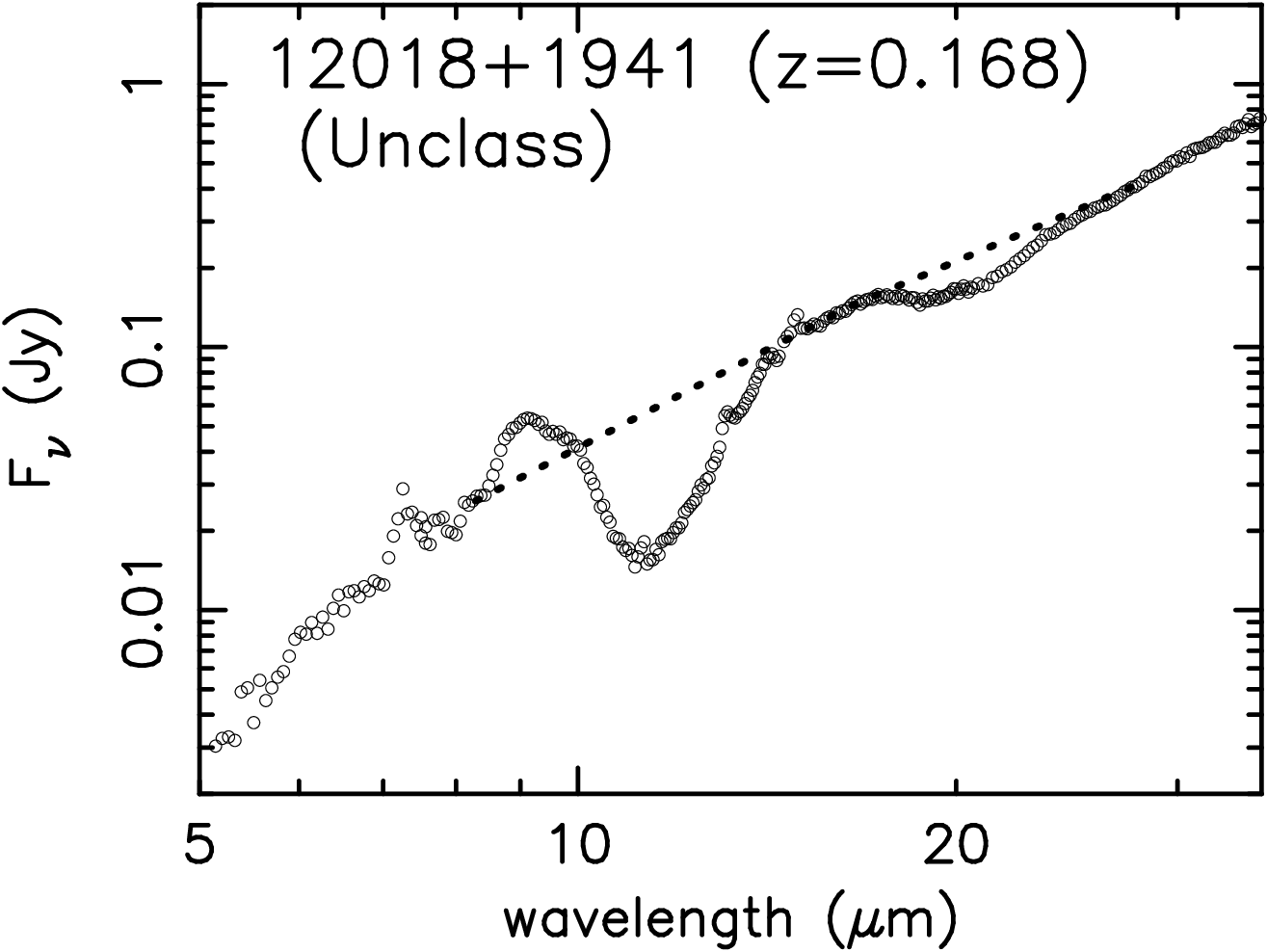} \\
\caption{
Infrared 5--35 $\mu$m ULIRG spectra obtained with 
{\it Spitzer} IRS. The abscissa and
ordinate are, respectively, the observed wavelength in $\mu$m and flux
F$_{\nu}$ in Jy, both shown in decimal logarithmic scale.  
For all objects, the ratio of the
uppermost to lowermost scale in the ordinate is a factor of 1000, 
showing the variation of overall spectral energy distribution.  
Dotted line: power-law continuum determined from data points at
$\lambda_{\rm rest}$ = 7.1 $\mu$m and 14.2 $\mu$m for 9.7 $\mu$m
silicate absorption, and at $\lambda_{\rm rest}$ = 14.2 $\mu$m and
24 $\mu$m for 18 $\mu$m silicate absorption. 
}
\end{figure}

\begin{figure}
\includegraphics[angle=-90,scale=.35]{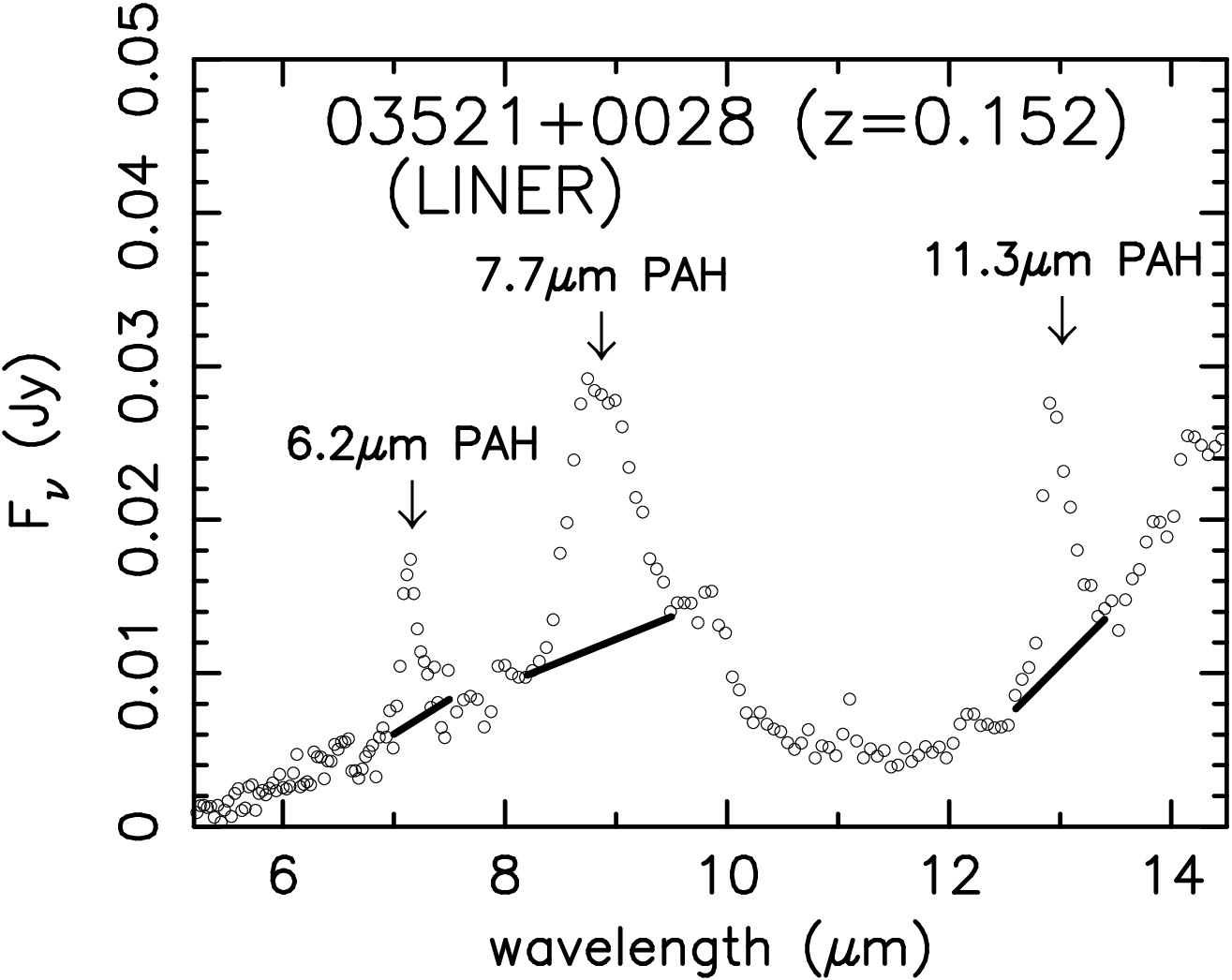} \hspace{0.3cm}
\includegraphics[angle=-90,scale=.35]{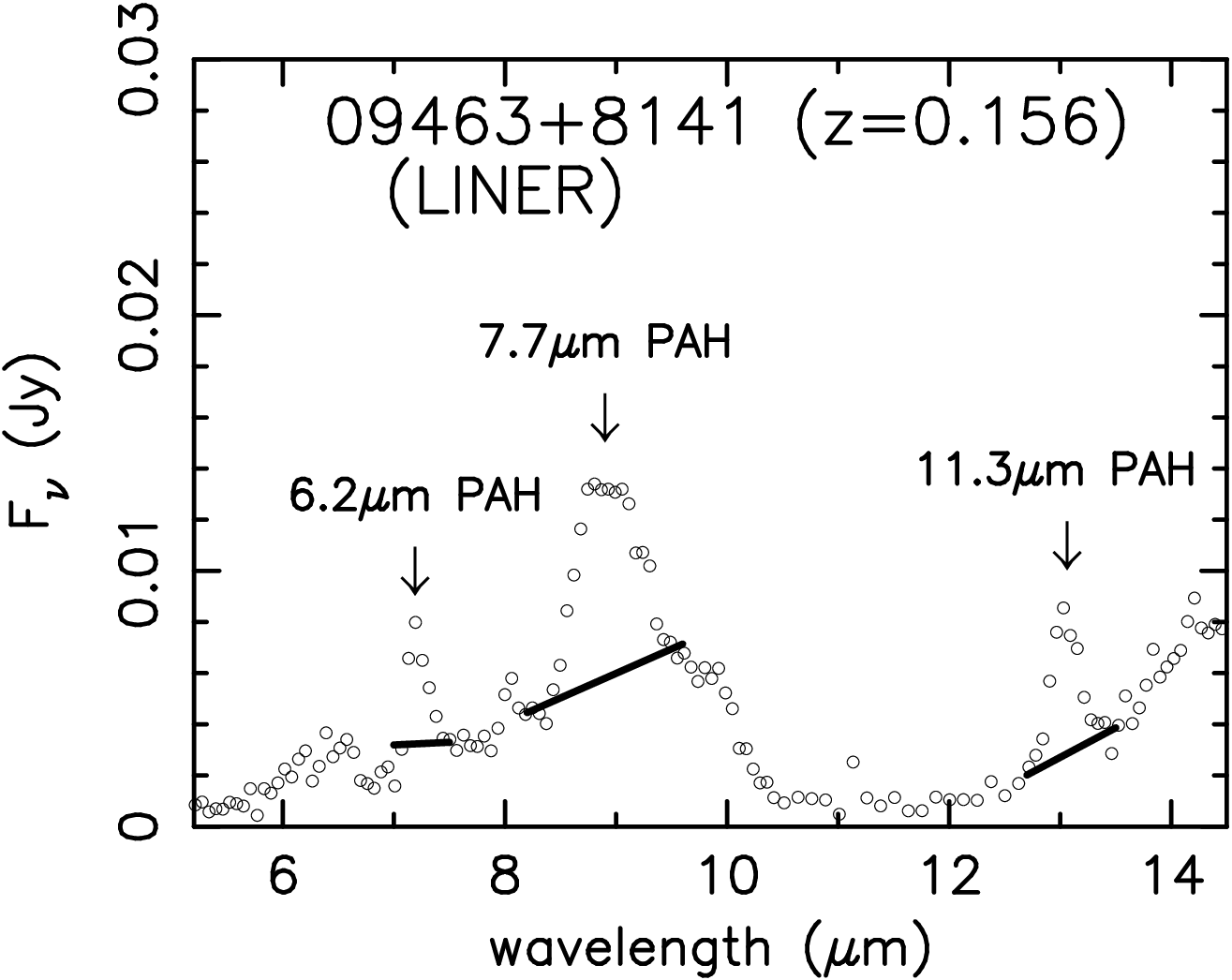} \\
\includegraphics[angle=-90,scale=.35]{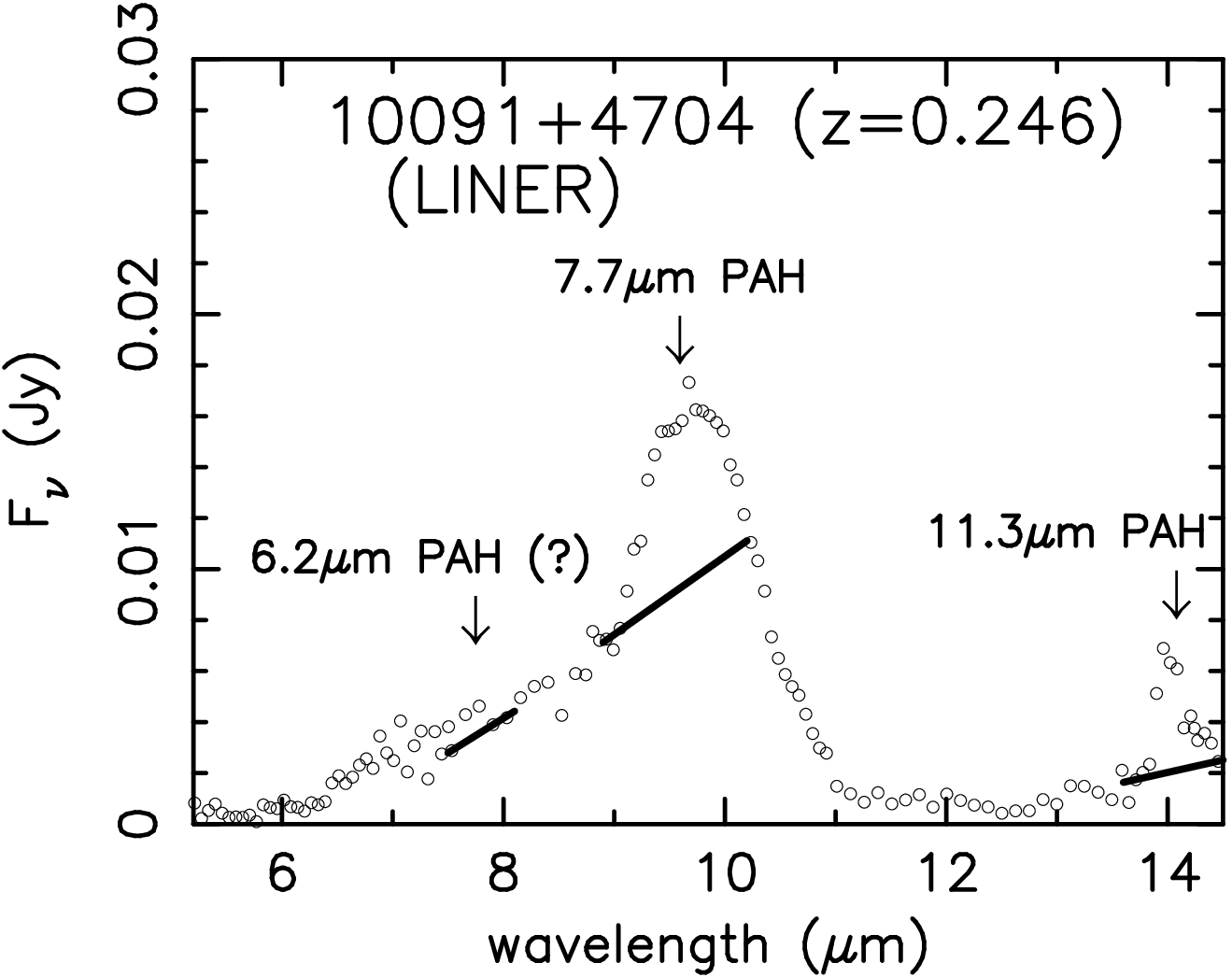} \hspace{0.3cm}
\includegraphics[angle=-90,scale=.35]{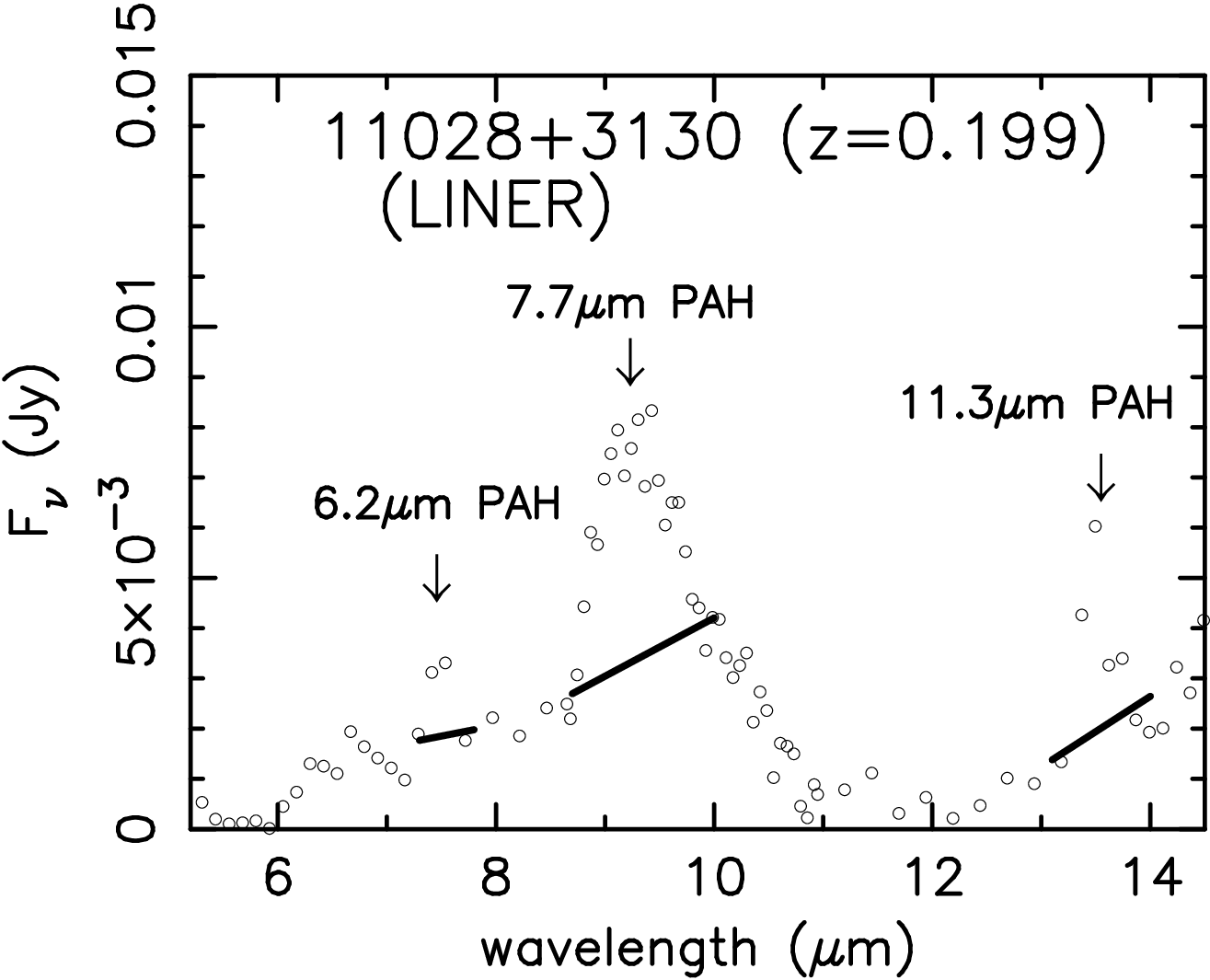} \\
\includegraphics[angle=-90,scale=.35]{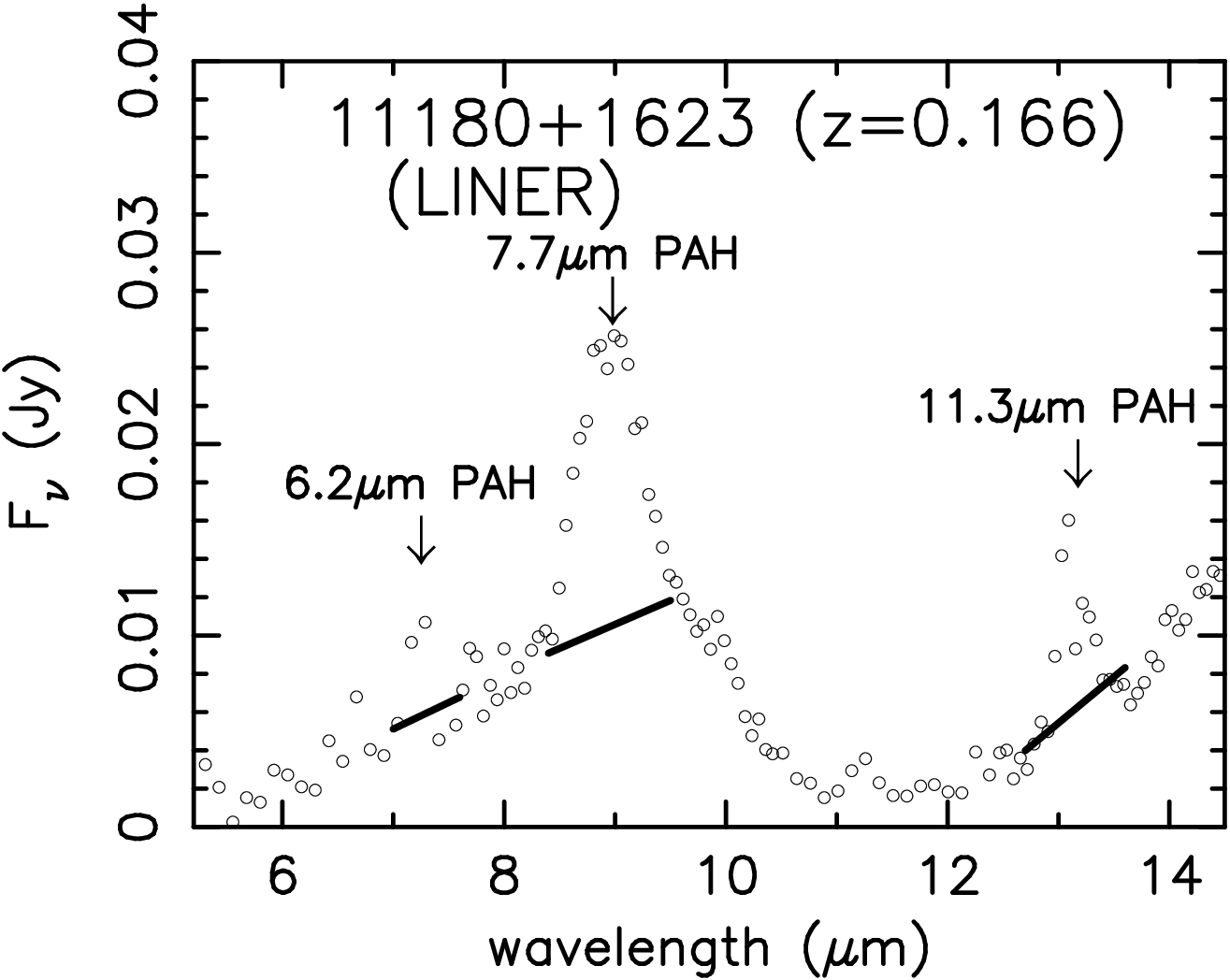} \hspace{0.3cm} 
\includegraphics[angle=-90,scale=.35]{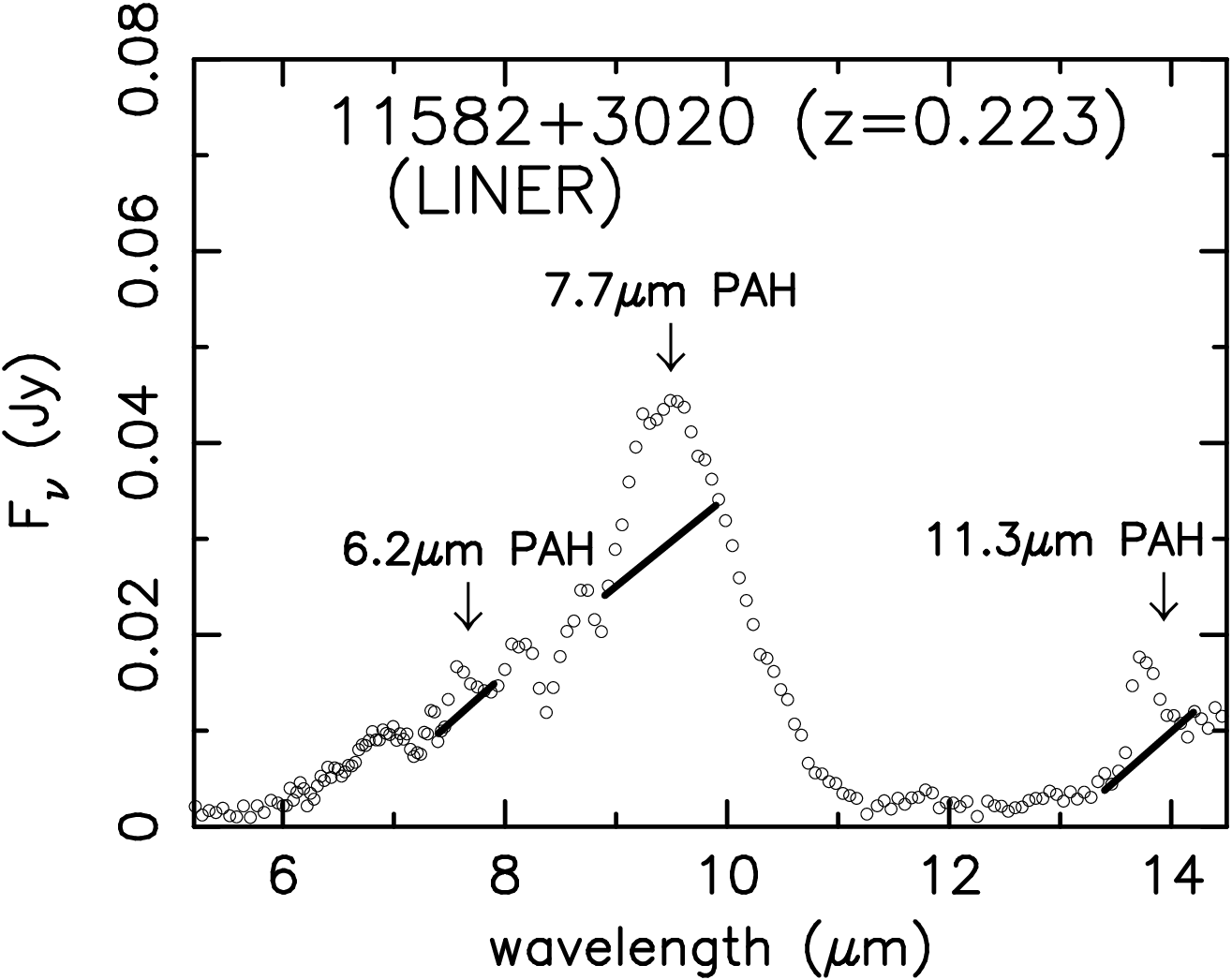} \\
\includegraphics[angle=-90,scale=.35]{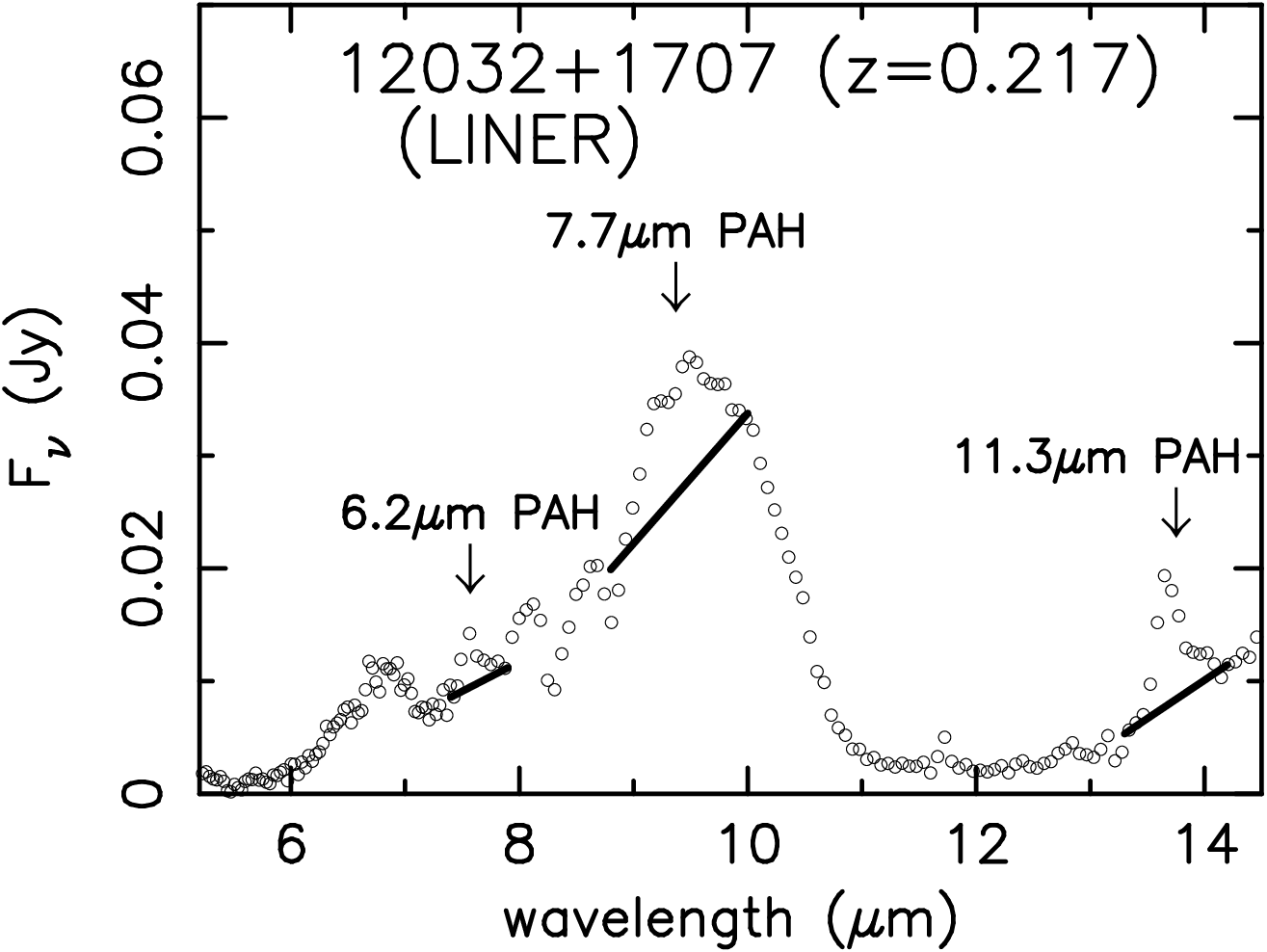} \hspace{0.3cm} 
\includegraphics[angle=-90,scale=.35]{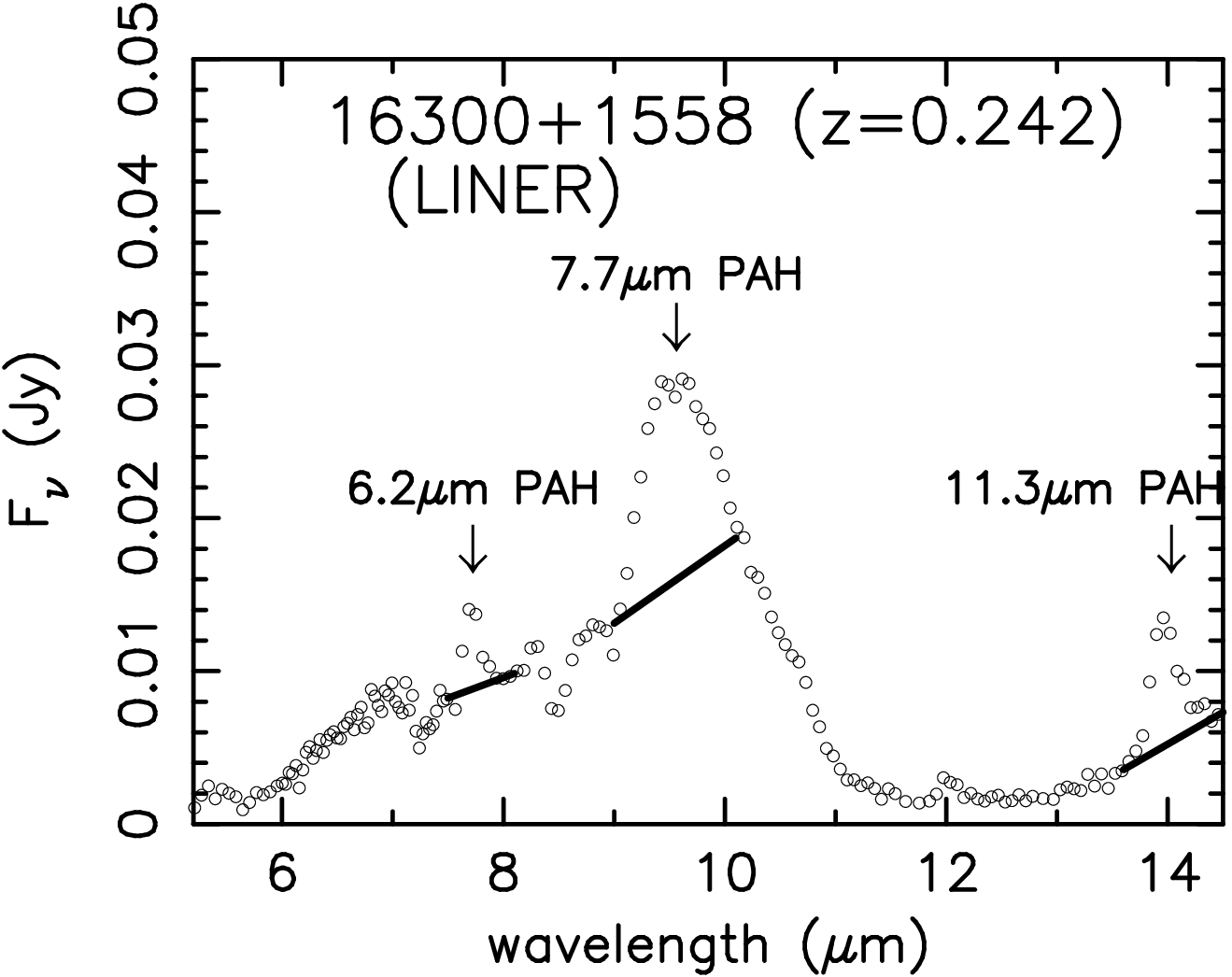} \\
\end{figure}

\begin{figure}
\includegraphics[angle=-90,scale=.35]{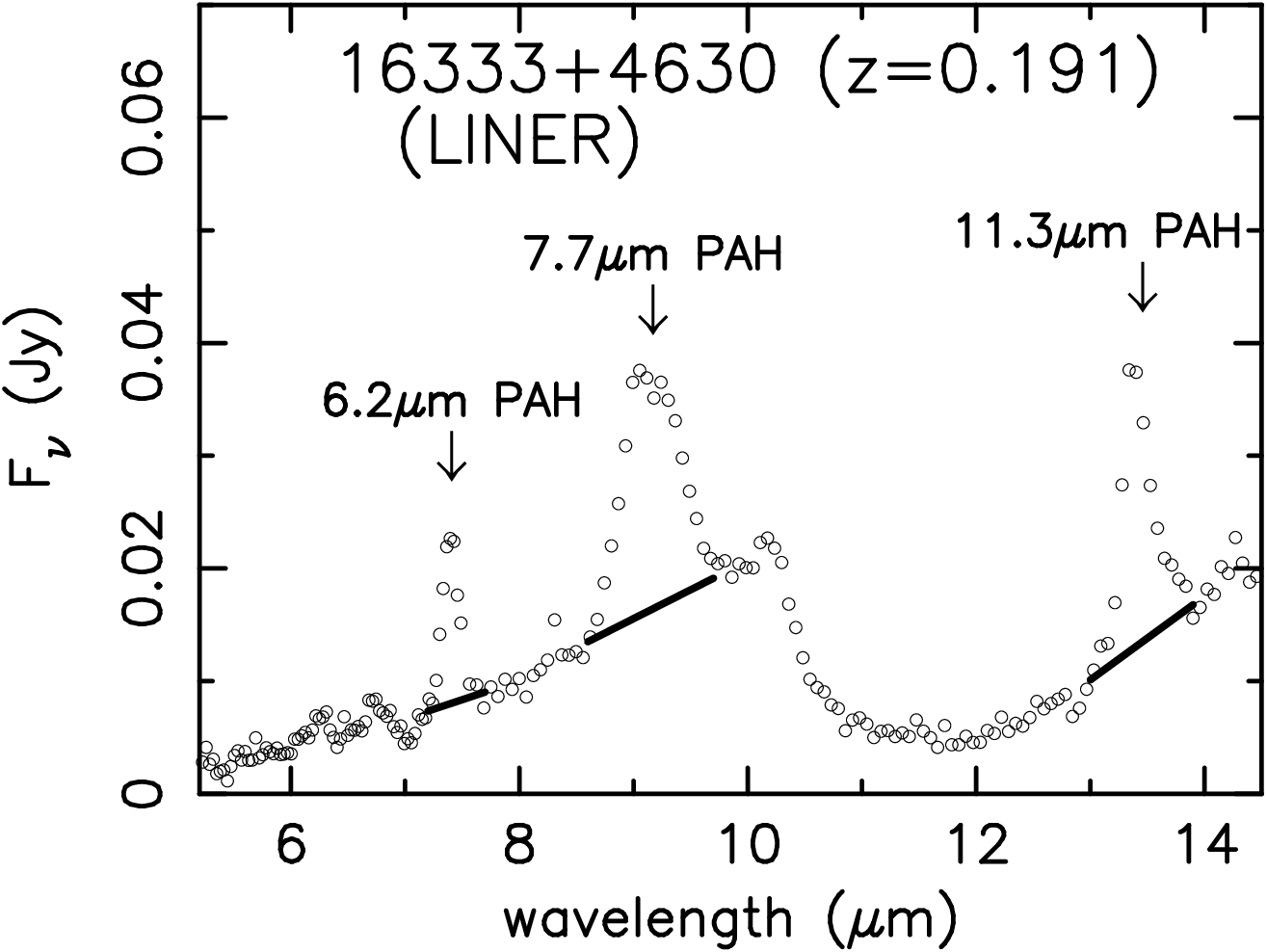} \hspace{0.3cm} 
\includegraphics[angle=-90,scale=.35]{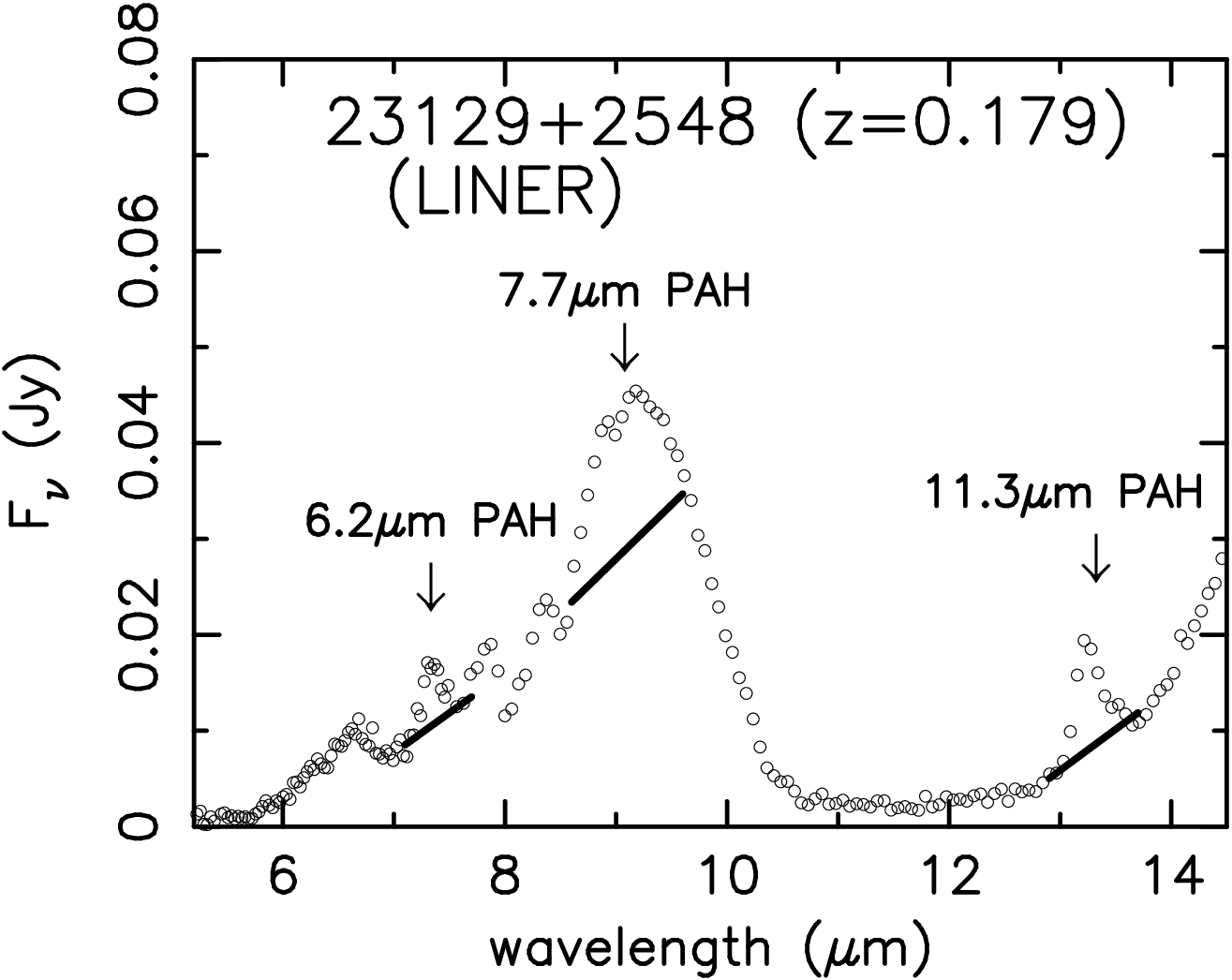} \\ 
\includegraphics[angle=-90,scale=.35]{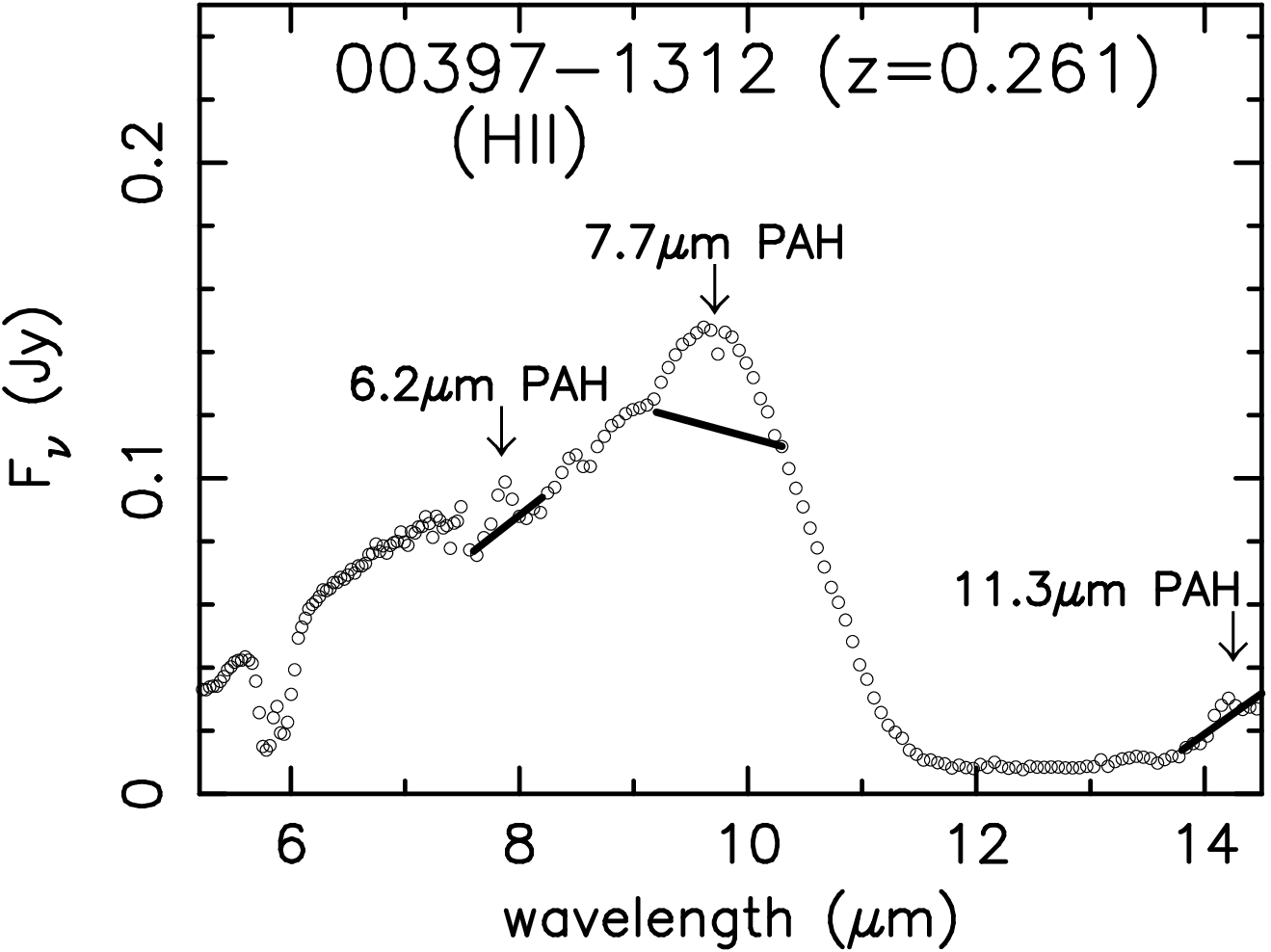} \hspace{0.3cm}
\includegraphics[angle=-90,scale=.35]{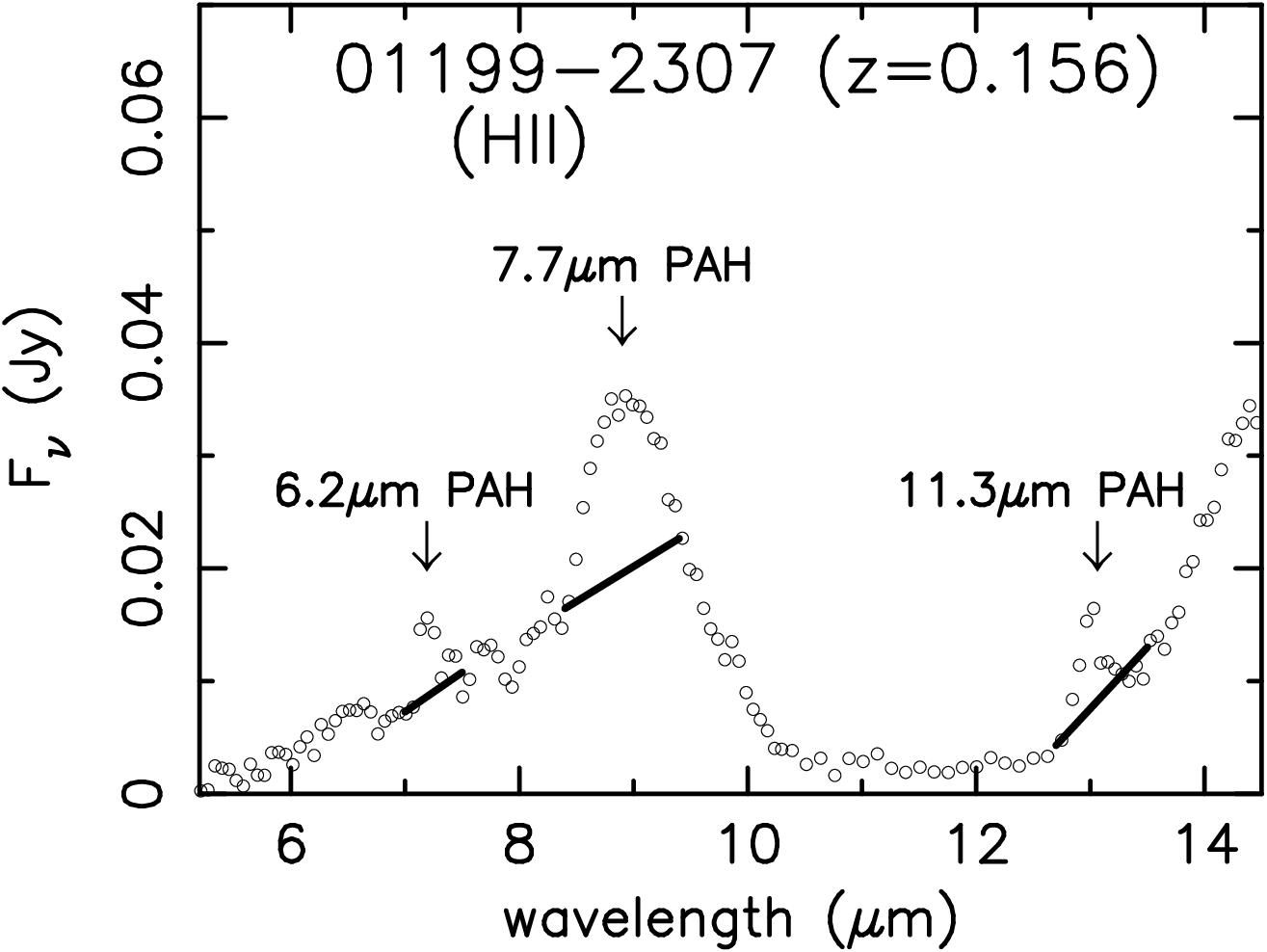} \\
\includegraphics[angle=-90,scale=.35]{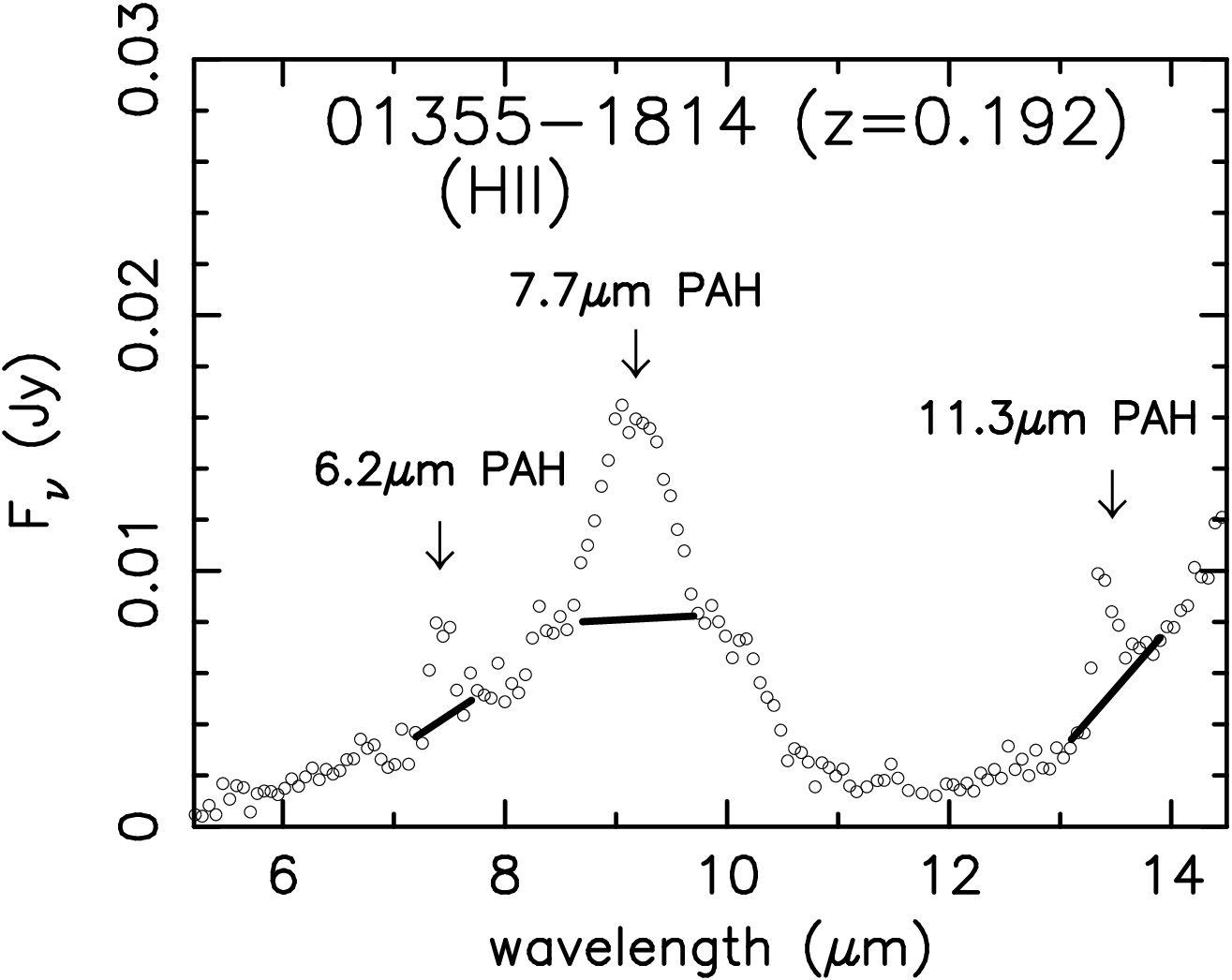} \hspace{0.3cm}
\includegraphics[angle=-90,scale=.35]{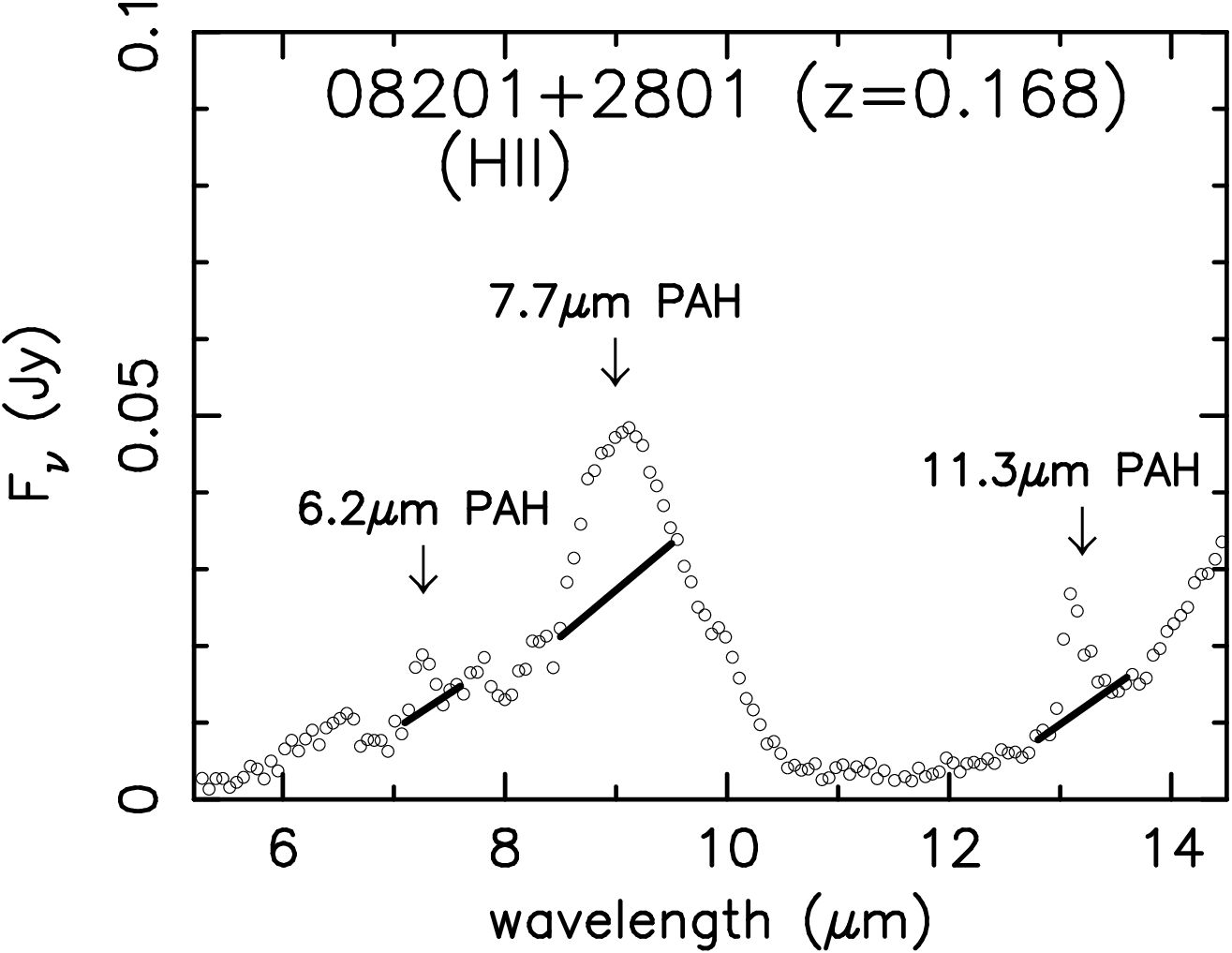} \\
\includegraphics[angle=-90,scale=.35]{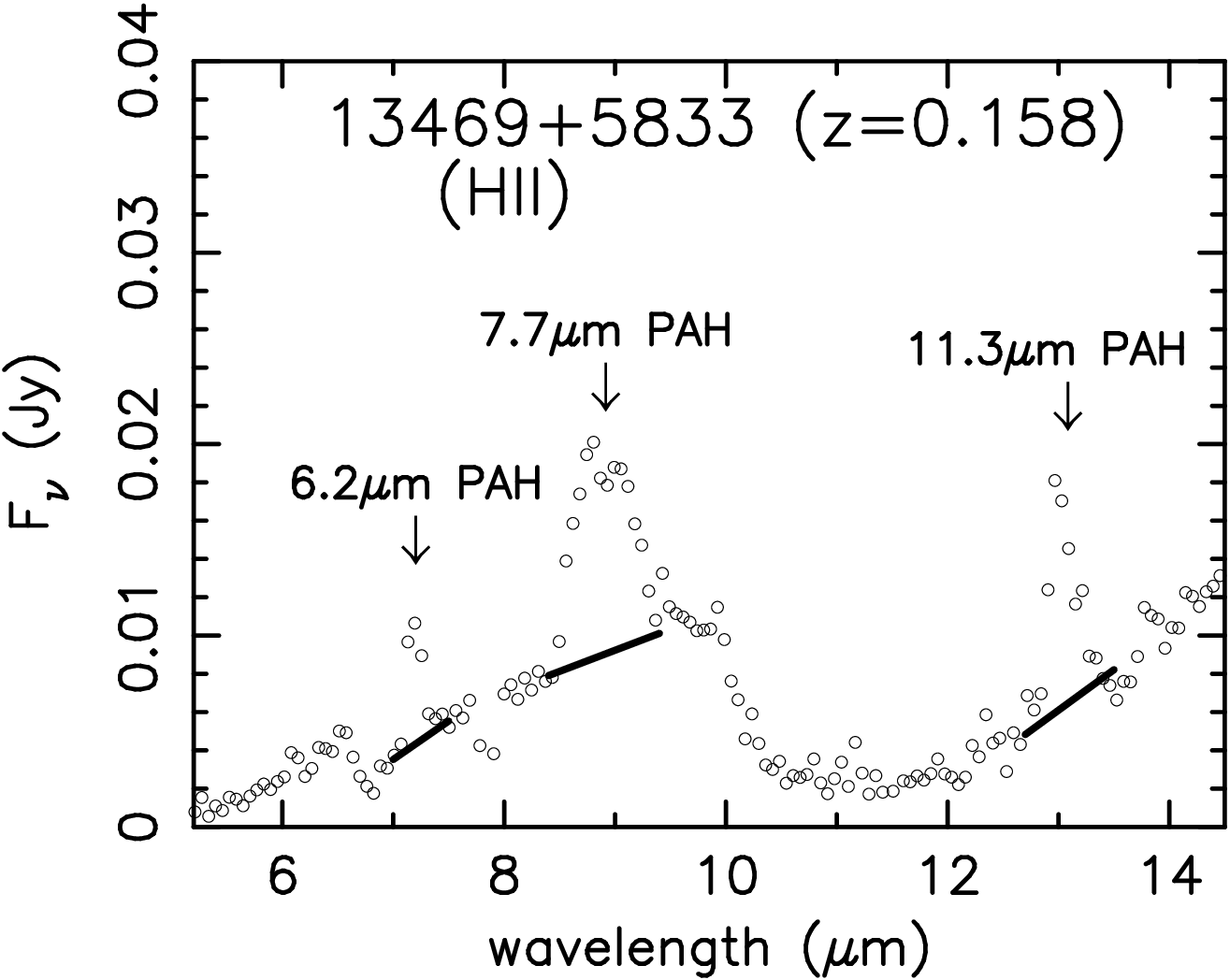} \hspace{0.3cm}
\includegraphics[angle=-90,scale=.35]{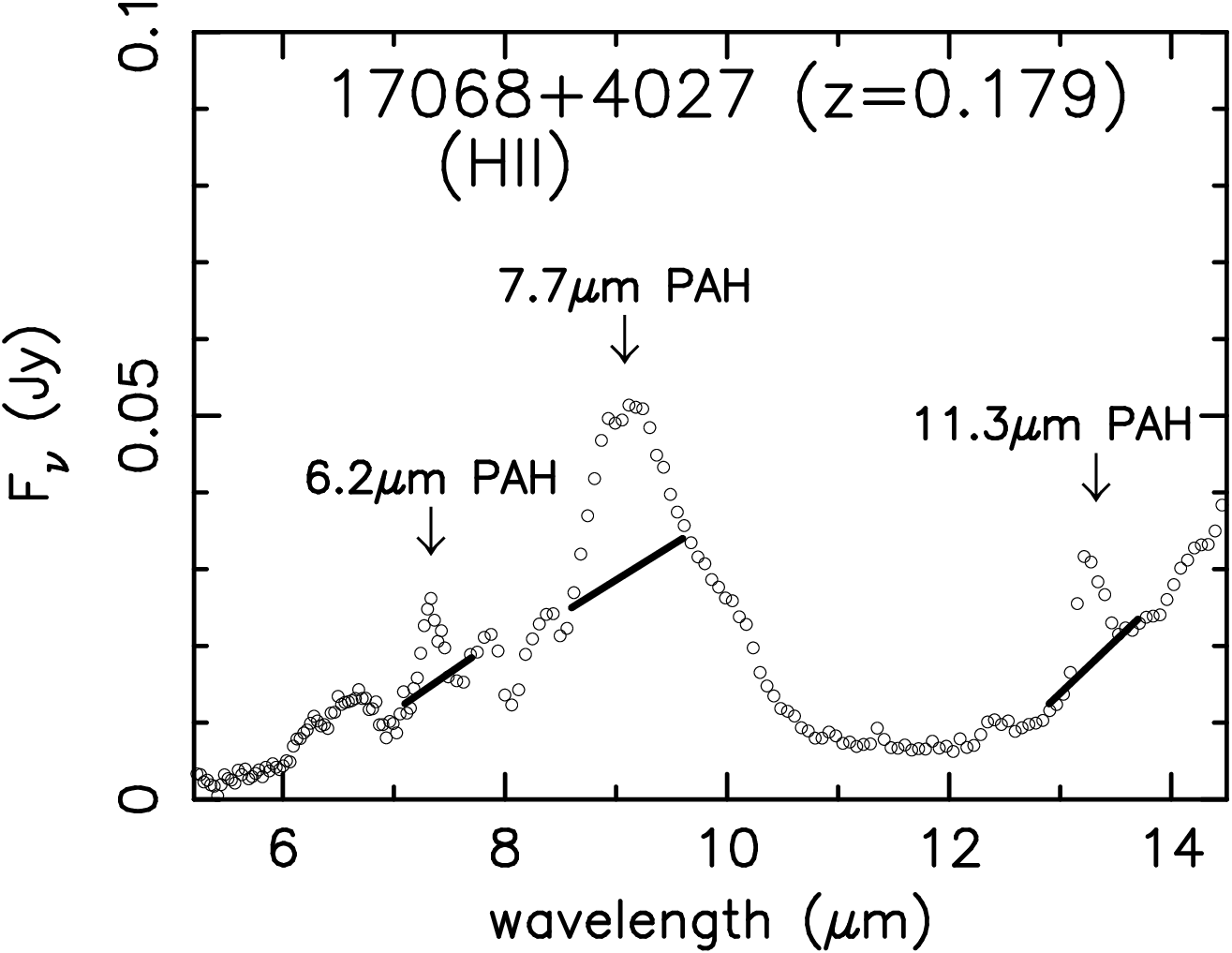} \\
\end{figure}

\clearpage

\begin{figure}
\includegraphics[angle=-90,scale=.35]{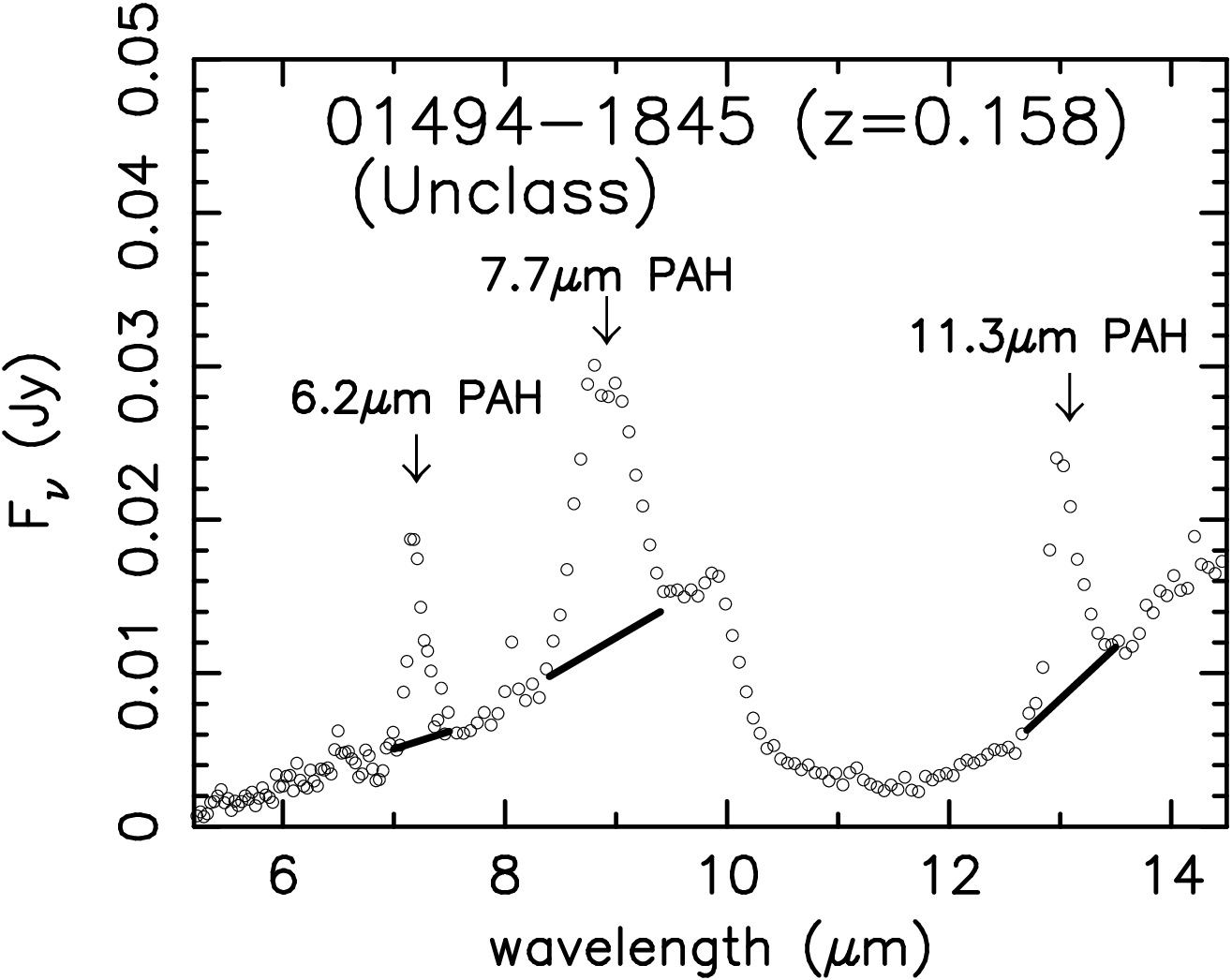} \hspace{0.3cm}
\includegraphics[angle=-90,scale=.35]{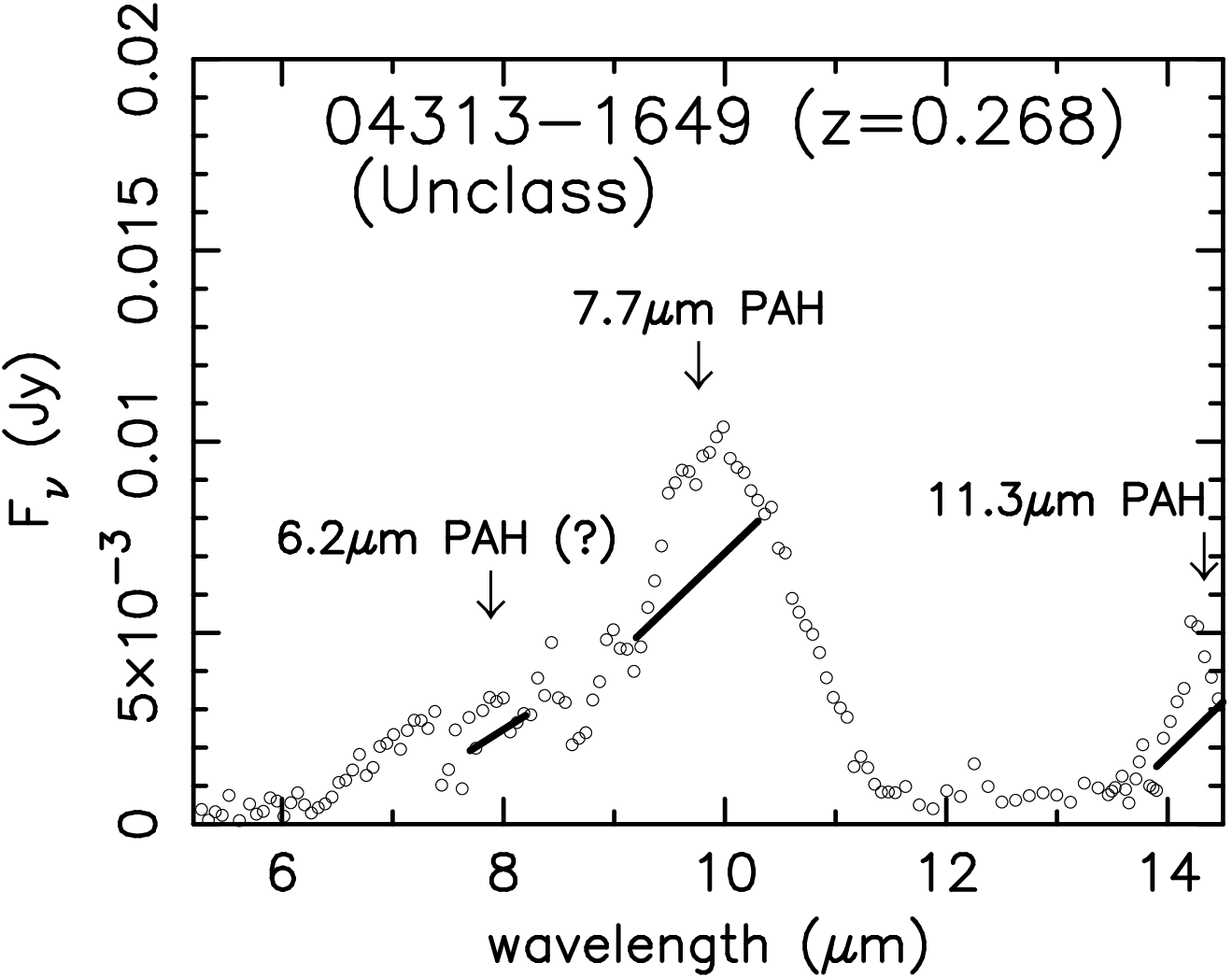} \\
\includegraphics[angle=-90,scale=.35]{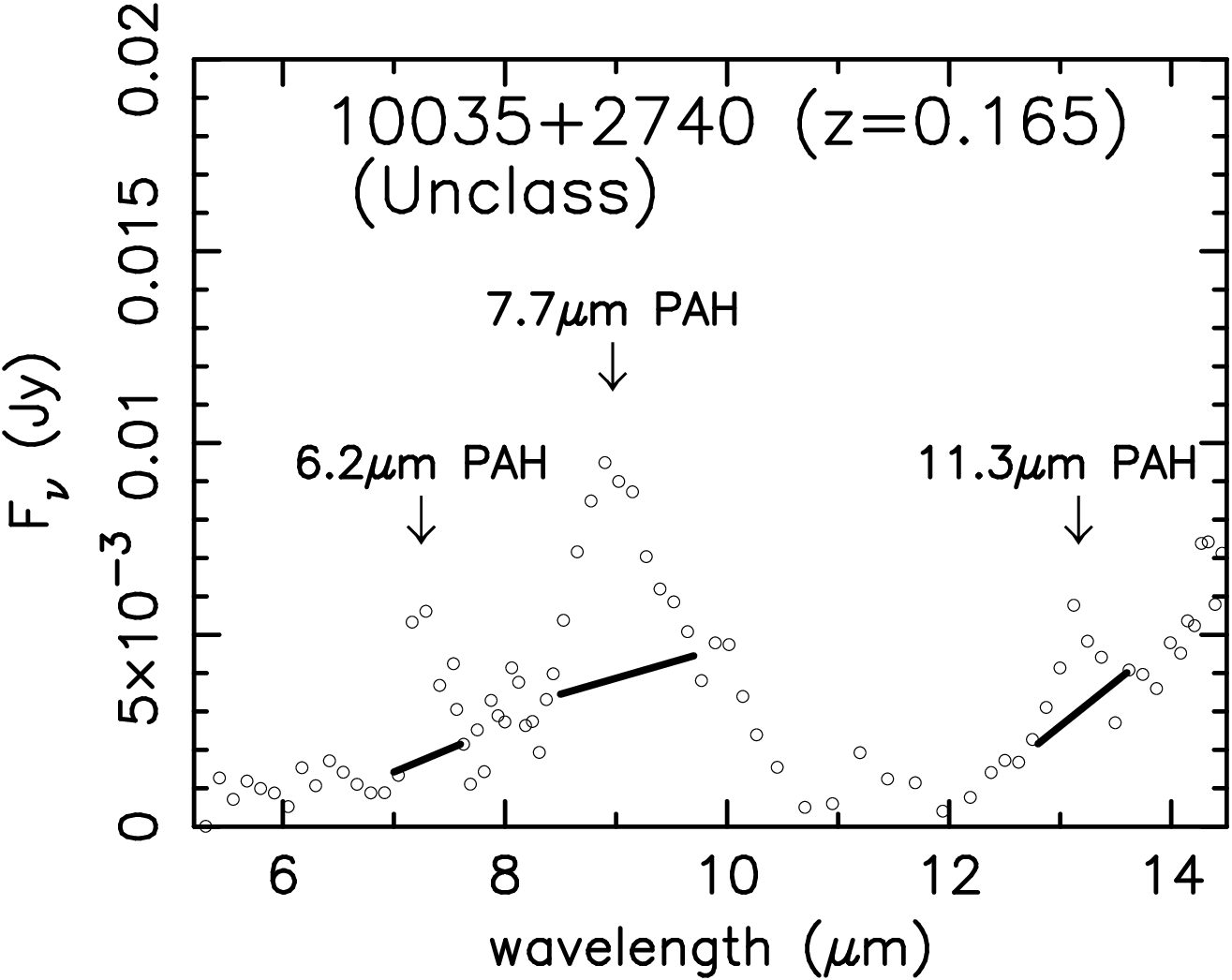} \hspace{0.3cm}
\includegraphics[angle=-90,scale=.35]{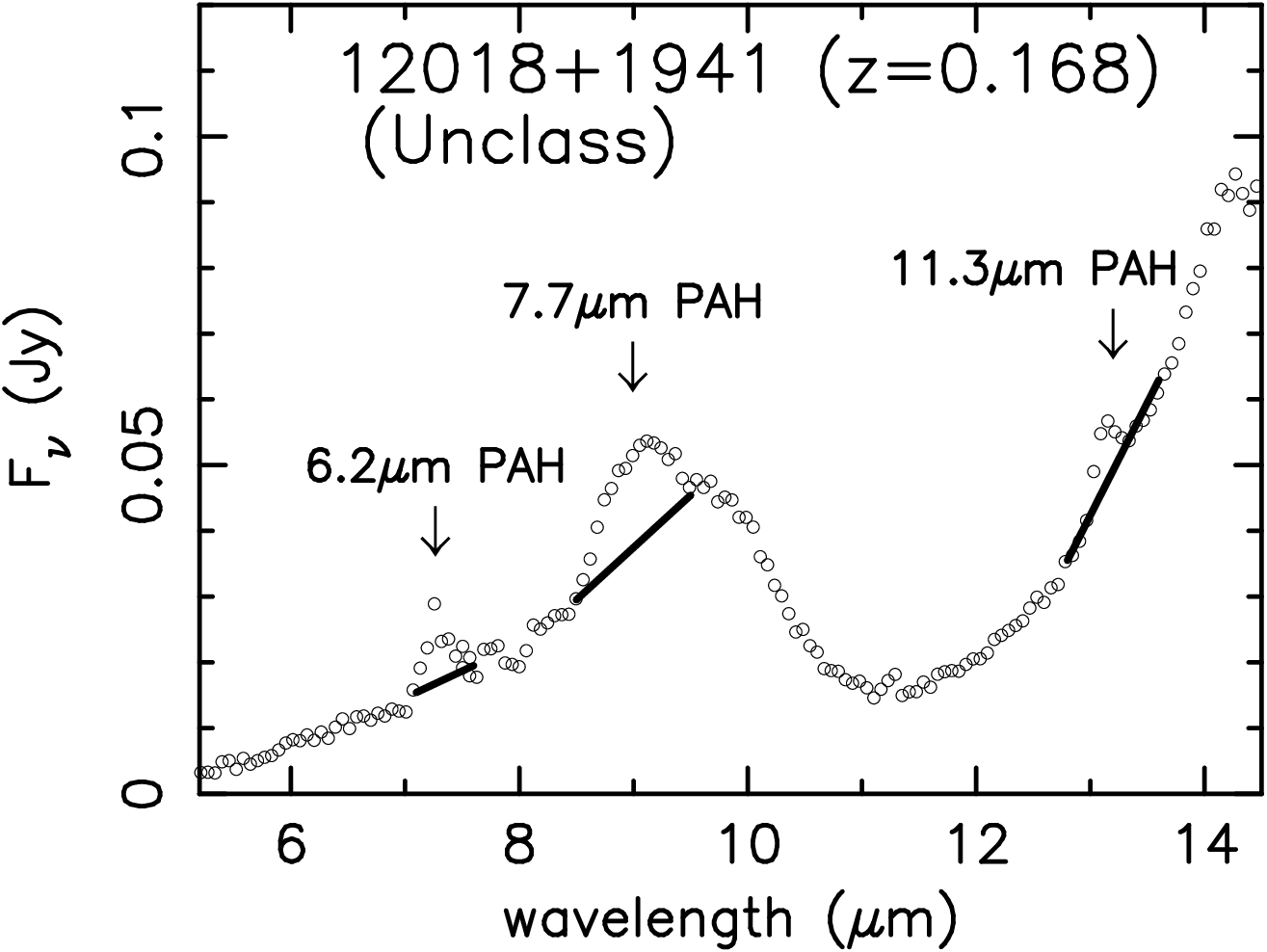} \\
\caption{
{\it Spitzer} IRS spectra of all ULIRGs at $\lambda_{\rm obs}$ = 5.2--14.5
$\mu$m, selected to display the PAH emission features in
detail.  The abscissa and ordinate are, respectively, observed
wavelength in $\mu$m and flux in Jy, both shown on a linear scale.
The expected wavelengths of the 6.2 $\mu$m, 7.7 $\mu$m, and 11.3 $\mu$m
PAH emission features are indicated as down arrows.  
The mark ``(?)'' is added when detection is unclear.  
The solid lines are adopted continuum levels to estimate the strengths 
of the PAH emission features. 
}
\end{figure}

\begin{figure}
\includegraphics[angle=-90,scale=.35]{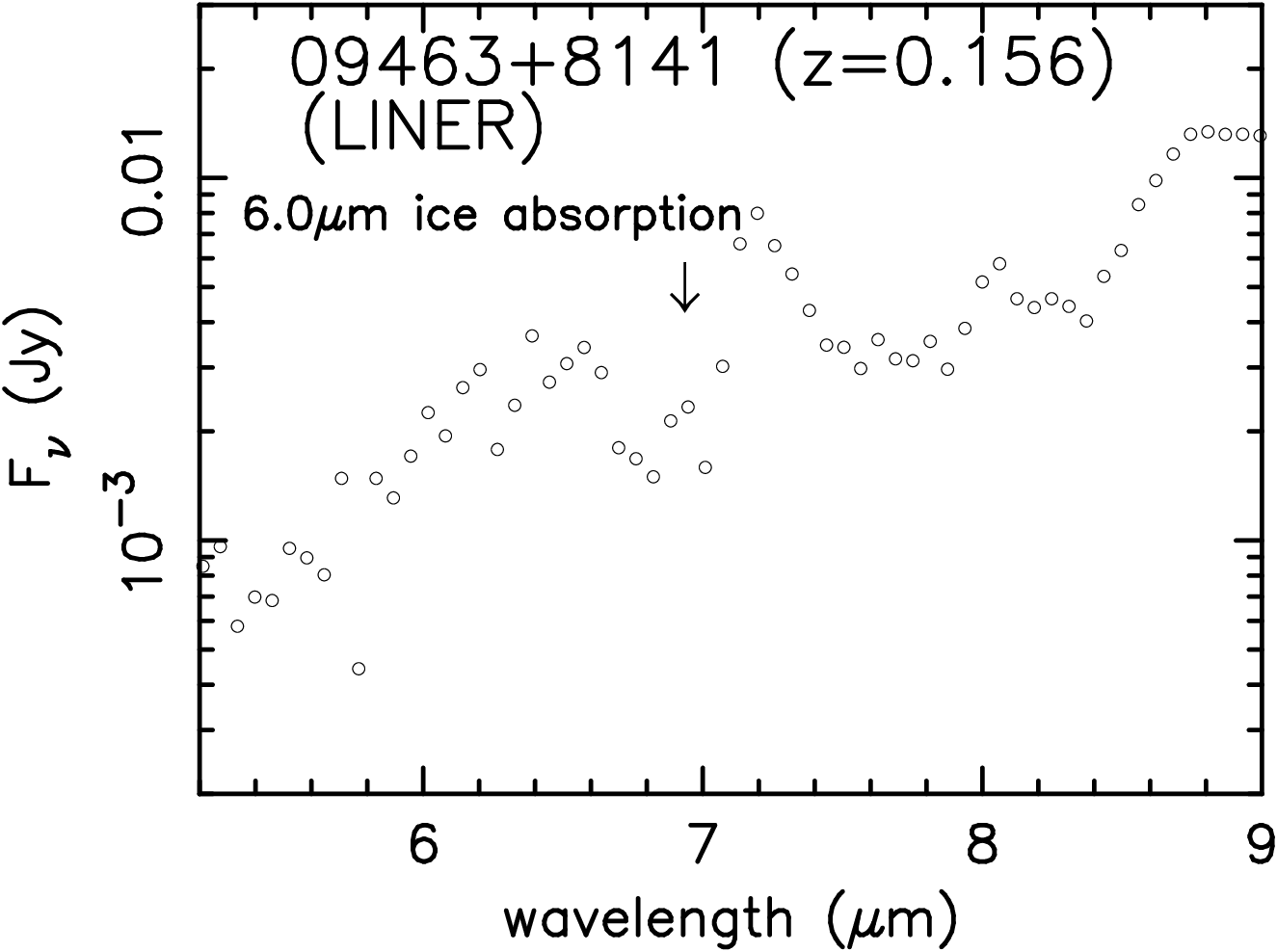} \hspace{0.3cm} 
\includegraphics[angle=-90,scale=.35]{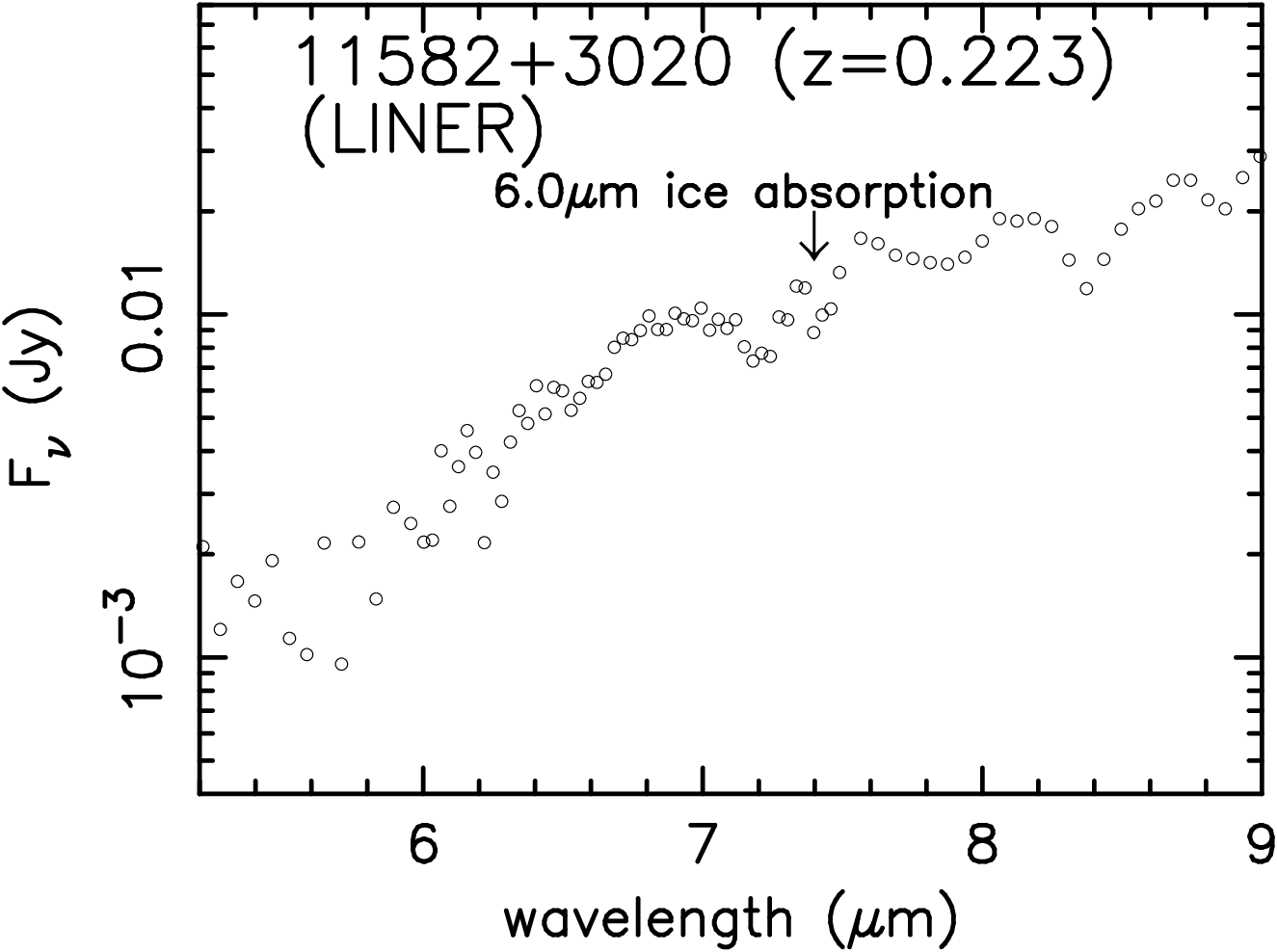} \\
\includegraphics[angle=-90,scale=.35]{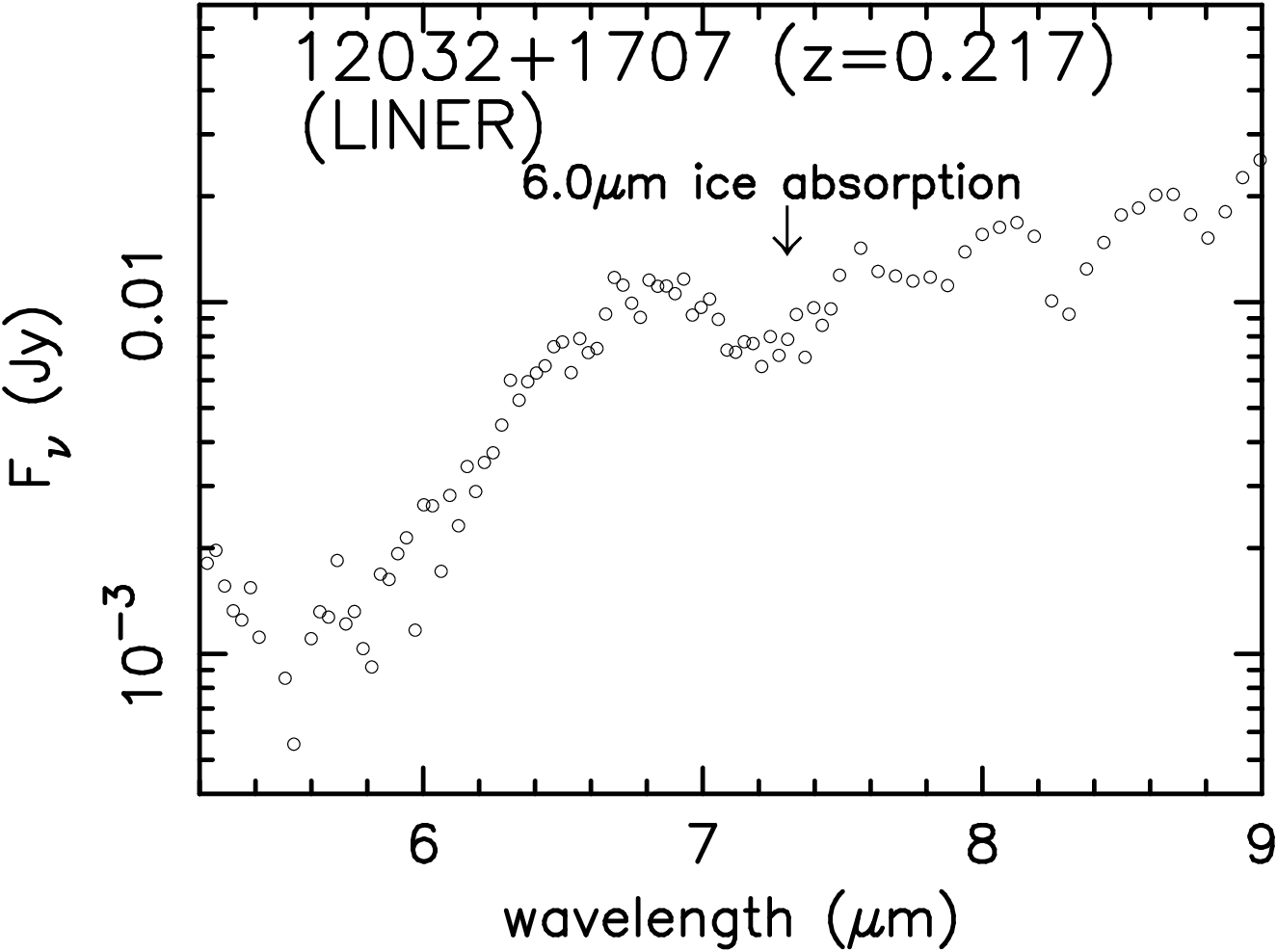} \hspace{0.3cm} 
\includegraphics[angle=-90,scale=.35]{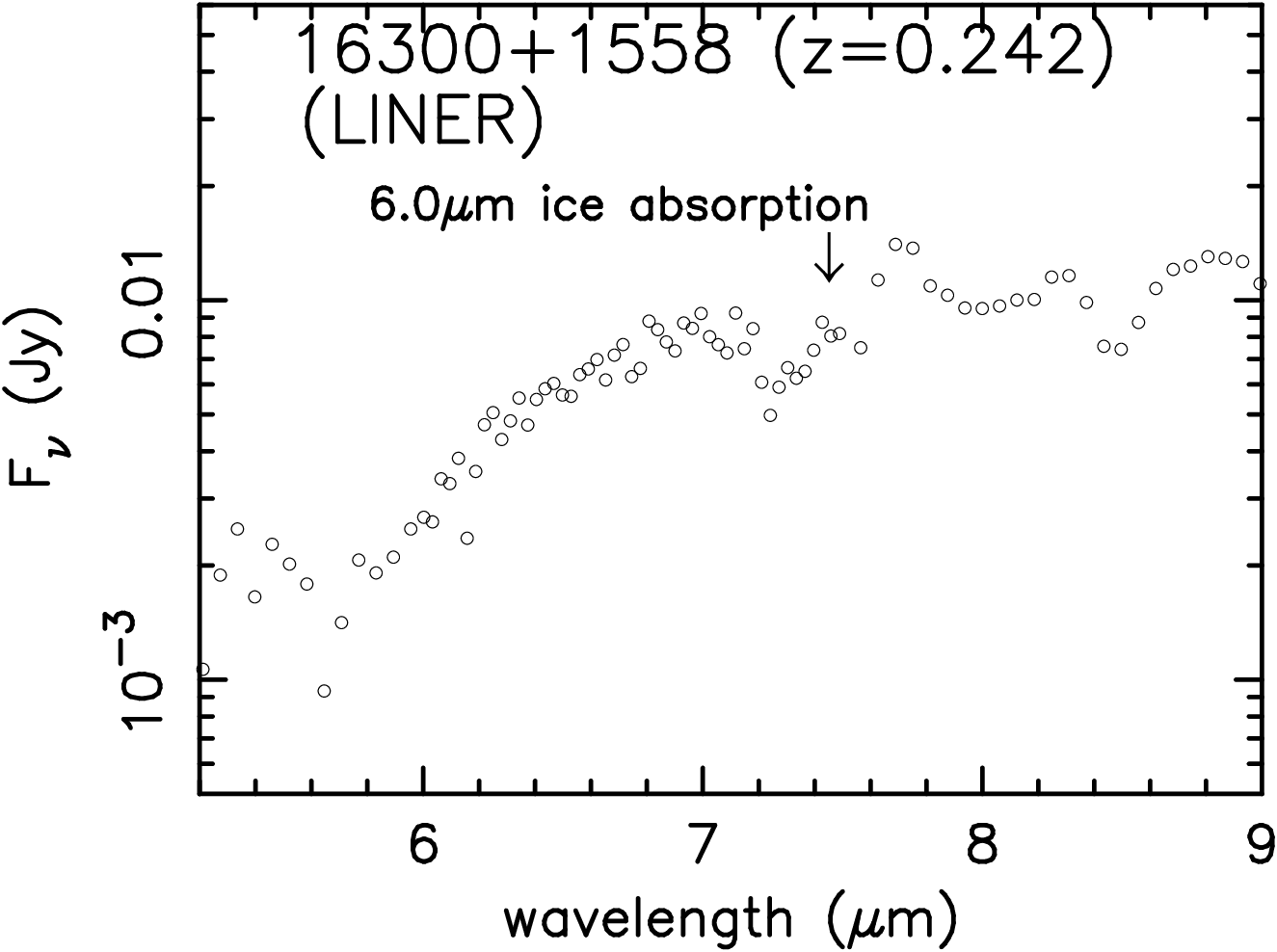} \\
\includegraphics[angle=-90,scale=.35]{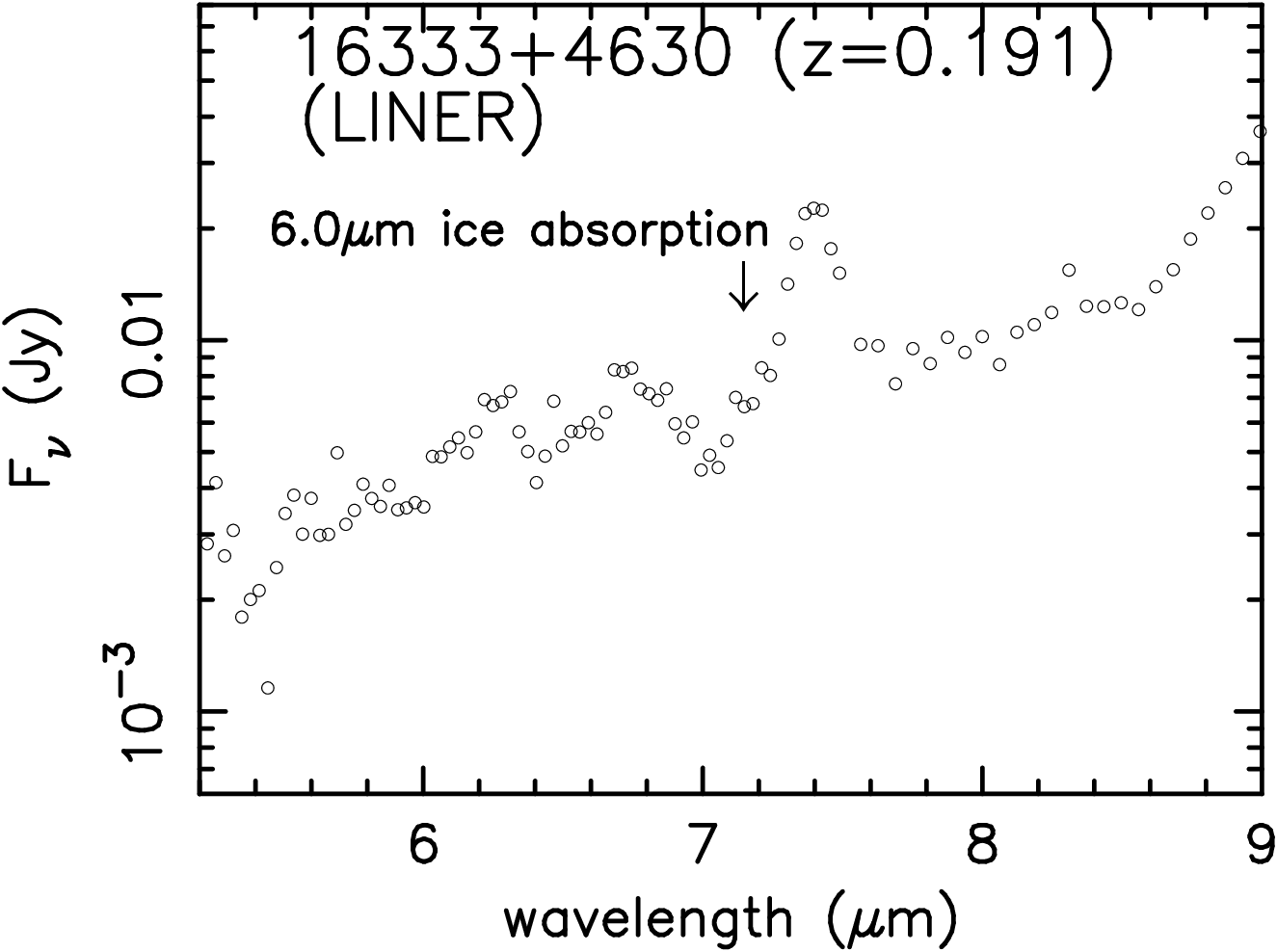} \hspace{0.3cm} 
\includegraphics[angle=-90,scale=.35]{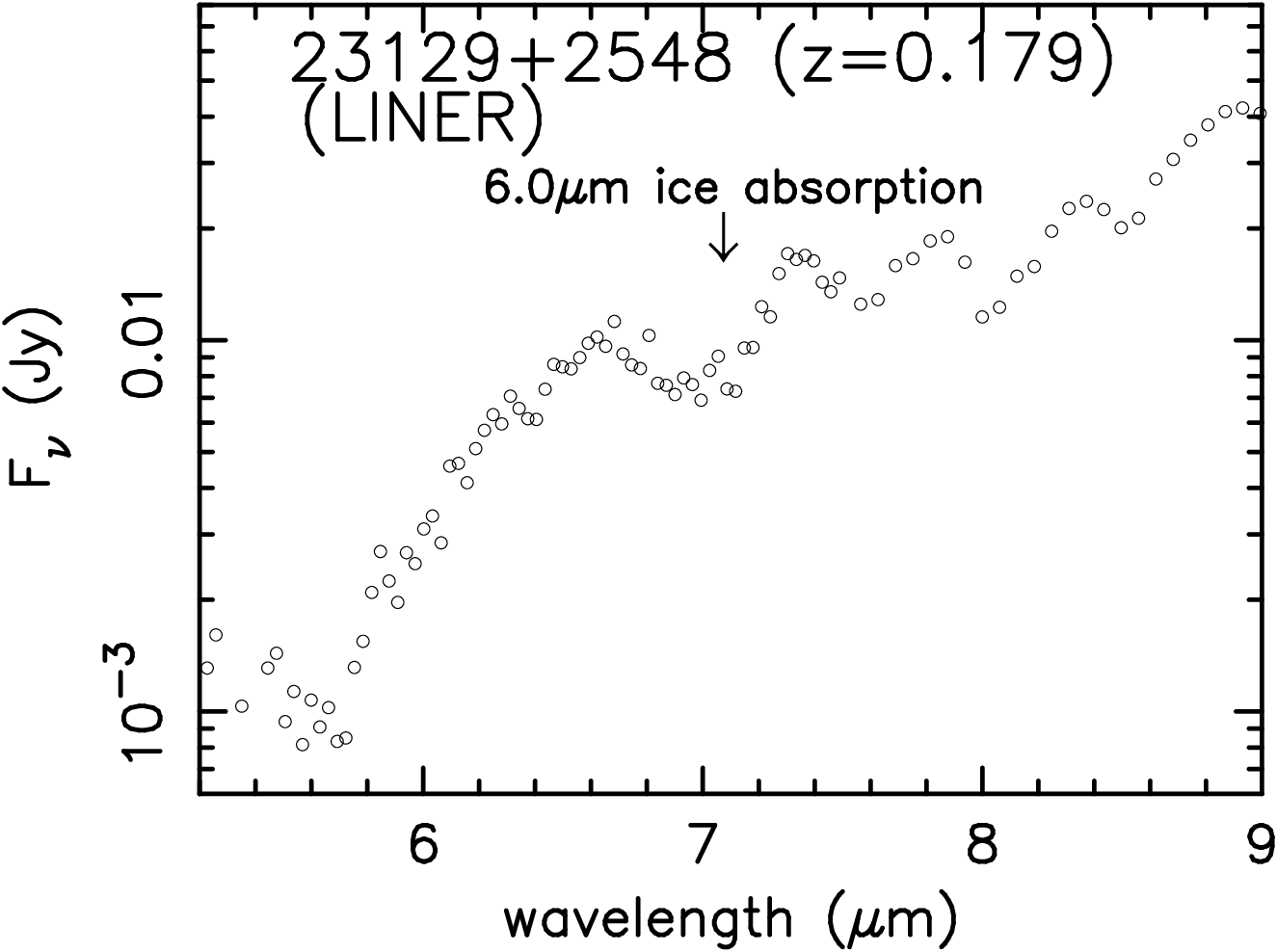} \\ 
\includegraphics[angle=-90,scale=.35]{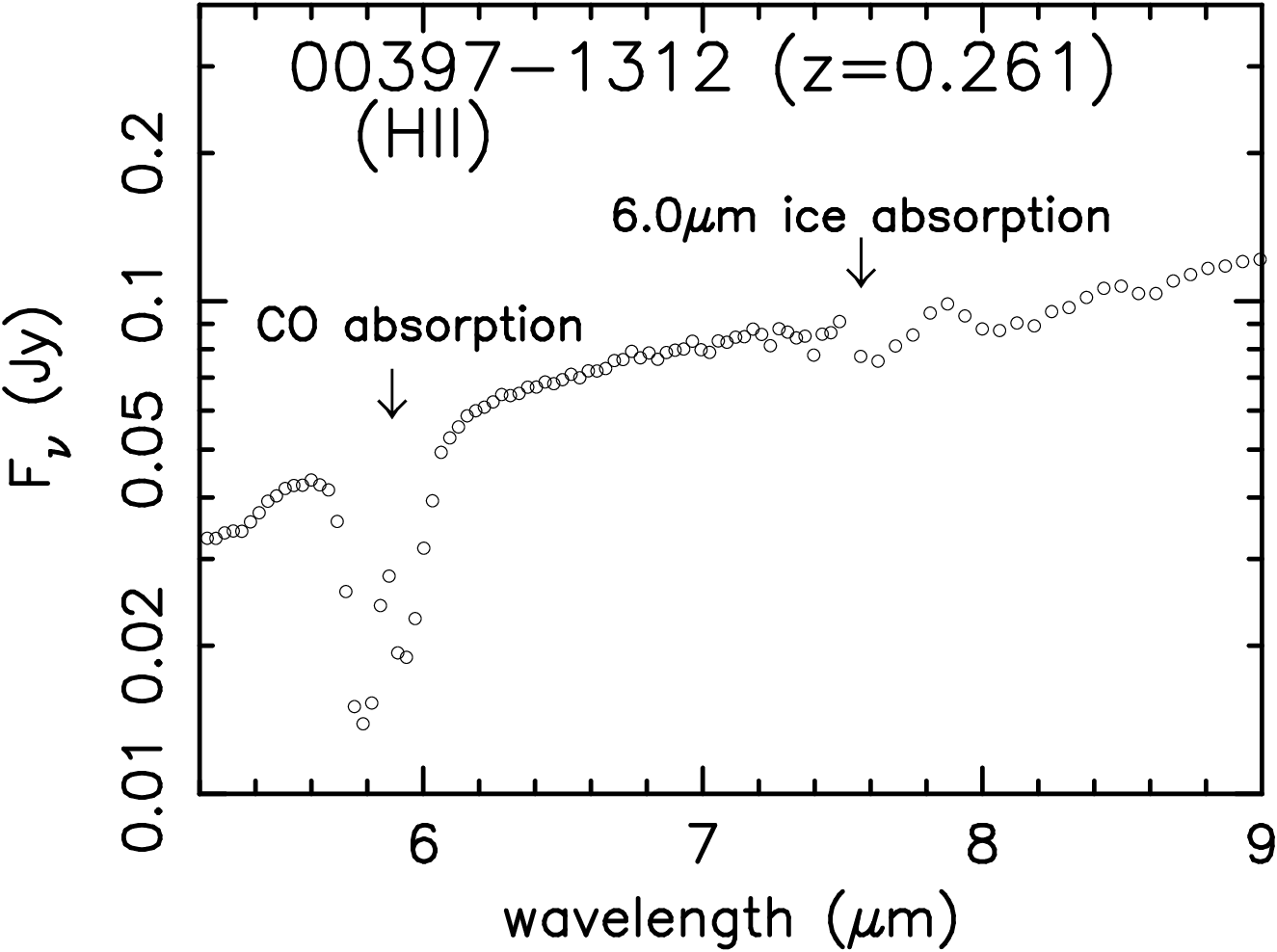} \hspace{0.3cm}
\includegraphics[angle=-90,scale=.35]{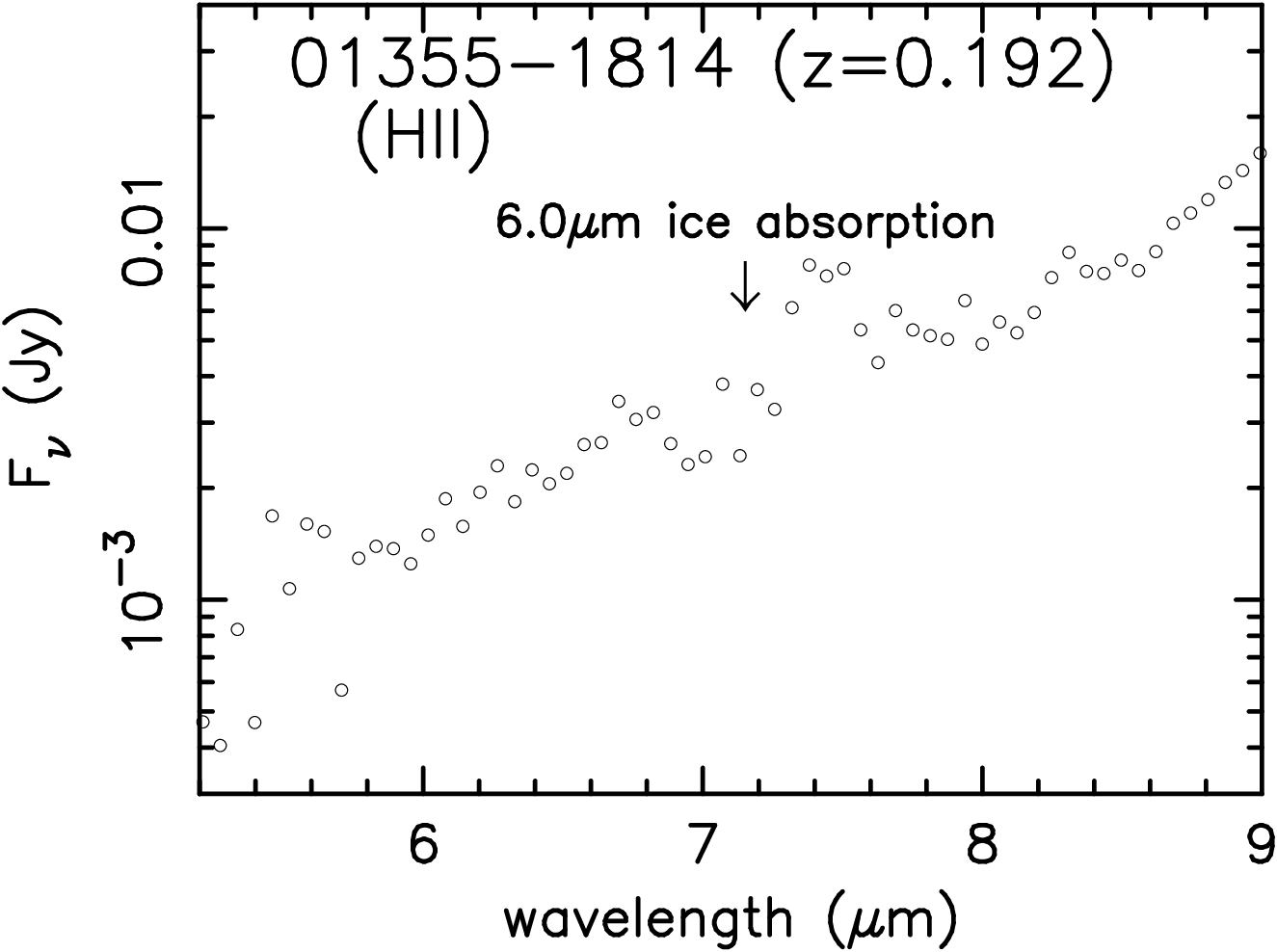} \\ 
\end{figure}

\clearpage

\begin{figure}
\includegraphics[angle=-90,scale=.35]{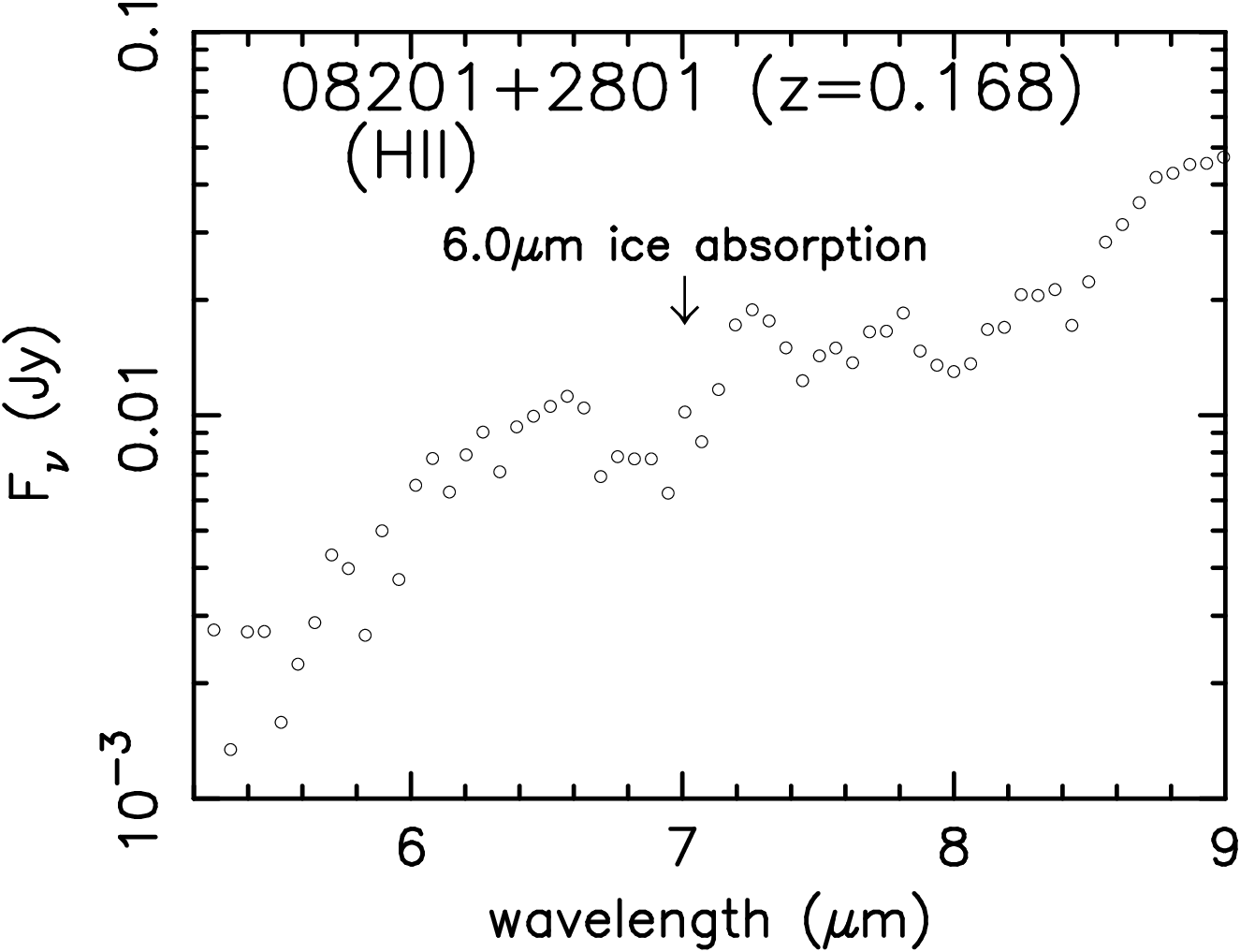} \hspace{0.3cm}
\includegraphics[angle=-90,scale=.35]{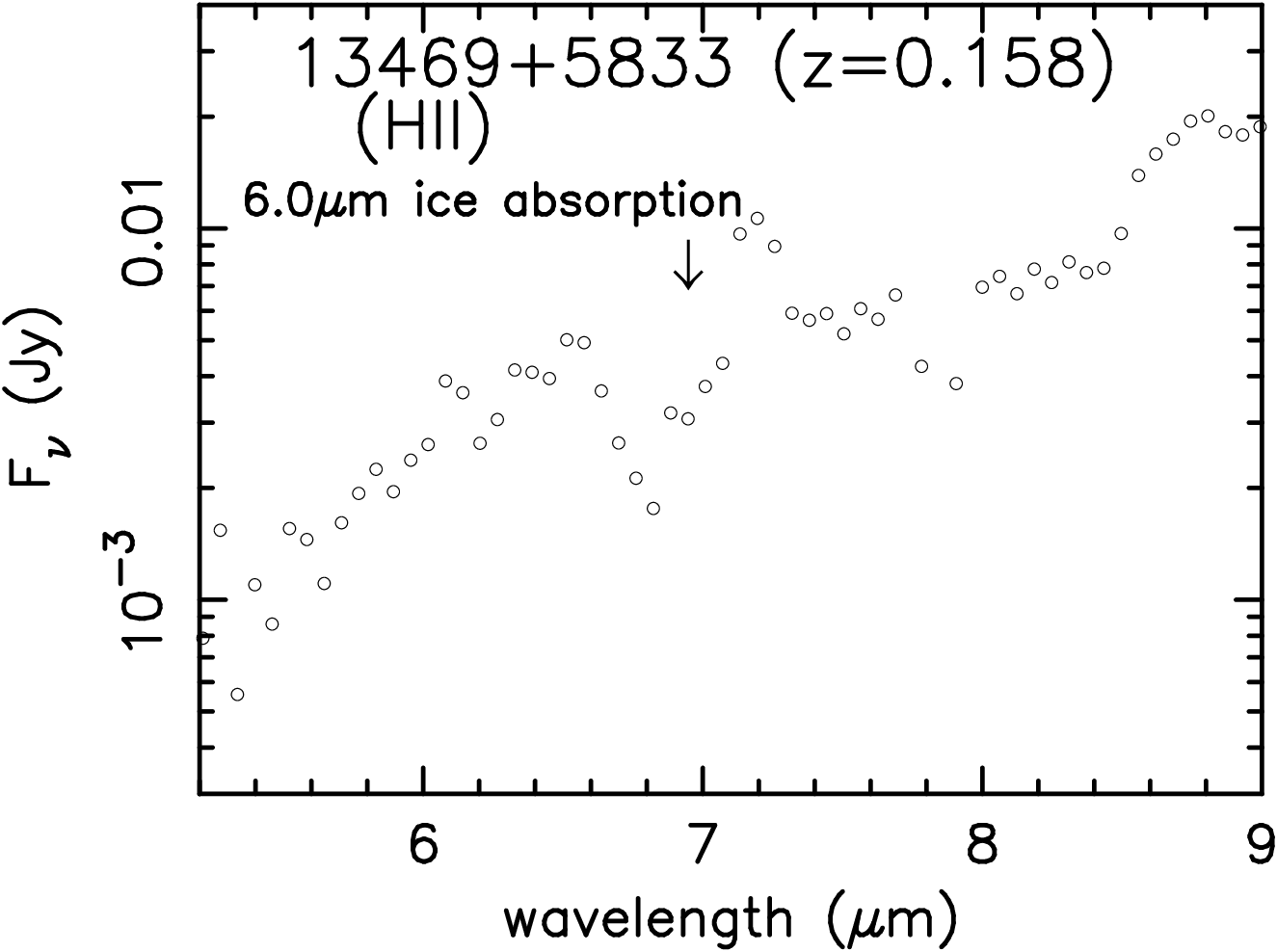} \\ 
\includegraphics[angle=-90,scale=.35]{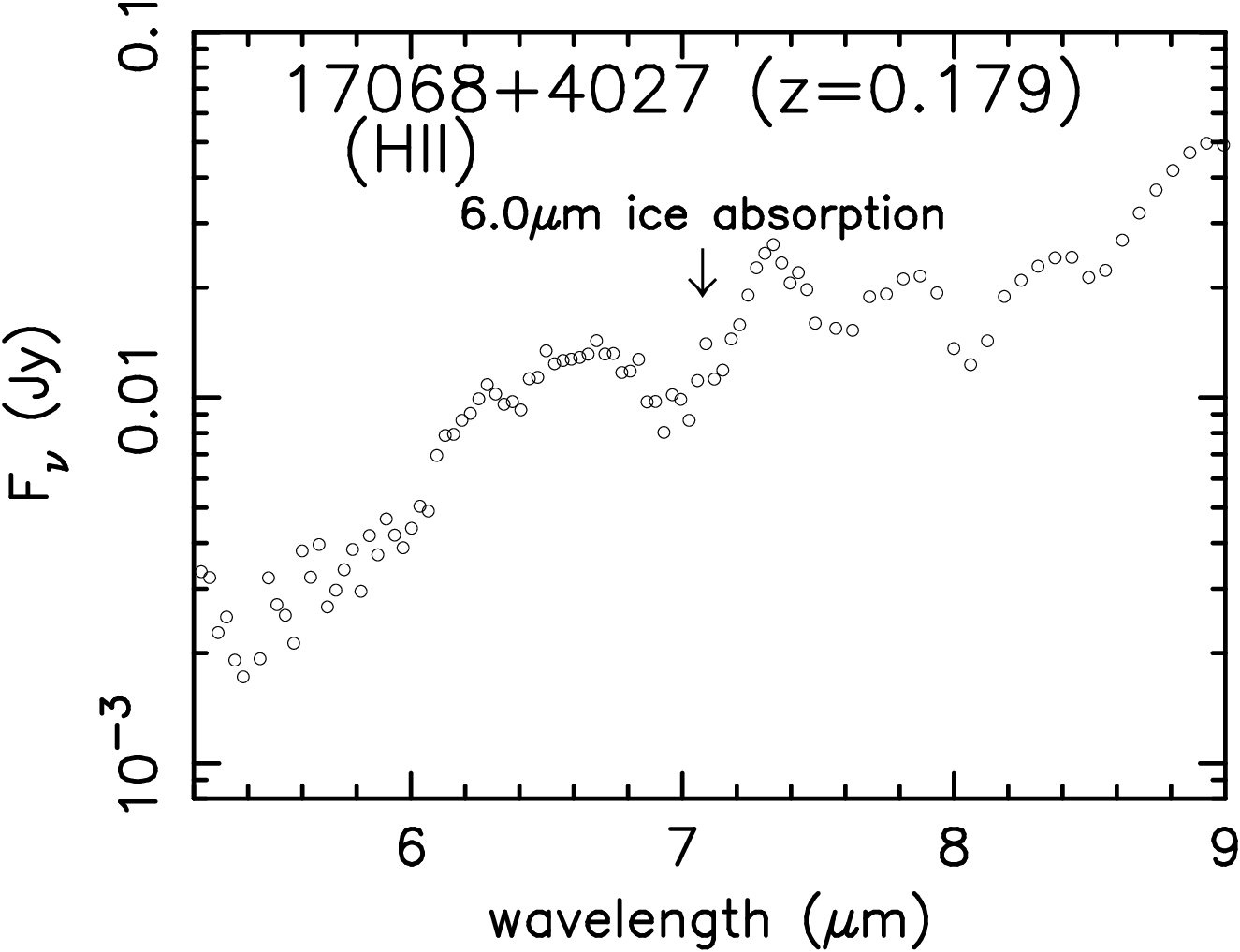} \hspace{0.3cm}
\includegraphics[angle=-90,scale=.35]{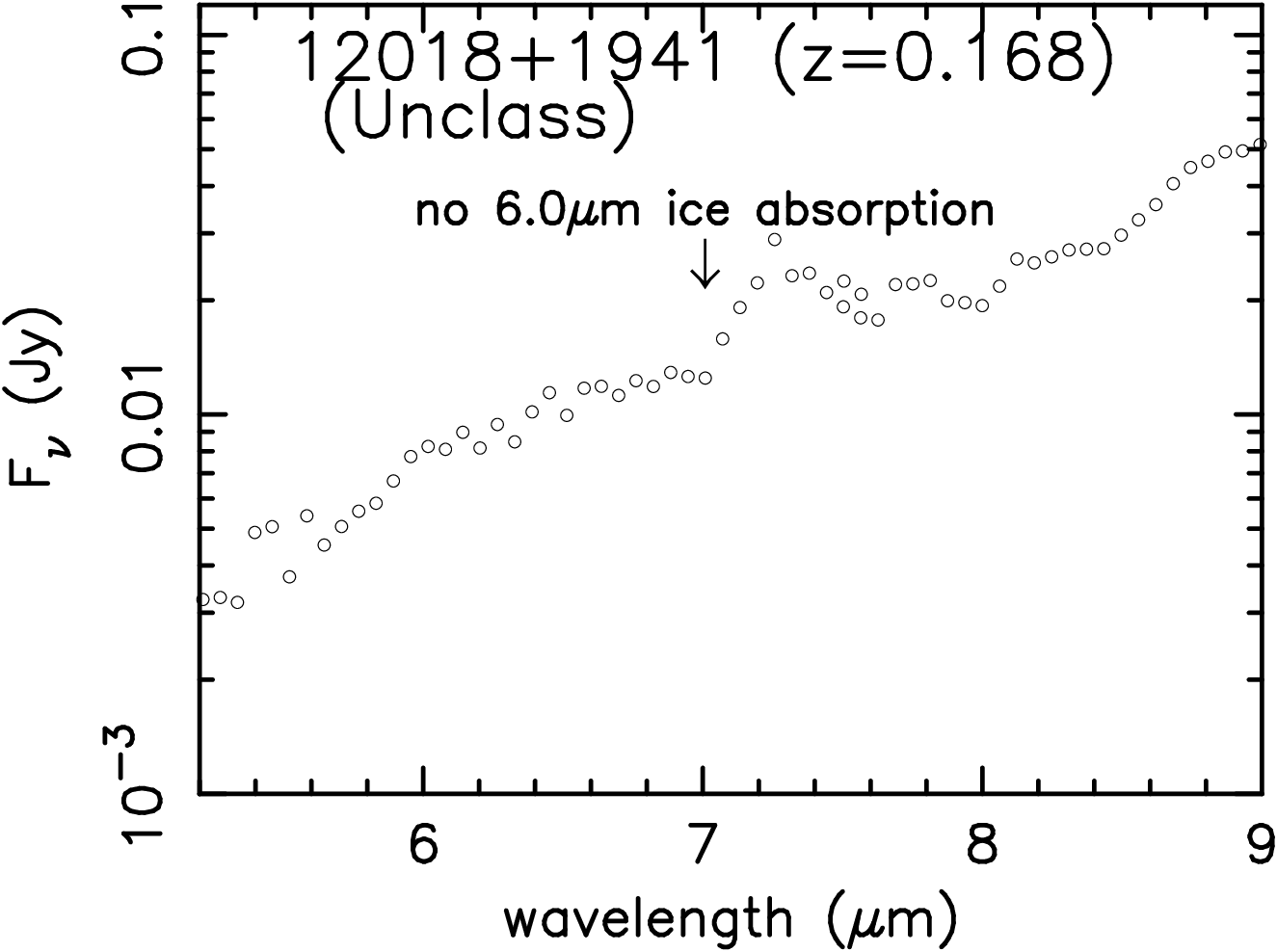} \\ 
\caption{
{\it Spitzer} IRS spectra at $\lambda_{\rm obs}$ = 5.2--9 $\mu$m 
for ULIRGs
displaying clear 6.0 $\mu$m H$_{2}$O ice absorption features (marked with
``6.0$\mu$m ice absorption'' in the first 11 plots).  
The spectrum of IRAS 12018+1941, marked ``no 6.0$\mu$m ice
absorption'', is shown as an example of an undetected ice absorption
feature.  The abscissa is observed wavelength in $\mu$m on a linear
scale, and the ordinate is flux in Jy in decimal logarithmic scale.}
\end{figure}

\begin{figure}
\includegraphics[angle=-90,scale=.35]{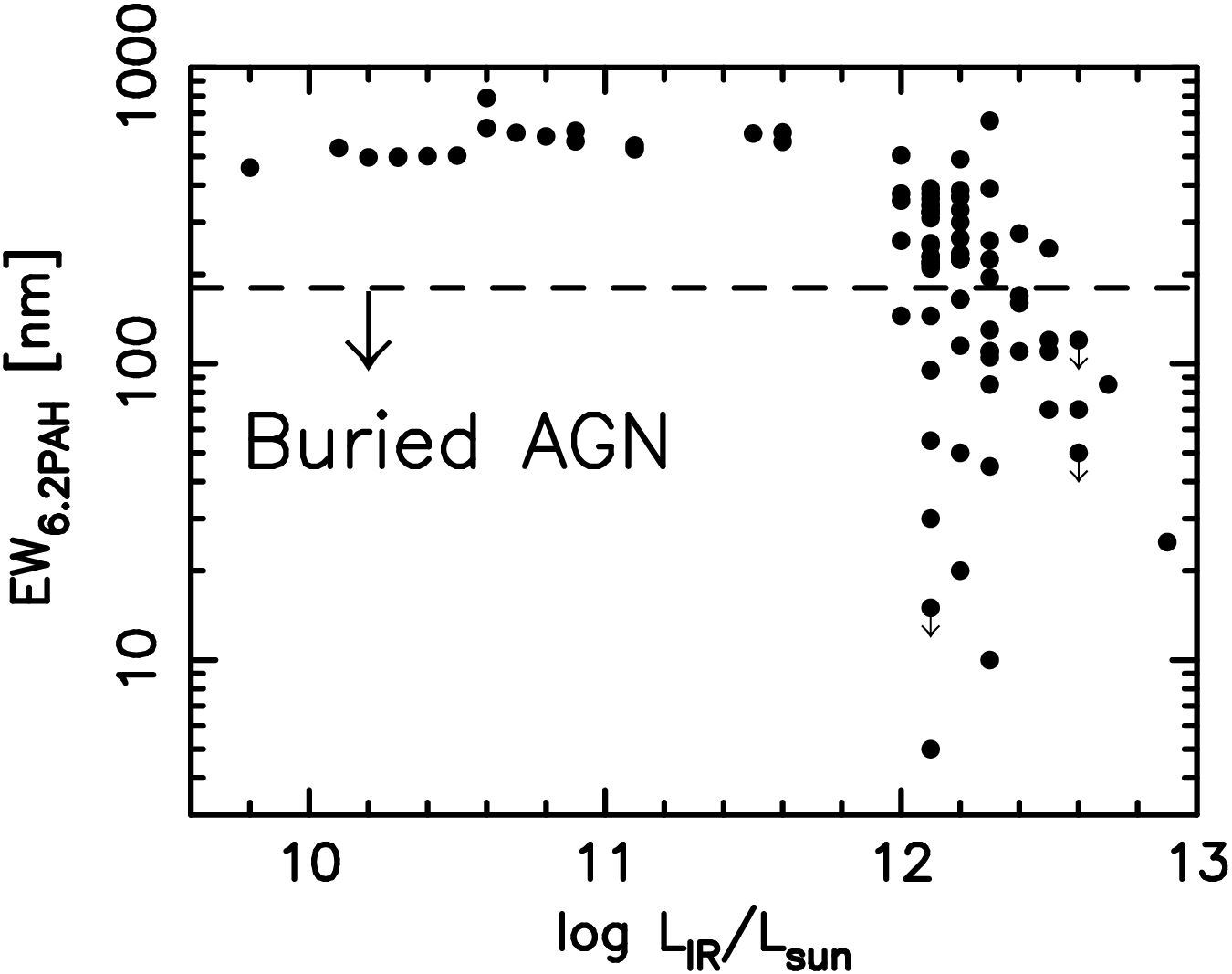}  \hspace{0.3cm}
\includegraphics[angle=-90,scale=.35]{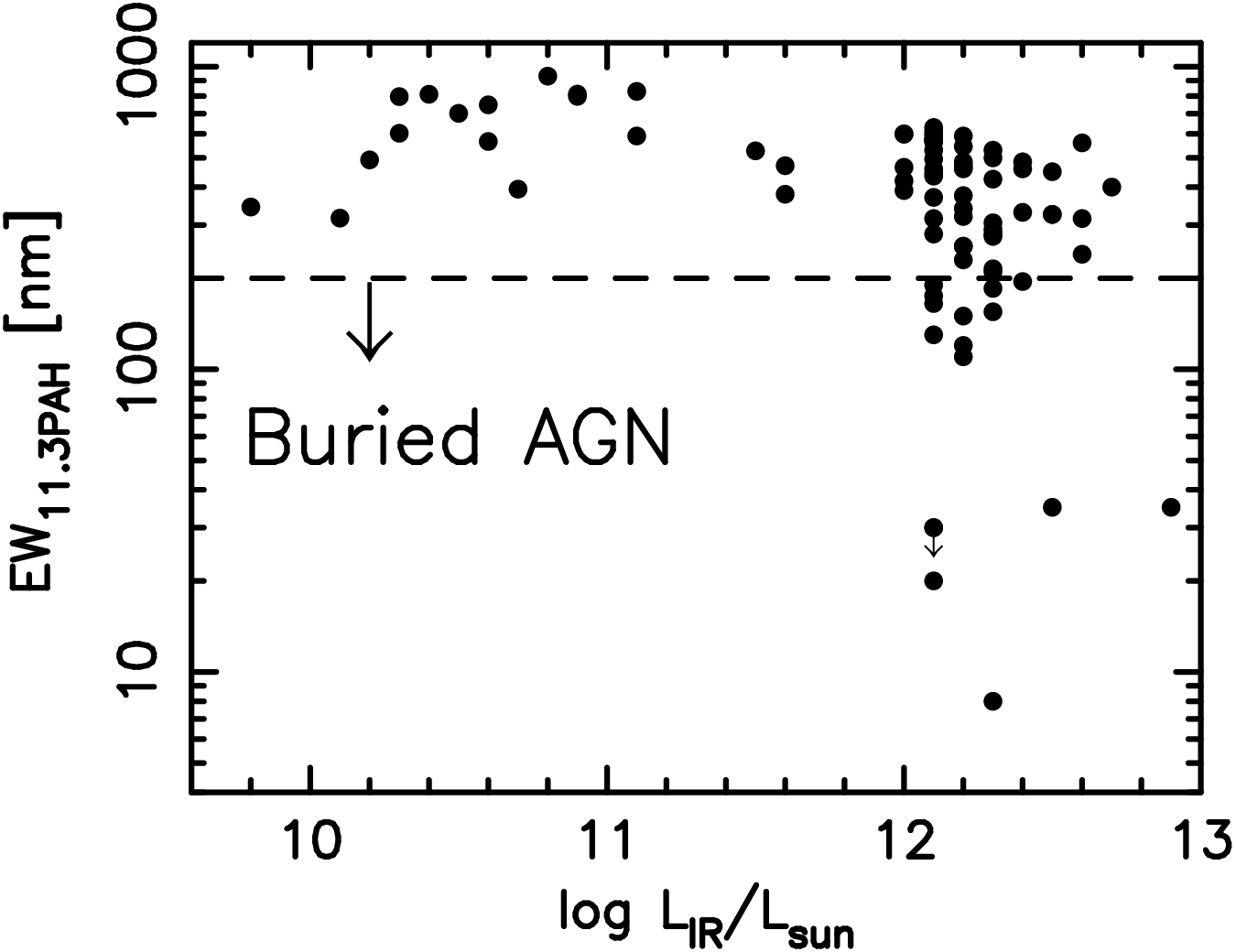} \\
\includegraphics[angle=-90,scale=.35]{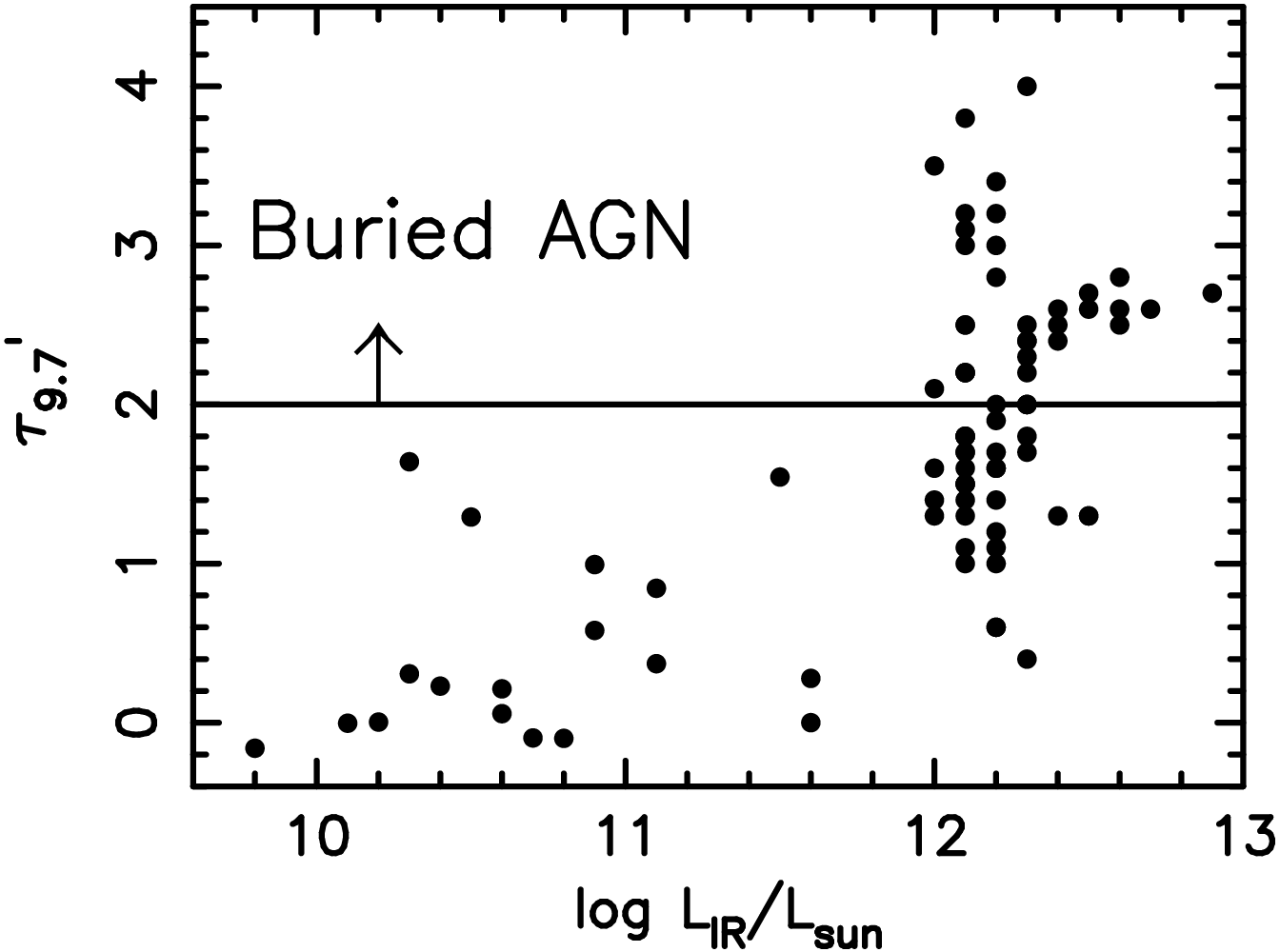} 
\caption{
Distribution of {\it (a)} EW$_{\rm 6.2PAH}$, 
{\it (b)} EW$_{\rm 11.3PAH}$, and {\it (c)} $\tau_{9.7}'$, as a function
of galaxy infrared luminosity. 
Plotted sources are optically non-Seyfert ULIRGs (This paper; Imanishi 
et al. 2007) and optically non-Seyfert galaxies with lower infrared
luminosities \citep{bra06}. 
The horizontal dashed lines indicate the threshold to become buried AGN
candidates ($\S$5.2.1 and 5.2.2).
} 
\end{figure}

\begin{figure}
\includegraphics[angle=-90,scale=.6]{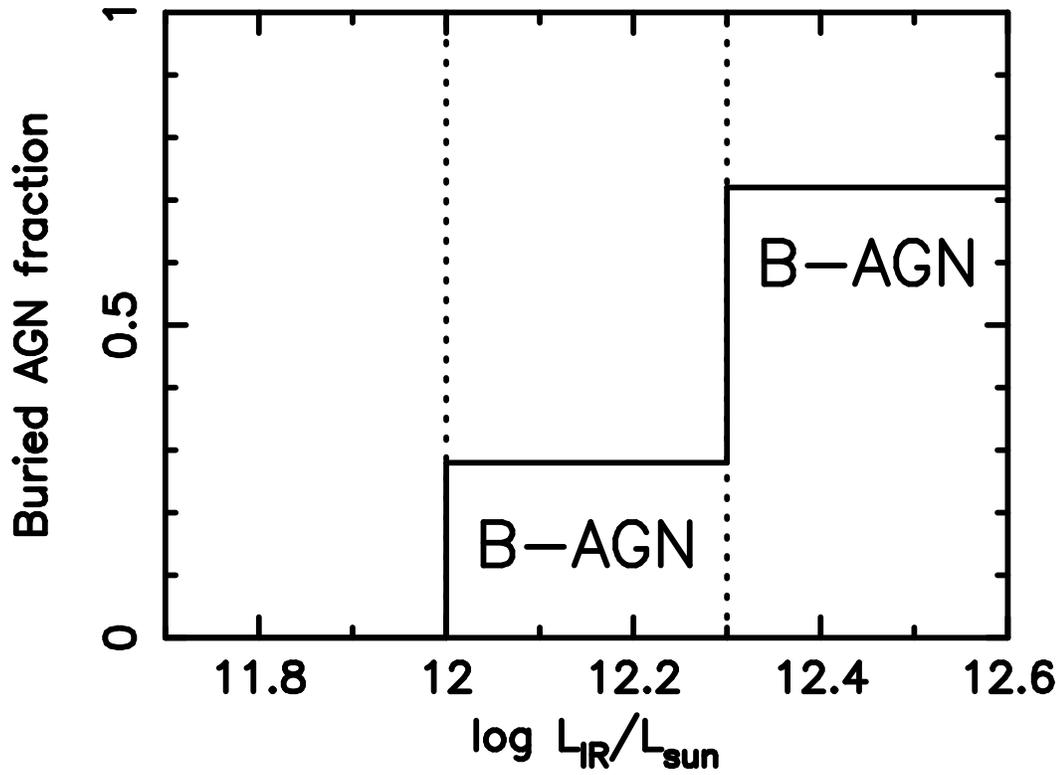} 
\caption{
Fraction of sources with clearly detectable buried AGN signatures as a
function of galaxy infrared luminosity. 
For ULIRGs (L$_{\rm IR}$ $>$ 10$^{12}$L$_{\odot}$), the fraction is
derived from this paper. 
For galaxies with L$_{\rm IR}$ $<$ 10$^{12}$L$_{\odot}$, it is from the
limited number of sample in \citet{bra06}, and so does not necessarily
mean that no buried AGNs are present.
} 
\end{figure}

\end{document}